\documentclass[11pt]{amsart}
\usepackage{amsfonts}
\usepackage{amsmath,amsthm}
\usepackage{latexsym}
\usepackage{array}
\usepackage{amssymb}
\usepackage{enumerate}
\usepackage{comment}
\usepackage{graphicx}
\usepackage{tikz}

\usepackage{hyperref}
\hypersetup{colorlinks,
	citecolor=blue,
	linkcolor=blue,
	urlcolor=blue}

\DeclareFontFamily{U}{wncy}{}
\DeclareFontShape{U}{wncy}{m}{n}{<->wncyr10}{}
\DeclareSymbolFont{mcy}{U}{wncy}{m}{n}
\DeclareMathSymbol{\Sh}{\mathord}{mcy}{"58}

\usepackage[english]{babel}
\usepackage{fdsymbol}
\usepackage{color}
\usepackage[latin1]{inputenc}

\numberwithin{equation}{section}
\theoremstyle{plain}
\newtheorem{theorem}{Theorem}[section]
\newtheorem*{theorem*}{Theorem}
\newtheorem{lemma}[theorem]{Lemma}
\newtheorem{proposition}[theorem]{Proposition}

\newtheorem{definition}[theorem]{Definition}
\newtheorem{problem}[theorem]{Problem}

\theoremstyle{remark}
\newtheorem{remark}[theorem]{Remark}

\newtheorem*{lem*}{Lemma}
\newtheorem*{sublem*}{Sublemma}
\newtheorem*{remark*}{Remark}
\newtheorem*{NB*}{NB}

\newcommand{\Sst}{{}^s\Sigma_*}

\newcommand{\R}{ \mathbb{R} }

\newcommand{\C}{ \mathbb{C} }
\newcommand{\Z}{ \mathbb{Z} }
\newcommand{\N}{ \mathbb{N} }

\newcommand{\T}{ \mathbb{T} }

\newcommand{\mI}{\mathbb{I}}

\newcommand{\gD}{\mathfrak D}
\newcommand{\gF}{\mathfrak F}
\newcommand{\gC}{\mathfrak C}

\newcommand{\Del}{\Delta}

\newcommand{\cA}{ \mathcal{A} }
\newcommand{\cB}{ \mathcal{B} }

\newcommand{\cY}{ \mathcal{Y} }

\newcommand{\cK}{ \mathcal{K} }
\newcommand{\cF}{ \mathcal{F} }

\newcommand{\EE}{ {\mathbb E}}

\newcommand{\Lc}{ \mathcal{L} }
\newcommand{\cZ}{ \mathcal{Z} }

\newcommand{\cN}{ \mathcal{N} }

\newcommand{\Rc}{ \mathcal{R} }

\newcommand{\cJ}{ \mathcal{J} }
\newcommand{\I}{ \mathcal{I} }

\newcommand{\om}{ \omega }

\newcommand{\Om}{ \Omega }

\newcommand{\ga}{\gamma }

\newcommand{\s}{ \sigma }

\newcommand{\ka}{ \kappa }

\renewcommand{\phi}{ \varphi }
\newcommand{\oms}{ \omega^{12}_{3s}  }
\newcommand{\omch}{ \omega^{12}_{34}  }

\newcommand{\eps}{\varepsilon}

\newcommand{\de}{ \delta }
\newcommand{\al}{ \alpha }

\newcommand{\la}{ \lambda }

\newcommand{\dess}{\delta'^{12}_{3s}}

\newcommand{\be}{\begin{equation}}
\newcommand{\ee}{\end{equation}}
\newcommand{\ben}{\begin{equation*}}
\newcommand{\een}{\end{equation*}}

\newcommand{\ov}{ \overline }
\newcommand{\z}{ \overline{z} }

\newcommand{\lan}{ \langle }

\newcommand{\ran}{ \rangle}

\newcommand{\p}{ \partial}

\newcommand{\wt}{ \widetilde }

\newcommand{\abs}[1]{\big\vert #1 \big\rvert}

\newcommand{\lbl}{\label}
\newcommand{\non}{\nonumber}
\newcommand{\qu}{\quad}
\newcommand{\qmb}{\quad\mbox}
\newcommand{\qnd}{\qmb{and}\qu}

\newcommand{\cS}{\mathcal{S}}

\newcommand{\volna}{\thicksim}

\newcommand{\az}[1]{a^{(0)}_{#1}}
\newcommand{\ao}[1]{a^{(1)}_{#1}}

\newcommand{\bz}[1]{\bar a^{(0)}_{#1}}
\newcommand{\bo}[1]{\bar a^{(1)}_{#1}}

\newcommand{\Ga}{\Gamma}
\newcommand{\fa}{\frak{a}}

\newcommand{\sm}{\setminus}

\newcommand{\dep}{\delta'^{12}_{3s}}

\title[Stochastic model for wave turbulence 2: diagrams]
{Formal expansions in stochastic model for wave turbulence 2: method of diagram decomposition
 \\ \normalfont extended version}

\author{Andrey  Dymov}
\address{Steklov Mathematical Institute of RAS, 119991 Moscow, Russia 
	\& National Research University Higher School of
	Economics, 119048 Moscow, Russia
\& Skolkovo Institute of Science and Technology, 143005 Skolkovo, Russia} \email{dymov@mi-ras.ru}
\author{Sergei Kuksin}
\address{Universit\'e Paris-Diderot (Paris 7), UFR de Math\'ematiques - Batiment Sophie Germain, 5 rue Thomas Mann, 75205 Paris,
	France  \& School of Mathematics, Shandong University, Jinan, PRC }
\email{ Sergei.Kuksin@imj-prg.fr}

\begin{document}

	\begin{abstract}  
	In this paper we 
	continue to study  small amplitude solutions of the damped cubic NLS equation, driven by a random force (the study was 
	initiated in our previous work  \cite{DK} and continued in \cite{DKMV}). 
	We write solutions of the equation as formal series  in the amplitude and discuss the behaviour of this series under
	the wave turbulence limit, when the amplitude goes to zero, while the space-period goes to infinity. 
\end{abstract}

\maketitle
\date{}

%\tableofcontents
%	\newpage

% \section{Introduction  }\label{s1}

\section{Introduction and results}\label{s2}

We continue the  study of the   nonlinear cubic Shr\"odinger equation (NLS) 
on a torus of large period, dumped by a viscosity and driven by a random force, initiated  in  our papers  \cite{DK, DKMV}. 
Results of this work are crucially used in \cite{DK, DKMV} and are of independent interest for further study of 
wave turbulence  (WT) in the equation which we consider and in similar ones. 
In particular, our approach applies to the NLS equations with higher order nonlinearities.

Below we recall the setup, state the main theorem and then discuss its relevance for works \cite{DK, DKMV} and for the theory of WT. 
%Mwave turbulence in the equation which which consider.
The theorem's proof is outlined in Section~\ref{s:p and o} and is given in details
in Sections~\ref{s:D}-\ref{est-imroved}. In Section~\ref{s_add} we establish some lemmas, used in the proof, 	and discuss a number of
related results. 

For a general theory of  wave turbulence we refer to works \cite{Naz11,ZLF92, NR}, and for an additional discussion of the relevance of
our results for WT -- to the introduction in \cite{DK}.

\subsection{The setting}

Let ${\mathbb{T}}^d_L={  \mathbb{R}   }^d/(L{  \mathbb{Z}   }^d)$ be a $d$-dimensional torus, $d\geq 1$, 
of  period $L\geq 1$. 
We denote by $\|u\|$ the normalized $L_2$-norm of a complex function $u$ on $\T_L^d$, 
$
\|u\|^2 =
L^{-d}\int_{{{\mathbb T}}^d_L} |u(x)|^2\,dx\,,
$
and write the Fourier series of $u$ as \footnote{ 
	The scaling factor $ L^{-d/2} $ is convenient for our calculations.}
\begin{equation}\lbl{Fourier-def}
u(x)= L^{-d/2} \,{\sum}_{s\in{{\mathbb Z}}^d_L} v_s e^{2\pi  i s\cdot x}, 
\qquad\qu{{\mathbb Z}}^d_L = L^{-1} {{\mathbb Z}}^d\,.
\end{equation}
Here  the vector of  Fourier coefficients  $v= (v_s)_{s\in\Z^d_L}$ 
is given by the Fourier transform of $u(x)$, 
$$
v=\cF(u), \quad v_s = L^{-d/2}\,\int_{\T^d_L} u(x) e^{-2\pi i s\cdot x}\,dx \quad \text{for}\; s\in \Z_L^d. 
$$
The Parseval identity reeds  
$
\|u\|^2 = L^{-d}\,{\sum}_{s\in\Z^d_L} |v_s|^2.
$
We will study  solutions $u(t,x)$ whose norms  satisfy  $\|u(t,\cdot)\|\sim 1$ as $L\to\infty$. 

Our goal is to study  the cubic NLS equation with modified nonlinearity 
\be\label{NLS}
\frac{{{\partial}}}{{{\partial}} t}  u +i\Delta u- i\la \,\big( |u|^2 -2\|u\|^2\big)u=0 , 
\qquad x\in {\mathbb{T}}^d_L,
\ee
where $u=u(t,x)$,  $ \Delta=(2\pi)^{-2}\sum_{j=1}^d ({{\partial}}^2 / {{\partial}} x_j^2)$ and 
$\la\in(0,1]$ is a small parameter. 
The modification of the nonlinearity by the term $2i\la\|u\|^2u$  keeps the main features of the standard cubic NLS equation, reducing 
some non-crucial technicalities, see the introduction to \cite{DK} for  a detailed discussion. 
% Note that, if $u(t,x)$ satisfies  equation \eqref{NLS}, then $u'= e^{2it\nu \| u \|^2}u$ is a solution to a  standard cubic NLS equation.

The objective of WT is to study solutions of \eqref{NLS}  under the wave turbulence limit $L\to\infty$ and $\la\to 0$ on long time intervals, 
usually while  "pumping the energy to low modes and dissipating it in high modes".  
To make this rigorous, following Zakharov-L'vov \cite{ZL75},
we consider the NLS equation \eqref{NLS} dumped by a (hyper) viscosity and driven by a random force:
\begin{equation}\label{ku3s}
\frac{{{\partial}}}{{{\partial}} t}  u +i  \Delta u  - i\la\,\big( |u|^2 -2\|u\|^2\big)u =  -\nu\frak A(u)    + \sqrt\nu\frac{{{\partial}}}{{{\partial}} t} \eta^\omega(t,x).
\end{equation}
Here $\nu\in(0,1/2]$ is a small parameter and $\frak A$ is the dissipative linear operator 
\be\lbl{diss_op}
\frak A(u(x)) = L^{-d/2}{\sum}_{s\in \Z^d_L} \ga_s v_s e^{2\pi is\cdot x},  \quad v= \cF(u),\;\;
\ga_s =\ga^0(|s|^2).
\ee
The real function  $\ga^0(y)\ge 1$ is  smooth and 	 has at most a polynomial growth at infinity, together with all  its derivatives. 
\footnote{For example  $\frak A= (1-\Delta)^{r^*}$ for some $r^*\ge0$.} 
The random noise  $\eta^\omega$  is given by the Fourier series
$$
\eta^\omega(t,x)=
L^{-d/2} \, {\sum}_{s\in\Z^d_L} b(s) \beta^\omega_s(t) e^{2\pi  i s\cdot x},
$$
where
$\{\beta_s(t), s\in{{\mathbb Z}}^d_L\}$ are standard independent complex Wiener processes\,\footnote{
	i.e. $\beta_s = \beta_s^1 + i\beta_s^2$, where $\{\beta_s^j, s\in{{\mathbb Z}}^d_L, j=1,2  \}$ are
	standard independent real Wiener processes. 	} and   $b(x)$  is a Schwartz  function on ${{\mathbb R}}^d\supset \Z^d_L$.
\footnote{Often it is assumed that  the intensity  $b(x)$ of the noise $\eta^\omega$ is non-negative, but we do not impose this condition. 
	Still note that if $b(x)\equiv 0$, then our results become trivial since below we will provide \eqref{ku3s} with the zero initial conditions.}
If the function $\ga^{0}(y)$ grows at infinity as $|y|^{r_*}$ with $r_*$ sufficiently big, then 	 eq.~\eqref{ku3s} is well posed. In difference with
\cite{DK,DKMV}, where we assumed $d\geq 2$, in this work we allow $d=1$ and impose less restrictions on the function $\ga^{0}$.

We will study the equation on 	time intervals of order  $\nu^{-1}$, so it is convenient to pass  from $t$ 
to the slow time $\tau=\nu t$. Then eq.~\eqref{ku3s} takes the form
\begin{equation}\label{ku3}
\begin{split}
\dot u +i \nu^{-1} \Delta u  - i\rho\,\big( |u|^2 -2\|u\|^2\big)u &= -\frak A(u)    + \dot \eta^\omega(\tau,x),\\
\eta^\omega(\tau,x)&=
L^{-d/2} \, {\sum}_{s\in\Z^d_L} b(s) \beta^\omega_s(\tau) e^{2\pi  i s\cdot x},
\end{split}
\end{equation}
where $\rho=\la\nu^{-1}$, the upper-dot stands for $d/d\tau$  
%the derivative with respect to  the slow time \nolinebreak{$\tau$} 
and $\{\beta_s(\tau), \, s\in\Z^d_L\}$ is another set of standard independent complex Wiener processes.  
Below we use   $\rho$, $\nu$ and $L$ 	 as the parameters of the equation.

In the context of eq.~\eqref{ku3}, 
the objective of WT is to study its solutions $u$ when
\be\label{lim}
L\to\infty \;\;\; \text{and}\;\; \; \nu\to 0,
\ee
as well as  main characteristics of the solutions 
such as the {\it energy spectrum}  
\be\lbl{spec_en}
n_s(\tau): =  {{\mathbb E}} |v_s (\tau) |^2,   \qquad s\in \Z^d_L, \;\; \text{
	where }v=\cF(u).
\ee 
In  \cite{DK, DKMV} and in 	 this paper we study formal decompositions in $\rho$ of solutions to eq. \eqref{ku3},  regarding $\rho$ 
as an independent parameter, and those of the energy spectrum $n_s$. 
In  Section~\ref{s:mot}, where  we present the  results of \cite{DK} and \cite{DKMV}, 
we assume that $\rho\sim\nu^{-1/2}$ or $\rho~\sim L$ correspondingly, since due to  \cite{DK} and \cite{DKMV} exactly under these scalings  the  quadratic in $\rho$ part of the decomposition of the energy spectrum
has a non-trivial behaviour under the  limit \eqref{lim}, taken in the regimes, considered  in \cite{DK} and \cite{DKMV}.

\subsection{The result}

Let us  take the Fourier transform  of eq.~\eqref{ku3}:

\begin{equation}\label{ku33}
\dot v_s -i\nu^{-1}|s|^2v_s + \gamma_s v_s
=  i\rho  L^{-d}  \Big({\sum}_{s_1,s_2,s_3\in\Z^d_L} \dep v_{s_1} v_{s_2} \bar v_{s_3}- |v_s|^2v_s\Big)+b(s) \dot\beta_s \,,
\end{equation}
$s\in{{\mathbb Z}}^d_L$, where $|s|$ stands for  the Euclidean norm of a vector $s$ and 

\be\lbl{def_del}
\dep={\delta'}^{s_1s_2}_{s_3s} =
\left\{\begin{array}{ll}
	1,& \text{ if $s_1+s_2=s_3+s$ and $\{s_1, s_2\} \ne \{s_3, s\}$}
	\,,
	\\
	0,  & \text{otherwise}.
\end{array}\right.
\ee
%\begin{comment}
Note that 
\be\label{notethat}
\text{ if  $\dess =1$, 
	%$\{s_1, s_2\}\cap \{s_3, s_4\}\ne\emptyset$,
	then % ${\delta'}^{1 2}_{3 4} =0$. }
	$\{s_1, s_2\}\cap \{s_3, s\} \not\equal \emptyset$.}
\ee
In  view of the factor  ${\delta'}^{12}_{3s} $,
in  the   sum in \eqref{ku33} the variable  $s_3$ is a function of $s_1, s_2, s$.  That is why below we write the sum as
$$
{\sum}_{s_1,s_2,s_3\in\Z^d_L} {\delta'}^{12}_{3s}=:
{\sum}_{s_1}{\sum}_{s_2} {\delta'}^{12}_{3s} .
$$
%\end{comment} 
We fix $0 \leq  T\leq  +\infty$ and study equation \eqref{ku3}=\eqref{ku33} with  zero initial condition  at $\tau=-T$,  so
\be\lbl{ini_cond}
v_s(-T)=0 \qmb{for all}\qu s\in\Z^d_L.
\ee 
Let us write solutions of \eqref{ku33}, \eqref{ini_cond}  as formal series in $\rho$: 
\footnote{As we  mentioned above, in paper \cite{DK} we choose  $\rho\sim \nu^{-1/2}$ to be large. However, the first $d$ coefficients $v_s^{(i)}$ of series \eqref{decomp} are in fact of order  $\lesssim \nu^{i/2}$ and we expect that  the others also satisfy this estimate, see in Section~\ref{s:meaning}. So there is a hope that the series converges. A similar situation takes place in paper \cite{DKMV}, where we assume $\rho\sim L$.}
\be\label{decomp}
v_s=\sum_{i=0}^\infty \rho^i v_s^{(i)},\qquad v_s^{(i)}(-T)=0,
\ee
where $v_s^{(i)}=v_s^{(i)}(\tau;\nu,L)$.  The processes $v_s^{(0)}$ satisfy the linear equations
\be\lbl{v^0}
\dot v_s^{(0)}(\tau) -i\nu^{-1}|s|^2v_s^{(0)}(\tau) + \gamma_s v_s^{(0)}(\tau)
=  b(s) \dot\beta_s(\tau), \qquad v^{(0)}_s(-T)=0,
\ee
so 	these are independent  Gaussian processes. 
The linear in $\rho$ term $v_s^{(1)}$ satisfies 
\be\lbl{v^1}
\dot v^{(1)}_s -i\nu^{-1}|s|^2v^{(1)}_s + \gamma_s v^{(1)}_s
=  i L^{-d}  \Big( {\sum}_{s_1,s_2,s_3} \dep v^{(0)}_{s_1} v^{(0)}_{s_2} \bar v^{(0)}_{s_3} - |v_s^{(0)}|^2v_s^{(0)}\Big),
\ee
and $v^{(1)}_s(-T)=0$. Other terms $v_s^{(n)}$ can be found recursively in a similar way.

Our goal  is to study correlations of the terms of   series \eqref{decomp}  under the limit \eqref{lim}. 
Repeating the argument of  Lemma~2.3 from \cite{DK} it is straightforward to show that  
$\EE v_s^{(m)}(\tau_1) v_{s'}^{(n)}(\tau_2)=0$ for any $s,s'\in\Z^d_L$ and that 
$\EE v_s^{(m)}(\tau_1) \bar v_{s'}^{(n)}(\tau_2)=0$  for $s\neq s'$.
Our main result, given below in Theorem~\ref{t:mainpr}, establishes  upper bounds for the  correlations with $s=s'$. 
Its 	 particular case  is crucially used  in \cite{DK}, 	 see below in  Section~\ref{s:mot}.
Our second main result is Theorems~\ref{p:aman}, \ref{th:Ea^ma^n-cont}, which are steps in the proof of 
%forward to prove 
Theorem~\ref{t:mainpr}. There we develop an
instrumental representation of the correlations above in a form of explicit sums, convenient 
for further analysis. This result is crucially used in \cite{DKMV}. 

To state Theorem~\ref{t:mainpr} we use the function 
\be\lbl{chi^N_d}
\chi_d^{N}(\nu)=\left\{
\begin{array}{cl}
	-\ln\nu &\qu\mbox{if} \qu(N,d)=(3,2) \mbox{ or }(2,1)  \\
	1 &\qu\mbox{otherwise} 
\end{array}
\right.
\ee
(we recall that $0<\nu\le1/2$). 	For  an  $x\in\R$ we denote  
$
\lceil x \rceil  = \min\{n\in\Z: n\ge x\};
$
and denote by $C^\#(\cdot)$ various 
positive continuous functions on $\R^k$, $k\in\N$, fast decaying at infinity: 
%That is, $C^\#:\R^k\mapsto (0,\infty)$ is such that for any $n\geq 0$ 
\be\lbl{c_sharp}
C^\#(x) \leq c_n \,(1+|x|)^{-n} \qu \forall x\in\R^k,
\ee
for every $n\in\N$, with  a suitable constant $c_n>0$.

\begin{theorem}
	\lbl{t:mainpr}
	Assume that $d\geq 2$. Then for any integers $m,n\geq 0$ there exists a function $C^\#$ such that for each $s\in\Z^d_L$ and $\tau_1,\tau_2\geq -T$ we have
	\be\lbl{mainpr}
	\abs{\EE v_s^{(m)}(\tau_1) \bar v_s^{(n)}(\tau_2)}
	\leq C^{\#}(s)\big(\nu^{-2}L^{-2} + \nu^{\min(\lceil N/2\rceil,d)}\chi_d^{N}(\nu)\big),
	\ee
	where $N=m+n$.  
	% Moreover, the correlation $\EE v_s^{(m)}(\tau_1) \bar v_s^{(n)}(\tau_2)$ extends to a Schwartz function of $s\in\R^d\supset\Z^d_L$ which satisfies the same estimate \eqref{mainpr}.
	For  $d=1$  the assertion above remains true if we modify the r.h.s.  of \eqref{mainpr} by adding  the term $C^{\#}(s)L^{-1}$.
\end{theorem}

In our work  \cite{DK} estimate \eqref{mainpr} (with $d\geq 2$) is used in the situation  when $\nu^{-2}L^{-2} \leq  \nu^{\min(\lceil N/2\rceil,d)}$ -- then it 
implies that $\EE |v_s^{(m)}(\tau)|^2 \leq C^{\#}(s)\nu^{\min(m,d)}$, uniformly in $L$. 
The above restriction on the parameters 
well agrees with a postulate, usually accepted by the WT community,  that in \eqref{lim} 
$L$ should go fast to infinity when $\nu\to 0$.

To prove the theorem  we approximate, with  accuracy $\nu^{-2}L^{-2}$, the expectation
$\EE v_s^{(m)} \bar v_s^{(n)}$ by a finite sum of  integrals,  independent from $L$ and  parametrized by a suitable class $\gF_{m,n}$ of   Feynman diagrams  (the cardinality of the set  $\gF_{m,n}$   grows factorially with $m+n$).  Next we  study the behaviour of each integral as $\nu\ll 1$. 
To this end, in Section~\ref{s:ch}, for every diagram $\gF\in\gF_{m,n}$ we find a special coordinate system such that in these coordinates the integral, 
corresponding to $\gF$, has the  simple explicit form: 
\be\lbl{intro_ints}
\wt J_s(\tau_1,\tau_2;\gF) = \int_{\R^N} dl\int_{\R^{dN}} dz\, F^\gF_s(\tau_1,\tau_2;l,z) e^{i\nu^{-1}\Om^\gF(l,z)}
\ee 
with explicit  functions $\Om^\gF$ and $F^\gF_s$,
see Theorem~\ref{th:Ea^ma^n-cont}.
Here the  phase function $\Om^\gF$ is a real quadratic form in $z$ (usually degenerate) which 
linearly depends  on $l$. 
The density function $F^\gF_s$ is Schwartz in $(s,z)$, while as a function of $l$ it is piecewise smooth and fast decays as $|l|\to \infty$.  Then
in Sections~\ref{s:est} and \ref{est-imroved} we  show  that 
\be\non
|\wt J_s(\gF)| \le C^\#(s) \nu^{\min(\lceil N/2\rceil,d)}\chi_d^{N}(\nu)\qquad \forall \gF,
\ee
which implies \eqref{mainpr}.  
More precisely, there we prove abstract Theorem~\ref{t:est wt_J} where we get an upper estimate for a class of fast oscillating integrals of the form \eqref{intro_ints}.

The diagram decomposition and transformation of coordinates from Sections~\ref{s:ch} can be rather straightforwardly generalized to NLS equations with higher order polynomial nonlinearities. Together with abstract Theorem~\ref{t:est wt_J} this allows to generalize Theorem~\ref{t:mainpr} for these equations.
\medskip

Behaviour of  energy spectra  of small-amplitude solutions for nonlinear equations when the space-period of the  solution  grows to infinity 
%under the wave turbulence limit 
has been  intensively studied by physicists since the 1960's. Last decade this problem draw attention of mathematicians starting \cite{LS}, see 
\cite{BGHS18, BGHS, CG1, CG2, DH, DH1,  Faou, FGH}
and discussions in introduction to \cite{CG1, DK, DH1}. In some of the mentioned works 
decompositions of  energy spectra   to series of the form \eqref{decomp} where examined, evoking the 
techniques of Feynman diagrams,   inspired by the works \cite{EY00,ESY07} (where the diagram 
techniques were used to analyse expansions of solutions in different, but related situations).  
Our diagram presentation is based on similar ideas,  but is rather different from those in the mentioned works. 
In particular, as explained above, our approach leads to an explicit  presentation of the correlations $\EE v_s^{(m)}\bar v_s^{(n)}$ which
may be    treated  by asymptotical methods, developed in the present paper and in \cite{DK, DKMV}.

In the next two subsections we briefly present the results of \cite{DK, DKMV} and explain the role of the 
present paper for the research started in \cite{DK, DKMV}.

\subsection{Kinetic limit}
\lbl{s:mot}
Let  $u(\tau,x)$ be a  solution of equation~\eqref{ku3}, so that $v(\tau)=\cF\big( u(\tau)\big)$ is
a solution of \eqref{ku33},  
and $n_s(\tau)$ be its energy spectrum (see \eqref{spec_en}). 
One of 	objective of  WT is to study the 
behaviour  of the energy spectrum 
under the limits \eqref{lim},  with a properly scaled $\rho$, see \cite{ZL75,ZLF92,Naz11}. 
The order of  limits is not quite clear   but, as we mentioned above, 
in WT it is usually accepted that $L$   grows to infinity fast when $\nu$ goes to zero 
(for a discussion of other possibilities see introductions to \cite{DK} and \cite{DKMV} and see below).
Accordingly  in \cite{DK} we assume  that
\be\lbl{L,nu}
L\geq \nu^{-2-\de} \qu\mbox{for some}\qu \de>0,
\ee	
or that first $L\to\infty$ and then $\nu\to 0$. It is traditional in WT to study the quadratic in $\rho$ truncation of  series \eqref{decomp}, 
postulating that it well approximates small amplitude solutions. Thus motivated, in \cite{DK} we considered the process 
\be\lbl{quasisol}
\wt v_s(\tau) := v_s^{(0)}(\tau) +\rho v_s^{(1)}(\tau)  +\rho^2 v_s^{(2)} (\tau) , \quad s\in\Z^d_L, 
\ee
which we called a {\it quasisolution} of equation \eqref{ku33}, \eqref{ini_cond},
and examined its energy spectrum $\wt n=(\wt n_s)_{s\in\Z^d_L}$, $\wt n_s= \EE|\wt v_s|^2$,
\be\lbl{spec cut off}
\wt n_s(\tau)= n_s^{(0)}(\tau) + \rho n_s^{(1)}(\tau) +\rho^2 n_s^{(2)}(\tau) + \rho^3 n_s^{(3)}(\tau) + \rho^4 n_s^{(4)}(\tau),
\ee
	where 
$$
\displaystyle{n_s^{(k)} = \sum_{\substack{0\leq k_1, k_2 \leq 2: \\
			k_1+k_2=k
	}} \EE v_s^{(k_1)}\bar v_s^{(k_2)}}. 
$$
Assuming $d\geq 2$, 
we proved that $n_s^{(2)}\volna \nu$ as $\nu\to 0$ uniformly in $L$ satisfying \eqref{L,nu}, while a simple computation showed that 
$n_s^{(0)}\sim 1$ and $n_s^{(1)}=0.$ 
This indicates that the right scaling for $\rho$ is such that $\rho^2 \nu\volna 1$ as $\nu\to 0$, and  accordingly in 
\cite{DK}  
we 
chose $\rho$ to be of  the form 
\be\lbl{rho-scale}
\rho=\sqrt{\eps}\nu^{-1/2},\qmb{so that}\qu \rho^2\nu=\eps.
\ee  
Here $0<\eps\le1$ is a fixed small number (independent from $\nu$ and $L$). Using  estimate \eqref{mainpr} 
under the scaling \eqref{L,nu},~\eqref{rho-scale}   
we found that $|\rho^3 n_s^{(3)}|,$ $ |\rho^4 n_s^{(4)}|\lesssim\eps^2$.  
Accordingly, 
$$
\wt n_s=n^{(0)}_s + \eps \wt n^{(2)}_s(\eps), \quad\mbox{where}\quad \wt n_s^{(2)} = \nu^{-1} n_s^{(2)} + O(\eps)\,\sim1
$$
when $\nu\to0$, since $n_s^{(2)}  \sim\nu$. 

Thus, the  parameter $\eps$ measures the properly scaled amplitude of  oscillations, described by eq.~\eqref{ku3},  so
indeed it should be 
small for the methodology  of WT   apply to  the solutions.
Next we establish that for  $\nu$ small in terms of $\eps$
the energy spectrum $\wt n_s(\tau)$ is $\eps^2$-close to a function  $m(\tau,s)$, which is a unique solution of the 
damped/driven {\it wave kinetic equation}:
\be\label{WKE}
\dot m(\tau,s) = -2\ga_s m(\tau,s)+ 2 b(s)^2+ \eps K(m(\tau,\cdot))(s) ,  \;\; s\in \R^d;  \;\;\;
% m(-T) =0. 
\ee
$ m(-T) =0. $ 
Here  $K(m)(s)$ is the {\it wave kinetic integral} 
$$
2\pi\int_{\Sst} \frac{ds_1\, ds_2\mid_{\Sst}\, m_{1}m_{2}m_{3}m_{s}}{ \sqrt{|s_1-s|^2+|s_2-s|^2}}\left(
\frac1{ m_s} +\frac1{m_3} -\frac1{ m_1}-\frac1{ m_2}
\right),
$$
where $m_{i}=m(s_i)$, $m_s=m(s)$ with 
$s_3=s_1+s_2-s$ (cf. \eqref{def_del}). The set 	 $\Sst$ is the resonant surface 
(cf. \eqref{omega}-\eqref{ku4})
\be\lbl{res_surf}
\Sst=\{(s_1,s_2)\in\R^{2d}:\, |s_1|^2+|s_2|^2=|s_3|^2+|s|^2,\qu s_3=s_1+s_2-s\},
\ee
and $ds_1\, ds_2\mid_{\Sst}$ stands for  the  volume element on $\Sst$, corresponding to the  euclidean structure in  $\R^{2d}$.
This is exactly the wave kinetic integral which appears in physical works on  WT to describe the 
4--waves  interaction, see \cite{ZLF92}, p.~71 and \cite{Naz11}, p.~91.

In particular from what was said  it follows that in the regime \eqref{L,nu},  the  scaling 
$\rho\sim\nu^{-1/2}$ is the only scaling of $\rho$ 
when the energy spectrum admits a 
non-trivial kinetic limit of order one; see \cite{DK} for a  more detailed discussion.

\smallskip

In \cite{DKMV} we consider 
the opposite to \eqref{L,nu} order of limits, when first $\nu\to 0$ and then $L\to\infty$.  As in \cite{DK}, we study the energy spectrum \eqref{spec cut off} of the quasisolution, but now analysis of the term $n_s^{(2)}$ shows that the correct scaling is 
$\rho = \sqrt{\eps} L$.\,\footnote{In \cite{DKMV} the notation is slightly different: $\eps$ is written as $\eps^2$. Besides when 
	$d=2$, $\rho$ should be multiplied by the factor  $(\ln L)^{-1/2}$. }
In this setting, crucially using Theorem~\ref{p:aman}, we establish a result, 
similar to the discussed above that in \cite{DK}, but the 
limiting 
WKE \eqref{WKE} should be  modified. In particular, the WKE in  \cite{DKMV}  is non-autonomous.

Results of the present paper  also are used in  our work in progress, where by means of the KAM-technique we are aiming  to show 
that the results in \cite{DK, DKMV} remain true  for the energy spectrum of the exact solution for eq.~\eqref{ku3}, under the 
corresponding limits 
(and not only for the quasisolution). See subsection {\it What next?} in  \cite[Section~1.3]{DKMV}.

\subsection{Discussion of  results. Higher order truncations}
\lbl{s:meaning}

Let us come back to the decomposition \eqref{decomp} and accordingly write the energy spectrum $n_s$, defined in \eqref{spec_en}, as formal series in $\rho$:
\be\label{spec_form}
n_s = \sum_{k=0}^\infty \rho^k n^k_s, \qu\qquad n_s^k(\tau)=\sum_{k_1+k_2=k} \EE v_s^{(k_1)}(\tau)\bar v_s^{(k_2)}(\tau).
\ee
Here the  terms $n_s^0$, $n_s^1$ and $n_s^2$ coincide with  $n_s^{(0)}$, $n_s^{(1)}$ and  $n_s^{(2)}$  in \eqref{spec cut off}, 
while the terms $n_s^3$ and $n_s^4$ are 
slightly different (but still  satisfy the  estimate \eqref{n_s-est} below). According  to Theorem~\ref{t:mainpr}, 
\be\lbl{n_s-est}
|n_s^k|\leq C^{\#}(s)\big((\nu^{-2}L^{-2} + \nu^{\min(\lceil k/2\rceil,d)}\chi_d^{k}(\nu)\big)
\leq C_1^\#(s)\big(\nu^{-2}L^{-2} + \nu^{\min(k/2,d)}\big),
\ee
where for definiteness we assumed that  $d\geq 2$. For $k\leq 2d$ in the exponent in r.h.s. 
the minimum with $d$ can be removed, so  \eqref{n_s-est} takes the form
\be\lbl{est_imp}
|n_s^k|\leq C^\#(s)\big(\nu^{-2}L^{-2} + \nu^{k/2}\big).
\ee 
In particular, \eqref{est_imp} holds for $k\leq 4$ since $d\geq 2.$
Using estimate  \eqref{est_imp} with $k\leq 2d$ it is straightforward to show that under the scaling \eqref{rho-scale} 
the kinetic limit for quasisolution, established in \cite{DK}
and discussed in the previous section, takes place for the truncation $\sum_{i=0}^m \rho^i  v_s^{(i)}$ of series \eqref{decomp} not only of order $m=2$ 
as in \eqref{quasisol} but of any order $2\leq m\leq d$ (with the same kinetic equation  \eqref{WKE} and at least with 
the same accuracy $\eps^2$, once $L$ is sufficiently large in terms of $\nu^{-1}$).

If it was true that  \eqref{est_imp} is valid for any $k$, so that  $|\rho^kn_s^k|\lesssim\eps^{k/2}$ under the scaling \eqref{rho-scale} with $L\gg\nu^{-1}$, then the kinetic limit would hold  for  truncations of any order $m\geq 2$.
On the contrary,  if for some $k$  we have  $| n^k|_1:=L^{-d}\sum_s |n^k_s| \gg \nu^{k/2}$,
then this  is not the case since the term $\rho^kn^k$ explodes as $\nu\to 0$.

% Since, as explained above, the moments $\EE v_s^{(m)}\bar v_s^{(n)}$ can be 
% approximated by finite sums of integrals $\sum_{\gF}\wt J_s(\gF)$ with accuracy 
%  $ L^{-2}\nu^{-2}$,  then 
% they admit  finite limits  when $L\to\infty$ and $\nu$ stays fixed.
% So the limit $n_s^{k,\infty}(\tau):=\lim_{L\to\infty} n_s^k(\tau)$ also exists and satisfies 
% $$|n_s^{k,\infty}|\leq C^\#(s)  \nu^{k/2},$$
%  if \eqref{est_imp} holds for any $k$. 
% Then  the series $\sum_{k=0}^\infty \rho^kn_s^{k,\infty}$  with $\rho=\sqrt{\eps/\nu}$ becomes  a formal series in 
%  $\sqrt\eps$  uniformly in $\nu$, and it would be plausible that  it is an asymptotic of the limiting 
%  energy spectrum $\lim_{L\to\infty} n_s(\tau)$, where $n_s$ is defined in \eqref{spec_en}. 
%On the contrary, if 
% $| n^{k,\infty}|_1 \gg \nu^{k/2}$ for some $k$
%  then the series is not a formal series  uniformly in $\nu$.

The question if estimate \eqref{est_imp} holds  for high $k$ is  complicated. 
Namely, estimate \eqref{n_s-est} follows from \eqref{mainpr}, and in Theorem~\ref{t:mainpr} we established the latter 
by showing  that each integral $\wt J_s(\gF)$ is bounded by the r.h.s. of   \eqref{mainpr}.
We are able to prove that this estimate  is optimal at least for some  integrals $\wt J_s(\gF)$
in the sense that for some of  them  the minimum with $d$ in the exponent cannot be removed, so  for these terms 
estimate \eqref{est_imp} fails.  However, in the sum
$\sum_{\gF} \wt J_s(\gF)$ that approximates  $\EE v_s^{(m)}\bar v_s^{(n)}$ there are some non-trivial cancellations in a few first orders of its decomposition in $\nu$ 
which could lead to the validity of   \eqref{est_imp}. In Section~\ref{app:3} we discuss this phenomenon in more detail.

\medskip
 	
 	\noindent
 	{\bf Notation.}
 	 By $\Z_+^n$ we denote the set of integer  $n$-vectors with non-negative components.  
	 For  vectors in  $\R^n$ we denote by $|v|$ their  Euclidean norms and by $v\cdot u$ --  the 	 Euclidean scalar product.  	 
 	 By  $C^\#, C_1^\#,\wt C_1^\#, \dots$ we denote various 
 	 positive continuous functions  	 satisfying \eqref{c_sharp} for any $n\in\N$.
 	 If we want to indicate that a function $C^\#$ depends on a parameter $m$, we write it as $C^\#_m$. The functions $C^\#,C_1^\#,\ldots$ and constants $C,C_1,\ldots$ never depend on the  parameters $\nu,L,\rho,\eps$ and the moments of time $T,\tau,\tau_1,\ldots$ unless otherwise stated.
 	In Section 1.4 of \cite{DK} it is shown that
 	for  any function $C^\#(x,y)$, where $(x,y)\in\R^{d_1+d_2}$, $d_1, d_2\ge1$, 	there exist functions $C_1^\#(x)$, $C_2^\#(y)$  
 	such that
 	\be\label{useme}
	C^\#(x,y)  \le C_1^\#(x) C_2^\#(y).
	\ee

 	 \smallskip
 	 
 	\noindent 
 	{\bf Acknowledgements.} AD was funded by a grant from the Russian Science Foundation (Project 20-41-09009), and SK -- 
	by  Agence Nationale de la Recherche through  the grant    17-CE40-0006.
%the grant  18-11-00032 of Russian Science Foundation. 
 	
\section{Preliminaries and outline of the proof}
\lbl{s:p and o}

 \subsection{Approximate equation}

We start by getting rid in the  r.h.s. of equation \eqref{ku33}
 of the term $-i\rho L^{-d}|v_s|^2v_s$  since the latter is small but rather inconvenient for  analysis. With this term being dropped, equation \eqref{ku33}, \eqref{ini_cond} reads
\begin{equation}\label{ku33_y}
\begin{split}
\dot y_s &-i\nu^{-1}|s|^2y_s + \gamma_s y_s\\
&=  i\rho  L^{-d} {\sum}_{s_1,s_2,s_3\in\Z^d_L} \dep y_{s_1} y_{s_2} \bar y_{s_3}+b(s) \dot\beta_s \,,\;\;\;
y_s(-T)=0 ,\;\, s\in{{\mathbb Z}}^d_L. 
\end{split}
\end{equation}
%where $y_s(-T)=0$, $s\in{{\mathbb Z}}^d_L$.  
As in \eqref{decomp}, we decompose its solution $y_s$ into a formal series in $\rho$:
\be\label{decomp_y}
y_s=\sum_{i=0}^\infty \rho^i y_s^{(i)},\qquad y_s^{(i)}(-T)=0.
\ee
Then $y_s^{(0)}=v_s^{(0)}$ is a Gaussian process satisfying equation \eqref{v^0}, $y_s^{(1)}(\tau)$ satisfies equation \eqref{v^1} without 
 the term $|v_s^{(0)}|^2 v_s^{(0)}$, etc.   

In Section~\ref{s:app_eq_proofs} we prove that correlations of the processes  $v_s^{(i)}$ and $ y_s^{(i)}$ are $L^{-d}$-close:
\begin{proposition}\lbl{l:approx_eq}
For any $m,n\geq 0$ there exists a function $C^\#$ such that for each $s\in\Z^d_L$ and $\tau_1,\tau_2\geq -T$,
\be\lbl{corr_v-y}
\big|\EE v_s^{(m)} (\tau_1)\bar v_s^{(n)}(\tau_2)-\EE y_s^{(m)}(\tau_1)\bar y_s^{(n)}(\tau_2)\big|\leq L^{-d}C^\#(s).
\ee
\end{proposition}
In view of Proposition \ref{l:approx_eq} it suffices to establish  inequality \eqref{mainpr} (and its 1d analogy)  when the processes 
$v_s^{(m)}$ are replaced by $y_s^{(m)}$.  Accordingly below 
we study the latter 
%processes $y_s(\tau)$ 
instead of the former.
%$v_s(\tau)$.

\begin{comment}
Proposition 2.1 from \cite{DK} implies that for any $n\geq 0$ the processes $v^{(n)}_s(\tau)$ and $y_s^{(n)}(\tau)$ are $L^{-d}$-close: 
\be\lbl{v-y} 
\EE|v_s^{(n)}(\tau)-y_s^{(n)}(\tau)|^2 \leq L^{-2d}C_n^\#(s),
\ee 
uniformly in $\tau$ and $\nu$.
Theorem~\ref{t:main''}, formulated below, in particular implies that $\EE|y_s^{(n)}(\tau)|^2\leq C_n^\#(s)$. Then $\EE|v_s^{(n)}(\tau)|^2\leq C_n^\#(s)$ according to \eqref{v-y}, so the Cauchy  inequality implies 
\be\lbl{corr_v-y}
\big|\EE v_s^{(m)} (\tau_1)\bar v_s^{(n)}(\tau_2)-\EE y_s^{(m)}(\tau_1)\bar y_s^{(n)}(\tau_2)\big|\leq L^{-d}C_{n,m}^\#(s),
\ee
uniformly in  $\tau_1,\tau_2$ and $\nu$. Thus, it suffices to establish  inequality \eqref{mainn} (and its 1d analogy)  when the processes 
 $v_s^{(m)}$ are replaced by $y_s^{(m)}$. Accordingly below 
  we study the processes $y_s(\tau)$ instead of $v_s(\tau)$.
\end{comment}

\subsection{Interaction representation} 
\lbl{s:int_repr}
It is convenient to work in the {\it interaction representation}, that is in the $a$-variables defined as
\be\lbl{interaction r}
a_s(\tau)=y_s(\tau) e^{-i\nu^{-1}\tau |s|^2}, \qu s\in\Z^d_L.
\ee
We set
\be\label{omega}
\oms=\omega^{s_1s_2}_{s_3s}= |s_1|^2+|s_2|^2-|s_3|^2 - |s|^2,
\ee
(cf. \eqref{res_surf}).
Then equation \eqref{ku33_y}, written in the $a$-variables reads 
\begin{equation}\label{ku4}
\begin{split}
&\dot a_s + \gamma_s a_s
=  i\rho \cY_s(a,\nu^{-1}\tau) +b(s) \dot\beta_s,\qquad a_s(-T)=0,\\
&\cY_s(a, t )=L^{-d}  \sum_{s_1,s_2,s_3\in\Z^d_L} \dep a_{s_1} a_{s_2} \bar a_{s_3}
e^{  i t \oms}, \qquad s\in{{\mathbb Z}}^d_L,
\end{split}
\end{equation}
where $\{\beta_s\}$ is another set of standard independent complex Wiener processes.
Formal expansion \eqref{decomp_y} takes the form 
\be\non
a_s=\sum_{j=0}^\infty \rho^j a_s^{(j)},\qquad\qquad a_s^{(j)}(-T)=0,
\ee
where $
a_s^{(j)}(\tau)=y_s^{(j)}(\tau) e^{-i\nu^{-1}\tau |s|^2}.
$
For a process $a_s^{(j)}(\tau)$ we call the integer $j\ge0$ its {\it degree},
\be\lbl{deg-def}
\text{deg}\, a_s^{(j)}(\tau) = \text{deg}\, \bar a_s^{(j)}(\tau) = j,
\ee 
and call the vector $s\in \Z_L^d$ its {\it index}. 
 Since $a_s^{(k)}(\tau)$ differs from $y_s^{(k)}(\tau)$ by the  factor $e^{-i\nu^{-1}\tau |s|^2}$, then
  $|\EE a_s^{(m)} \bar a_s^{(n)}|=|\EE y_s^{(m)} \bar y_s^{(n)}|$. Thus, in view of \eqref{corr_v-y},
it suffices to establish Theorem~\ref{t:mainpr} with
the processes $v_s^{(m)}$ replaced by $a_s^{(m)}$:

\begin{theorem}
	\lbl{t:main''}
Assume that $d\geq 2$. Then for any integers $m,n\geq 0$ there exists a function $C^\#$ such that for 
 $s\in\Z^d_L$ and any $\tau_1,\tau_2\geq -T$,
\be\lbl{mainn}
\abs{\EE a_s^{(m)}(\tau_1) \bar a_s^{(n)}(\tau_2)}
\leq C^{\#}(s)\big(\nu^{-2}L^{-2} + \nu^{\min(\lceil N/2\rceil,d)}\chi_d^{N}(\nu)\big),
\ee
where $N=m+n$ and $\chi_d^N$ is defined in \eqref{chi^N_d}. 
If  $d=1$,  the assertion above holds  if we modify the r.h.s.  of \eqref{mainn} by adding the term $C^{\#}(s)L^{-1}$.
Moreover, 
the correlation $\EE a_s^{(m)}(\tau_1) \bar a_s^{(n)}(\tau_2)$ extends to a Schwartz function of $s\in\R^d\supset\Z^d_L$ which satisfies the same estimate \eqref{mainn}.
\end{theorem}
Note that for $N=1$ the l.h.s. of \eqref{mainn} vanishes, see in Section~\ref{s:F_not} and Lemma~2.3 in \cite{DK}. 
 
 The processes $a_s^{(i)}$ can be computed inductively. Namely, $a_s^{(0)}$ satisfies the linear equation 
 $$
 \dot a_s^{(0)}(\tau) + \gamma_s a_s^{(0)}(\tau)
 = b(s) \dot\beta_s(\tau),\qquad\qquad a_s^{(0)}(-T)=0\,,
 $$
so 
 \be\label{a0}
 a^{(0)}_s(\tau) = b(s) \int_{-T}^\tau e^{-\gamma_s(\tau-l)}d\beta_s(l),
 \ee
 and $\{  a^{(0)}_s(\tau) , s\in \Z^d_L\}$, are independent  Gaussian processes. 
 The process $a^{(1)}$ satisfies
 $$
 \dot a^{(1)}_s (\tau)+ \gamma_s a^{(1)}_s (\tau)
 =  i \cY_s(a^{(0)}(\tau),\nu^{-1}\tau), \qquad a_s^{(1)}(-T)=0,
 $$
 so
 \be\label{a1}
 \begin{split}
 a^{(1)}_s(\tau) 
 = i \int_{-T}^\tau e^{-\ga_s(\tau-l)} \cY_s(a^{(0)}(l),\nu^{-1}l)\,dl\,,
\end{split}
 \ee
 is a Wiener chaos of third order (see \cite{Jan}). 
 For what follows, it is convenient to introduce in \eqref{ku4} a fictitious index $s_4$  and write the function $\cY_s$
 from \eqref{ku4}   as 
 $$
 \cY_s(a, t )=L^{-d}  \sum_{s_1, \ldots,s_4\in\Z^d_L} \de'^{12}_{34}\, \de^s_{s_4} \,a_{s_1} a_{s_2} \bar a_{s_3}\,
 \theta(\om^{12}_{34},t),
% e^{  i t \om^{12}_{34}},
 $$   
 where $\de_{s_4}^s$ is the Kronecker symbol and
 we denote 
 \be\lbl{theta-def}
 \theta(x,t)=e^{i t\,x }.
 \ee
 Then  \eqref{a1} takes the form 
 \be\lbl{a1.1}
 a^{(1)}_s(\tau)= i \int_{-T}^\tau L^{-d} \sum_{s_1, \ldots,s_4} \de'^{12}_{34}\,e^{-\ga_{s_4}(\tau-l)}  \,\de^s_{s_4}
  (a^{(0)}_{s_1} a^{(0)}_{s_2}{\bar a}^{(0)}_{s_3})(l)\; \theta(\om^{12}_{34}, \nu^{-1}l)
 \,dl.
 \ee
 Similar, for $m\ge1$, 
 \be\label{an}
 \begin{split}
 	a^{(m)}_s&(\tau) 
 	= \sum_{m_1+m_2+m_3=m-1} i \int_{-T}^\tau   dl \\
 	&\times L^{-d}\sum_{s_1, \ldots,s_4}
 	 \de'^{12}_{34}\,e^{-\ga_{s_4}(\tau-l)} \,
 	\de^s_{s_4}\big( a_{s_1}^{(m_1)} a_{s_2}^{(m_2)} {\bar a_{s_3}}^{(m_3)}\big)(l)\;
 	\theta(\om^{12}_{34},\nu^{-1}l).
 \end{split}
 \ee
 This is a Wiener chaos of order $2m+1$. 
 \medskip

	In what follows, till the end of Section~\ref{s:ff-exp}, we do not use the explicit form \eqref{theta-def} of the function $\theta$ (except the next section, where we give an outline of the proof of Theorem~\ref{t:main''}). Instead we assume that the processes $a^{(m)}_s$ are given by equation \eqref{an} where 
\be\label{theta}
\text{
	 $\theta:\R^2\mapsto\C$ is a bounded measurable  function.}
\ee 
	 In particular, we may choose  for $\theta$ the function $\theta(x,t)=\mI_{\{0\}}(x)$,
 where  $\mI_{\{0\}}(x)$ is the indicator function of the point $x=0$ (in this case $\theta$ does not depend on $t$). 
 This choice of  $\theta$ is used in our paper \cite{DKMV}, where we use that in some sense a
 properly scaled   function 
 $
 (s_1, s_2, s_3, s_4) \mapsto \exp(i \nu^{-1}\tau \omch)
 $
 with $\tau\ne0$, 
 regarded as a ``generalised function on $(\Z^d_L)^4$", converges to
 $\mI_{\{0\}}(\omch)$ as $\nu\to0$.

We will use  correlations of  the process $a_s^{(0)}$. By \eqref{a0}, 
for any $\tau_1,\tau_2\ge -T$ and $s,s'\in\Z^d_L$,
\be\label{corr_a_in_time}
 \begin{split}
&\EE a_s^{(0)}(\tau_1) a_{s'}^{(0)}(\tau_2)=0,\\
& \EE a_s^{(0)}(\tau_1) \bar a_{s'}^{(0)}(\tau_2)
= \delta^s_{s'} \,  \frac{b(s)^2}{\gamma_s} \big( e^{-\gamma_s|\tau_1-\tau_2|} -
e^{-\ga_s(2T + \tau_1+ \tau_2)}\big)
 \end{split}
\ee
(see (2.8) in \cite{DK} for the calculation). 

\subsection{Outline of the proof of Theorem \ref{t:main''}}
\lbl{s:outline}

We divide a quite detailed outline of the proof to 
four steps, approximately corresponding to Sections~\ref{s:D}, \ref{sec:FD}, \ref{s:ch} and \ref{s:est}+\ref{est-imroved}. In 
 the outline we assume that $\theta$ takes the form \eqref{theta-def}.

{\it Step 1.} Let us write the variable $a_s^{(m)}(\tau_1)$ in the form \eqref{an} and then apply the Duhamel formula \eqref{an} to the variables 
$a_{s_i}^{(m_i)}(l)$ in its r.h.s.  with the degrees $m_i\ge1$ (see \eqref{deg-def}).
Iterating the procedure, we represent  $a_s^{(m)}(\tau_1)$ by a sum of terms of the form 
\be\lbl{II1}
\wt I_s= \int\dots\int L^{-md}\sum_{s_1,\ldots,s_{4m}} (\dots) \,dl_1\dots dl_m;
\ee
here and below we call terms of this form {\it sums}.
The integrating zone in \eqref{II1} is a	 convex polyhedron in $[-T, \tau_1]^m$ and 
the summation is taken over the vectors $s_1,\ldots,s_{4m}\in\Z^d_L$
which are subject to certain linear relations, 
following from the factors $\de^{\prime12}_{34}$ and $\de^s_{s_4}$ in \eqref{an}. 
The summand in brackets is a product of functions 
$e^{-\gamma_{s'}(l_k-l_r)}$, $\exp(i\nu^{-1} \om^{s'_1 s'_2}_{s'_3 s'_4} l_q)$  and the processes $[a^{(0)}_{s^{''}}(l_j)]^*$  where $a^*$ is either $a$ or $\bar a$, 
with various indices $1\leq k,r,q,j \leq m$ and  $s',s'_i,s'' \in \{s_1,\ldots, s_{4m}\}$. 
This product is of degree $2m+1$ with respect to the process $a^{(0)}$  (see \eqref{a1.1} for the case $m=1$). 

\begin{figure}[t]
	\qquad a) \quad\parbox{3.5cm} {
		\begin{tikzpicture}[]
		\node at (-1,0) (c0) {$a_{s}^{(2)}(\tau_1)$};
		
		\node at (-2.1, -1.2)  {$a_{s_1}^{(0)}$};
		\node at (-1.4, -1.2)  {$a_{s_2}^{(0)}$};
		\node at (-0.7, -1.2) (c1) {$\bar a_{s_3}^{(1)}$};
		\node at (0, -1.3) (w2)  {$\bar w_{s_4}$};  
		
		\node at (-2.1, -2.5)  {$a_{s_5}^{(0)}$};
		\node at (-1.4, -2.6) (w4) {$w_{s_6}$};
		\node at (-0.7, -2.5)  {$\bar a_{s_7}^{(0)}$};
		\node at (0, -2.5)  {$\bar a_{s_8}^{(0)}$}; 
		
		\draw [line width=0.25mm] (c0.south)--(w2.north);
		\draw [line width=0.25mm](c1.south)--(w4.north);
		
		\node at (-2.9, -1.25) {$l_1:$};
		\node at (-2.9, -2.55) {$l_2:$};
		
		\end{tikzpicture}
	} 
	\hfill
	b) \quad \parbox{4.5cm} {
		\begin{tikzpicture}[]
		\node at (-1,0) (c0) {$a_{\xi_0}^{(2)}(\tau_1)$};
		
		\node at (-2.1, -1.2)  {$a_{\xi_1}^{(0)}$};
		\node at (-1.4, -1.2)  {$a_{\xi_2}^{(0)}$};
		\node at (-0.7, -1.2) (c1) {$\bar a_{\s_1}^{(1)}$};
		\node at (0, -1.3) (w2)  {$\bar w_{\s_2}$};  
		
		\node at (-2.1, -2.5)  {$a_{\xi_3}^{(0)}$};
		\node at (-1.4, -2.6) (w4) {$w_{\xi_4}$};
		\node at (-0.7, -2.5)  {$\bar a_{\s_3}^{(0)}$};
		\node at (0, -2.5)  {$\bar a_{\s_4}^{(0)}$}; 
		
		\draw [line width=0.25mm] (c0.south)--(w2.north);
		\draw [line width=0.25mm](c1.south)--(w4.north);
		
		\node at (-2.9, -1.25) {$l_1:$};
		\node at (-2.9, -2.55) {$l_2:$};
		
		\end{tikzpicture}
	}
	\caption{ A diagram $\gD\in\gD_2$. The notation "$l_i:$" means that the vertices $a^{(k)}$, $\bar a^{(k)}$ situated opposite to it are taken at the time $l_i$. }
	\lbl{f:F0}
\end{figure}

 To each sum $\wt I_s$ as in \eqref{II1} we associate a diagram $\gD$ as in fig.~\ref{f:F0}(a), constructed according to a rule, explained 
below; we then write $\wt I_s=\wt I_s(\gD)$. 
%equivalently, $\gD_m$ is a set  of all different diagrams, which can be constructed by the following rule. 
The root of a diagram $\gD$, associated with a sum $\wt I_s$, is the vertex $a^{(m)}_{s}(\tau_1)$.
If $a^{(p)}_{s'}(t)$, $p\geq 1$, is any vertex of the diagram $\gD$, then it generates  a  {\it block} of four vertices, three of which correspond
 to a choice of the three terms $a^{(m_1)}, a^{(m_2)} , \bar a^{(m_3)}$ in the 
decomposition \eqref{an} of $a^{(p)}_{s'}(t)$. These three  are called  {\it real vertices},  and the vertex $a^{(p)}_{s'}$ is
called the  {\it parent} of the block. 
The fourth vertex of the block, denoted by $\bar w_{s''}$, is called the {\it conjugated virtual vertex}. It corresponds to the factor $\de^{s'}_{s''}$ 
in \eqref{an},  is coupled with the parent $a^{(p)}_{s'}(t)$ by an edge  and one can think that its role 
 is to couple the block with its parent. 
Similarly, each vertex $\bar a^{(p)}_{s'}(t)$ with degree $p\geq 1$ generates a block, where the three real vertices correspond to a choice of the terms  $\bar a^{(m_1)}, \bar a^{(m_2)} , a^{(m_3)}$ in the  formula,  conjugated  to \eqref{an}, 
and the fourth vertex $w_{s''}$  is the {\it non-conjugated virtual vertex} (corresponding  to  factor $\de_{s''}^{s'}$). 

We denote by $\gD_m$ the set of all diagrams $\gD$ associated with different  terms $\wt I_s$ from \eqref{II1}, sum of which gives $a_s^{(m)}(\tau_1)$.
\footnote{As we will see, structure of the set $\gD_m$ depends on $a^{(m)}_s(\tau_1)$ through the index $m$ only. 
This will become more clear in Section~\ref{s:D}, where we will construct  $\gD_m$ carefully.} 
Then  in every diagram $\gD\in\gD_m$ there are $m$ blocks, and in
 each block there are two conjugated and two non-conjugated vertices. 
By convention, in each block we  draw and enumerate first the non-conjugated vertices and then the conjugated.
The virtual vertex is always positioned at the second or forth place, depending whether it is conjugated or not, see fig.~\ref{f:F0}(a).

We  denote by $\xi_1,\ldots,\xi_{2m}$ 
 the indices $s_j\in\Z^d_L$, enumerating the non-conjugated vertices, and by $\s_1,\ldots,\s_{2m}$ -- those 
 enumerating the conjugated vertices. Then, setting $\xi_0=s$, the diagram from fig.~\ref{f:F0}(a) takes the form as in fig.~\ref{f:F0}(b).
For more examples see fig.~\ref{f:F2}, \ref{f:F3}(a-c),
where we write $c$ instead of $a$ and omit indices $\xi$ and $\s$.

Now the 
 linear relations, imposed on the multi-indices  $\xi_i,\s_j$ with $i,j\ge1$   
 in a sum  \eqref{II1} take the following compact form:
\be\lbl{lin-rel} 
1)\ \de'^{\xi_{2j-1} \xi_{2j}}_{\s_{2j-1}\s_{2j}}=1 \;\forall j \qu
2)\ \mbox{indices of adjacent in $\gD$ vertices are equal,}
\ee
where the first relation comes from the factor $\de'^{12}_{34}$ in \eqref{an} while the second one~-- from the factor $\de_{s_4}^s$.
Moreover, we see that in this  notation the frequency of a fast rotating exponent, hidden in the term
 $(\dots)$ in \eqref{II1} and arising from the factor $ \theta(\om^{12}_{34},\nu^{-1}l)=e^{i\nu^{-1}\om^{12}_{34}l}$ in \eqref{an}, also takes the compact form $i\nu^{-1}\sum_{j=1}^m\,\om^{\xi_{2j-1}\xi_{2j}}_{\s_{2j-1}\s_{2j}}\,l_j$.

We have seen that
\be\non
a^{(m)}_s(\tau_1)=\sum_{\gD\in\gD_m} \wt I_s(\gD),
\ee
where each sum $\wt I_s(\gD)$ has the form  \eqref{II1}.
Similarly, 
$\bar a^{(n)}_s(\tau_2) = \sum_{\bar\gD\in\ov\gD_n} \wt I_s(\bar\gD)$, where  $\ov\gD_n$ is the set of diagrams with the root $\bar a^{(n)}_{\s_0}(\tau_2)$, $\s_0:=s$, constructed by the same inductive rule as diagrams from the set $\gD_m$ (see fig.~\ref{f:F3}(d)). 
\footnote{Equivalently, the set $\ov\gD_n$ can be obtained by conjugating diagrams from the set $\gD_n$, then exchanging positions of conjugated and non-conjugated vertices in each block (we recall that, by convention, the non-conjugated vertices should always situate before the conjugated vertices), and exchanging the indices $\xi_j$ with $\s_j$, so that again 
 the non-conjugated vertices are enumerated by the indices $\xi_j$ while the conjugated ~-- by the indices $\s_j$.}

Finally, we introduce the set $\gD_m\times\ov\gD_n$ of diagrams $\gD=\gD^1\sqcup\bar\gD^2$, which are obtained by 
drawing side by side
  pairs  $\gD^1$ and $\bar\gD^2$ with various 
 $\gD^1\in\gD_m$ and $\bar\gD^2\in\ov\gD_n$. Here in the diagrams $\bar\gD^2$ the vertices (except the root)
  are enumerated  by the indices   $\xi_{2m+1},\ldots,\xi_{2N},\s_{2m+1},\ldots,\s_{2N}$, $N:=m+n$. 
  Since the root of $\bar\gD^2$ is denoted $\sigma_0$, then all vertices of $\gD$ are enumerated by $4N+2$
  indices
  $
  \xi_i, \sigma_j,
  $
  where $0\le i,j \le 2N$ and   always $\xi_0=\sigma_0=s$;   see fig.~\ref{f:F4}.
Then  setting  $I_s(\gD)=\wt I_s(\gD^1)\wt I_s(\bar\gD^2)$, where $\gD=\gD^1\sqcup\bar\gD^2$, we find
 \be\lbl{i:Ea^ma^n}
\EE  a_s^{(m)}(\tau_1) \bar a_s^{(n)}(\tau_2) =\sum_{\gD\in \gD_m\times\ov\gD_n} \EE  I_s(\gD).
\ee
In fact, we do not write down the sums $\wt I_s$ but directly write the products $I_s$, see Lemma~\ref{l:a^ma^n} for their explicit form.

{\it Step 2.} 
To  estimate  the r.h.s.  of \eqref{i:Ea^ma^n} we study separately the expectation 
$\EE   I_s(\gD)$ for each diagram $\gD = \gD^1\sqcup\bar\gD^2
\in\gD_m\times\ov\gD_n$.  
 Due to \eqref{II1},
\be\lbl{II}
\EE   I_s(\gD) = \int\ldots\int L^{-Nd}\sum_{\xi_1,\ldots, \xi_{2N}, \s_1,\ldots,\s_{2N}}
\EE(\ldots)\,dl_1\ldots dl_{N},
\ee
where  the term in the brackets  is obtained by taking the product of the terms $(\ldots)$  in the  integrands of
$\wt I_s(\gD^1)$ and $\wt I_s(\bar\gD^2)$, corresponding to the leaves of the diagram $\gD$.
So, the brackets in \eqref{II} contain a product of $2N+2$ random variables of the form $[a_{s'}^{(0)}(t)]^*$, 
multiplied by a deterministic factor.
Since $ a_{s'}^{(0)}(t)$ are  independent complex 
Gaussian random variables whose correlations are given by \eqref{corr_a_in_time}, then
by the Wick theorem (see e.g. in \cite{Jan}) the expectation $\EE(\ldots)$ is a sum over different Wick-pairings of the non-conjugated 
variables $a_{\xi_j}^{(0)}(l_k)$ with  conjugated  $\bar a_{\s_r}^{(0)}(l_q)$, where
\be\lbl{WPnonv}
\EE a_{\xi_j}^{(0)}(l_k)\bar a_{\s_r}^{(0)}(l_q)\ne 0\qmb{only if}\qu \xi_j=\s_r.
\ee
The summands can be parametrised by Feynman diagrams $\gF$ (see e.g. in \cite{Jan}), 
obtained from the diagram $\gD$ by connecting with an edge every pair of the Wick-coupled vertices $a_{\xi_j}^{(0)}(l_k)$ and $\bar a_{\s_r}^{(0)}(l_q)$, see fig.~\ref{f:F6}. So  to every diagram $\gD$ correspond several Feynman diagrams $\gF$, one diagram for each Wick-pairing. 
By construction, every Feynman diagram decomposes to pairs of adjacent vertices, where in each pair either both vertices have zero (recall \eqref{deg-def}) degree, or one vertex is virtual and another one has positive degree. 

Therefore, by \eqref{i:Ea^ma^n},  
\begin{equation}\non
\EE  a_s^{(m)}(\tau_1) \bar a_s^{(n)}(\tau_2) =\sum_{\gF\in\gF_{m,n}} J_s(\gF),
\end{equation}
where the sum is taken over the set $\gF_{m,n}$ of Feynman diagrams $\gF$ obtained via all possible Wick-couplings in various diagrams $\gD\in\gD_m\times\ov\gD_n$.
Each (deterministic)  sum  $J_s(\gF)$ has the form 
\be\lbl{i:J}
J_s(\gF) = \int\ldots\int L^{-Nd}\sum_{\xi_1,\ldots,\xi_{2N},\s_1,\ldots, \s_{2N}}
(\ldots)\,dl_1\ldots dl_N,
\ee 
where  the term in the  brackets $(\dots)$ is obtained from those 
 in \eqref{II} by a Wick-pairing associated to the diagram $\gF$. See Lemma~\ref{l:Ea^ma^n} for an explicit form of the 
 sums $J_s(\gF)$.

{\it Step 3.}
The indices $\xi_i,\,\s_j$ in \eqref{i:J} are subject to the linear relations \eqref{lin-rel}, where $1\leq j\leq N$ and in \eqref{lin-rel}(2) the diagram $\gD$ is replaced by some  $\gF$  (see \eqref{WPnonv}). 
By \eqref{lin-rel}(2) (with $\gD$ replaced by $\gF$), the multi-index $\s=(\s_1,\ldots,\s_{2N})$  is a function of the multi-index
 $\xi=(\xi_1,\ldots,\xi_{2N})$, so it remains to study equation \eqref{lin-rel}(1), where 
 we substitute $\s=\s(\xi)$.
Analysing the diagrams $\gF$, forming the class $\gF_{m,n}$,  we find an $\gF$-dependent affine parametrization of 
solutions $\xi$ for this equation   by poly-vectors 
 $z=(z_1,\dots,z_N)$, $z_j\in\Z^d_L$. 
 Then we  write  the normalized sum  $L^{-Nd}\sum_{\xi,\s}$ in \eqref{i:J} as  $L^{-Nd}\sum_z\ $, 
approximate it  by an integral over $\R^{Nd}$
with accuracy  $L^{-2}\nu^{-2}$, and  accordingly find that  
\be\lbl{i:J_F}
J_s(\gF)=\int_{\R^N}dl \,\int_{\R^{Nd}}dz \,F_s^{\gF}(l,z) e^{i\nu^{-1}\Om^{\gF}(l,z)} + O(L^{-2}\nu^{-2}).
\ee
Here $l=(l_1,\ldots, l_N)\in\R^N$, 
the density function $F_s^\gF$ (which we write down explicitly) 
is Schwartz in the variables  $(s,\,z)$, piecewise smooth and fast decaying in $l$,
  %we emphasize that the fast rotating frequency has a nice form 
  while  the phase function $\Om^\gF$ is quadratic in $z$ 	and linear in $l$:
$
\Om^\gF(l,z)=\sum_{1\leq i,j\leq N}\al^\gF_{ij}\,z_i\cdot z_j (l_i-l_j),
$
where $z_i\cdot z_j$ is the scalar product in $\R^d$; see Theorem~\ref{th:Ea^ma^n-cont}.
The constant skew-symmetric 
matrix
$\al^\gF=(\al^\gF_{ij})$, $
\al_{ij}^\gF\in \{-1, 0, 1\},
$ is given by an explicit formula. Usually it  is degenerate, but all its rows and columns are non-zero vectors.

{\it Step 4.} The last step of the proof is to estimate  integral  \eqref{i:J_F} when $\nu\ll 1$.
To this end, inspired by the stationary phase method, we apply to 
 the latter the integral Parseval's identity, involving the fast oscillating Gaussian kernel $e^{i\nu^{-1}\Om^\gF}$.
The task is made complicated 
by the degeneracy of the quadratic form $\Om^\gF(l,\cdot)$. To handle it, 
in Sections~\ref{s:est}, \ref{est-imroved} and \ref{s:est int quadr}  we prove abstract theorems which allow to estimate integrals of a
 more general forms than \eqref{i:J_F}.

\section{Diagrams $\gD$ and formula for products $ a^{(m)}_{s} \bar a^{(n)}_{s}$}
\lbl{s:D}

In this section we construct the set of diagrams $\gD_m\times\ov\gD_n$, associated to the product 
$ a^{(m)}_{s}(\tau_1) \bar a^{(n)}_{s}(\tau_2)$, and express the latter
through the random processes $a^{(0)}_k$ and $\bar a^{(0)}_k$ (see Lemma~\ref{l:a^ma^n}). 
The construction is rather tedious but the main difficulties are notational: once the diagrammatic language is developed and convenient notation are introduced, proof of Lemma~\ref{l:a^ma^n} becomes a simple computation. 

\subsection{The set of diagrams $\gD_m\times\ov\gD_n$}

\begin{figure}[t]
	a)\qu
	\parbox{3cm}{ 
		\begin{tikzpicture}[]
		\node at (-1,0) (c0) {$c_0^{(2)}$};
		
		\node at (-2.1, -1.2) (c1) {$c_1^{(1)}$};
		\node at (-1.4, -1.2)  {$c_2^{(0)}$};
		\node at (-0.7, -1.2)  {$\bar c_1^{(0)}$};
		\node at (0, -1.3) (w2)  {$\bar w_{2}$};  
		
		\node at (-2.1, -2.5)  {$c_3^{(0)}$};
		\node at (-1.4, -2.5)  {$c_4^{(0)}$};
		\node at (-0.7, -2.5)  {$\bar c_3^{(0)}$};
		\node at (0, -2.6) (w4) {$\bar w_{4}$}; 
		
		\draw [line width=0.25mm] (c0.south)--(w2.north);
		\draw [line width=0.25mm](c1.south)--(w4.north);
		
		\end{tikzpicture}
	}
	\hfill
	b) \qu\parbox{3cm}{
		\begin{tikzpicture}[]
		\node at (-1,0) (c0) {$c_0^{(2)}$};
		
		\node at (-2.1, -1.2)  {$c_1^{(0)}$};
		\node at (-1.4, -1.2) (c2) {$c_2^{(1)}$};
		\node at (-0.7, -1.2)  {$\bar c_1^{(0)}$};
		\node at (0, -1.3) (w2)  {$\bar w_{2}$};  
		
		\node at (-2.1, -2.5)  {$c_3^{(0)}$};
		\node at (-1.4, -2.5)  {$c_4^{(0)}$};
		\node at (-0.7, -2.5)  {$\bar c_3^{(0)}$};
		\node at (0, -2.6) (w4) {$\bar w_{4}$}; 
		
		\draw [line width=0.25mm] (c0.south)--(w2.north);
		\draw [line width=0.25mm](c2.south)--(w4.north);
		
		\end{tikzpicture}
	}
	\hfill
	c) \qu \parbox{3cm}{
		\begin{tikzpicture}[]
		\node at (-1,0) (c0) {$c_0^{(2)}$};
		
		\node at (-2.1, -1.2)  {$c_1^{(0)}$};
		\node at (-1.4, -1.2)  {$c_2^{(0)}$};
		\node at (-0.7, -1.2) (bc1) {$\bar c_1^{(1)}$};
		\node at (0, -1.3) (w2)  {$\bar w_{2}$};  
		
		\node at (-2.1, -2.5)  {$c_3^{(0)}$};
		\node at (-1.4, -2.6) (w4) {$w_4$};
		\node at (-0.7, -2.5)  {$\bar c_3^{(0)}$};
		\node at (0, -2.5)  {$\bar c_{4}^{(0)}$}; 
		
		\draw [line width=0.25mm] (c0.south)--(w2.north);
		\draw [line width=0.25mm](bc1)--(w4.north);
		
		\end{tikzpicture}
	}
	\caption{The set of diagrams $\gD_2.$}
	\lbl{f:F2}
\end{figure}

For integers $m,n\geq 0$ we define the set $\gD_m\times\ov\gD_n$ as a direct product  of the sets of diagrams $\gD_m$ and $\ov\gD_n$, 
associated, correspondingly,  with the variables $a_s^{(m)}$ and $\bar a_s^{(n)}$.  

We start by constructing the set $\gD_m$, construction of  $\ov \gD_n$ is similar.
For example, the set $\gD_1$ consists of a unique diagram, which can be obtained from the diagram in fig.~\ref{f:F3}(a) by erasing the isolated vertex $\bar c_0^{(0)}$. 
Here each $c^{(p)}_j$ is regarded as an argument, which should be substituted by a variable $a_{\xi_j}^{(p)}(l)$, while $\bar c^{(p)}_j$ should be substituted by an $\bar a_{\s_j}^{(p)}(l')$.
E.g., this substitution, applied to the diagram in fig.~\ref{f:F2}(c), gives the diagram in fig.~\ref{f:F0}(b).
The set $\gD_2$ consists of three diagrams in fig.~\ref{f:F2}. The set $\gD_3$ consists of 12 diagrams, two of which are given in fig.~\ref{f:F3}(b,c).
Construction of the diagrams seems to be quite clear from these examples. Nevertheless, below we explain it in detail.

\subsubsection{Agreements and first notation }
Each diagram $\gD\in\gD_m$ consists of a number of vertices, some of them  are coupled by edges. 
Each vertex is either {\it non-conjugated} or {\it conjugated}. 
The non-conjugated vertices are denoted by $c_j$ while the conjugated ~-- by $\bar c_j$.
If we do not want to indicate whether a  vertex is conjugated or not, we write it as $\hat c_j$. Each vertex 
is characterised by its 
{\it degree} which we often write as an upper index: 
 $$
 \text{deg}\, \hat c_j^{(k)} =k,
 $$
and each vertex is either {\it virtual} or {\it real}. If a vertex $c_j,\bar c_j$ or $\hat c_j$ is virtual, we may 
 write it as $w_j, \bar w_j$ or $\hat w_j$.  Each virtual vertex $\hat w_j$ has zero degree, 
$$
\deg \hat w_j=0,
$$
so we do not write for it  the upper index. Real vertices of zero degree are called {\it leaves}. 

\smallskip

\begin{figure}[t]
	
	a)\qu
		\parbox{5cm}{
		\begin{tikzpicture}
		\path (-1,1.3) node(a0) {$c_0^{(1)}$}
		(0, 0) node(b_w2) {$\bar w_{2}$}; 
		\draw [line width=0.25mm] (a0.south)--(b_w2.north);
		\node at (-0.7, 0.1)  {$\bar c_1^{(0)}$};
		\node at (-1.4, 0.1)  {$c_2^{(0)}$};
		\node at (-2.1, 0.1) {$c_1^{(0)}$};
		\node at (0.5,1.3) {$\bar c_0^{(0)}$};
		\end{tikzpicture}
	}
	\hfill
	b)\qu
	\parbox{4cm}{ 
		\begin{tikzpicture}[]
		\node at (-1,0) (c0) {$c_0^{(3)}$};
		
		\node at (-2.1, -1.2) (c1) {$c_1^{(2)}$};
		\node at (-1.4, -1.2)  {$c_2^{(0)}$};
		\node at (-0.7, -1.2)  {$\bar c_1^{(0)}$};
		\node at (0, -1.3) (w2)  {$\bar w_{2}$};  
		
		\node at (-2.1, -2.5) (c3) {$c_3^{(0)}$};
		\node at (-1.4, -2.5)  (c4) {$c_4^{(1)}$};
		\node at (-0.7, -2.5)  {$\bar c_3^{(0)}$};
		\node at (0, -2.6) (w4) {$\bar w_{4}$}; 
		
		\node at (-2.1, -3.8)  {$c_5^{(0)}$};
		\node at (-1.4, -3.8)  {$c_6^{(0)}$};
		\node at (-0.7, -3.8)  {$\bar c_5^{(0)}$};
		\node at (0, -3.9) (w6) {$\bar w_{6}$}; 
		
		\draw [line width=0.25mm] (c0.south)--(w2.north);
		\draw [line width=0.25mm](c1.south)--(w4.north);
		\draw [line width=0.25mm](c4.south)--(w6.north);
		
		\end{tikzpicture}
	}
	\hfill \\
	\vspace{1cm}
	c) \parbox{6cm}{
		\begin{tikzpicture}[]
		\node at (-1,0) (c0) {$c_0^{(3)}$};
		
		\node at (-2.1, -1.2) (c1) {$c_1^{(1)}$};
		\node at (-1.4, -1.2)  {$c_2^{(0)}$};
		\node at (-0.7, -1.2) (bc1) {$\bar c_1^{(1)}$};
		\node at (0, -1.3) (w2)  {$\bar w_{2}$};  
		
		\node at (-4.4, -2.5)  {$c_3^{(0)}$};
		\node at (-3.7, -2.5)  {$c_4^{(0)}$};
		\node at (-3, -2.5)  {$\bar c_3^{(0)}$};
		\node at (-2.3, -2.6) (w4) {$\bar w_{4}$}; 
		
		\node at (-1.2, -2.5)  {$c_5^{(0)}$};
		\node at (-0.5, -2.6) (w6) {$w_6$};
		\node at (0.2, -2.5)  {$\bar c_5^{(0)}$};
		\node at (0.9, -2.5)  {$\bar c_{6}^{(0)}$}; 
		
		\draw [line width=0.25mm] (c0.south)--(w2.north);
		\draw [line width=0.25mm] (c0.south)--(w2.north);
		\draw [line width=0.25mm](c1)--(w4.north);
		\draw [line width=0.25mm](bc1.south)--(w6.north);
		
		\end{tikzpicture}
	}
\hfill
 d)\qu \parbox{4cm}{
	\begin{tikzpicture}
	\path (-1,1.3) node(a0) {$\bar c_0^{(1)}$}
	(-1.4, 0) node(b_w2) {$w_{2}$}; 
	\draw [line width=0.25mm] (a0)--(b_w2.north);
	\node at (-0.7, 0.1)  {$\bar c_1^{(0)}$};
	\node at (0, 0.1)  {$\bar c_2^{(0)}$};
	\node at (-2.1, 0.1) {$c_1^{(0)}$};
	\end{tikzpicture}
}
\hfill
	\caption{a) The unique diagram of the set $\gD_1\times \ov\gD_0$. b,c) Some diagrams from the set $\gD_3$. d) The unique diagram of the set $\ov \gD_1$.}
	\lbl{f:F3}
\end{figure}
The vertices of a diagram are organised in blocks of four, where the $k$-th block,  $k\geq 1$, is 
$
\{c_{2k-1}, c_{2k}, \bar c_{2k-1}, \bar c_{2k}\}.
$
Among these four 
 vertices either the vertex $c_{2k}$ or the vertex $\bar c_{2k}$ is virtual, while the other  are real. 
If the vertex $c_{2k}$ is virtual then we denote this block by $B_k$,
\be\lbl{B_kkk}
B_k=\{c_{2k-1}, w_{2k}, \bar c_{2k-1}, \bar c_{2k}\},
\ee
and call it the {\it non-conjugated  $k$-th block}.
If the vertex $\bar c_{2k}$ is virtual, we denote
\be\lbl{barB_kkk}
\bar B_k=\{c_{2k-1}, c_{2k}, \bar c_{2k-1}, \bar w_{2k}\},
\ee
and call it the {\it conjugated $k$-th block}. 
If we do not want to emphasize whether a block is conjugated or not, we denote it by $\hat B_k$.
We define the degree of each block as the sum of degrees of the  forming it vertices. So
$$
\deg B_k:=\deg c_{2k-1} + \deg \bar c_{2k} +\deg \bar c_{2k-1}
$$
(since deg$\,\bar w_{2k}=0$),  and similarly for $\deg \bar B_k$.
E.g., the diagram from fig.~\ref{f:F2}(a) has two conjugated blocks $\bar B_1$ and $\bar B_2$, where $\deg \bar B_1=1$ and $\deg \bar B_2=0.$

\subsubsection{Construction of the diagrams}
\lbl{s:constr}
If a vertex $c_j$ is coupled with a virtual vertex $\bar w_{2k}$ of a block $\bar B_k$
then we say that  $c_j$ {\it generates} the block $\bar B_k$ or that $c_j$ is a {\it parent} of the block $\bar B_k$.
The notion that a  vertex $\bar c_j$ 
generates a block $B_k$ is defined similarly.

The set of diagrams $\gD_m$ is constructed by the following inductive procedure. 
The set $\gD_0$ is formed by a unique diagram which consists of a unique (real) vertex $c_0^{(0)}$. 
If $m\geq 1$, the root of each diagram $\gD\in\gD_m$ is a real vertex $c_0^{(m)}$. 
Each vertex $\hat c_j$ of the diagram $\gD$ with $\deg \hat c_j\geq 1$ (i.e. which is real and is not a leave) 
generates a block $\hat B_k$ satisfying
\be\lbl{rule_diag}
\deg \hat B_k=\deg \hat c_j -1.
\ee
Non-conjugated vertices $c_j$ generate conjugated blocks $\bar B_k$ while conjugated vertices $\bar c_j$ generate non-conjugated blocks $B_k$. 
The numeration of the blocks (i.e. the dependence of $k$ as a function of  $(\hat c_j)$) is chosen in arbitrary way. 
E.g., from the left to the right and from the top to the bottom, as in fig.~\ref{f:F2},\ref{f:F3}.

Rule \eqref{rule_diag} does not determine uniquely the degrees of vertices in the block $\hat B_k$, but fixes only their sum. So, for $m\neq 0,\,1$ the diagram $\gD$ obtained by the inductive procedure above is not unique. 
We define  $\gD_m$ as the set of all possible diagrams obtained by this procedure.
%By construction, each diagram $\gD\in\gD_m$ has $m$ blocks, $2m+1$ non-conjugated vertices and $2m$ conjugated vertices.
 
The set of diagrams $\ov \gD_n$ is constructed by a similar procedure. The only difference is that in 
 diagrams $\bar\gD\in\ov\gD_n$ the root vertex is conjugated and has the form $\bar c_0^{(n)}$, see fig.~\ref{f:F3}(d).
  The set of diagrams $\ov\gD_n$ equals to the set $\gD_n$ in which we conjugate each diagram, and in every block of each diagram we exchange positions of conjugated and non-conjugated vertices, so that the non-conjugated vertices are situated first.
\begin{figure}[t]
a)\qu
\parbox{4cm}{ 
	\begin{tikzpicture}[]
	\node at (-1,0) (c0) {$c_0^{(2)}$};
	\node at (1,0)  {$\bar c_{0}^{(0)}$};
	
	\node at (-2.1, -1.2) (c1) {$c_1^{(1)}$};
	\node at (-1.4, -1.2)  {$c_2^{(0)}$};
	\node at (-0.7, -1.2)  {$\bar c_1^{(0)}$};
	\node at (0, -1.3) (w2)  {$\bar w_{2}$};  
	
	\node at (-2.1, -2.5)  {$c_3^{(0)}$};
	\node at (-1.4, -2.5)  {$c_4^{(0)}$};
	\node at (-0.7, -2.5)  {$\bar c_3^{(0)}$};
	\node at (0, -2.6) (w4) {$\bar w_{4}$}; 
	
	\draw [line width=0.25mm] (c0.south)--(w2.north);
	\draw [line width=0.25mm](c1.south)--(w4.north);
	
	\end{tikzpicture}
}
\hfill
b)\qu \parbox{6cm}{
	\begin{tikzpicture}
	\path (-1.2,0) node(a0) {$c_0^{(1)}$}
	(-0.2, -1.3) node(b_w2) {$\bar w_{2}$}; 
	\draw [line width=0.25mm] (a0.south)--(b_w2.north);
	\node at (-0.9, -1.2)  {$\bar c_1^{(0)}$};
	\node at (-1.6, -1.2)  {$c_2^{(0)}$};
	\node at (-2.3, -1.2) {$c_1^{(0)}$};

		\node at (1.8,0) (c0) {$\bar c_0^{(2)}$};
		
		\node at (0.8, -1.2)  (c1) {$c_3^{(1)}$};
		\node at (1.5, -1.3) (w2) {$w_4$};
		\node at (2.2, -1.2)  {$\bar c_3^{(0)}$};
		\node at (2.9, -1.2)  {$\bar c_{4}^{(0)}$};  
		
		\node at (0.8, -2.5)  {$c_5^{(0)}$};
		\node at (1.5, -2.5)  {$c_6^{(0)}$};
		\node at (2.2, -2.5)  {$\bar c_5^{(0)}$};
		\node at (2.9, -2.6) (w4) {$\bar w_{6}$}; 
		
		\draw [line width=0.25mm] (c0)--(w2.north);
		\draw [line width=0.25mm](c1.south)--(w4.north);
	\end{tikzpicture}
}
	\caption{A diagram from the set a) $\gD_2\times\ov\gD_0$ and b) $\gD_1\times\ov\gD_2$.}
	\lbl{f:F4}
\end{figure}

Finally, for a pair of diagrams $\gD^1\in\gD_m$ and $\bar\gD^2\in\ov\gD_n$ we consider the  diagram $\gD^1\sqcup\bar\gD^2$
 (see fig.~\ref{f:F3}(a) and \ref{f:F4}), obtained by drawing  the diagrams $\gD^1$ and $\bar\gD^2$ side by side 
and enumerate the vertices of $\bar\gD^2$, except the root $\bar c_0$, not from $1$ to $2n$ but from $2m+1$ to $2m+2n$. 
Thus, in  $\gD^1\sqcup\bar\gD^2$ we denote the  two roots as  $c_0$ and $\bar c_0$, then  enumerate the remaining vertices of the diagram $\gD^1$, and finally those of the diagram $\bar\gD^2$.
We define
$$
\gD_m\times\ov\gD_n=\{\gD^1\sqcup\bar\gD^2:\, \gD^1\in\gD_m, \bar\gD^2\in\ov\gD_n\}.
$$
%-- this is the set, formed by  all possible pairs of diagrams from the sets $\gD_m$ and $\ov \gD_n$. 

\subsubsection{Properties of the set $\gD_m\times\ov\gD_n$ and some terminology}
\lbl{s:D-prop}
 
Throughout the paper we use the notation
$$
N:=m+n.
$$
The following properties of  a diagram $\gD\in\gD_m\times\ov\gD_n$ are obvious. 

$\bullet$	The number of blocks of $\gD$ equals to $N$.

$\bullet$ The diagram $\gD$ has $4N+2$ vertices,  half of them  are conjugated while another half are non-conjugated. 
	
 $\bullet$	If vertices $\hat c_i$ and $\hat c_j$ are coupled by an edge, we call them {\it adjacent} and write $\hat c_i\volna \hat c_j$.
	Each pair of adjacent vertices in $\gD$ consists of a conjugated and  non-conjugated 
	 vertex, and has either the form 
	$\{c^{(p)}_{i}, \bar w_j \}$
	or 
	$\{\bar c^{(p)}_i, w_{j} \}$, where the degree $p\geq 1$.

	$\bullet$	An edge  which joins $c_i$ with $\bar c_j$ is denoted $(c_i,\bar c_j)$ (either $c_i$ or $\bar c_j$ is a virtual vertex). Each  diagram $\gD\in\gD_m\times\ov\gD_n$  has $N$ edges.

 $\bullet$	Recall that a real vertex $\hat c_j^{(0)}$ of zero degree is called a leaf. 
 %E.g., in the diagram fig.~\ref{f:diag1}(8) the leaves are $c_1,c_2,\bar c_1,\bar c_3,\bar c_4,c_5,c_6,\bar c_5$.
	There is no vertex in $\gD$,  adjacent to a leaf. 
	The number of conjugated leaves equals to the number of non-conjugated leaves and equals to $N+1$.

\subsubsection{Values of  vertices $\hat c_j$}

Consider a diagram  $\gD\in\gD_m\times\ov\gD_n$. 
To every vertex $\hat c_i$ of a block $\hat B_k$, $1\leq k\leq N$,   we assign the same  time variable  $l_k\geq -T$.
E.g., in the diagram from fig.~\ref{f:F4}(b) the times assigned to the vertices $c_2$ and $\bar c_6$ are $l_1$ and $l_3$, respectively.
To the roots $c_0$ and $\bar c_0$ we assign the times $\tau_1$ and $\tau_2$.  The vectors
\be\label{time}
l=(l_1,\ldots, l_N) \qu\qnd\qu {\bf l}=(\tau_1,\tau_2,l)
\ee
are called {\it the vectors of times}.
Let 
\be\non
\xi=(\xi_0,\ldots,\xi_{2N}), \qu \s=(\s_0,\ldots,\s_{2N}),
\ee
where $\xi_i,\s_j\in\Z^d_L$, so $(\xi,\s) \in\Z^{(4N+2)d}_L$,
and we recall that $4N+2$ is the number of vertices in   $\gD$.
These will be the indices for summation in a formula for  $a_s^{(m)} (\tau_1)\bar a_s^{(n)} (\tau_2)$, 
where  $\xi_j$'s are  indices for  the non-conjugated variables $a_s$
 and $\s_j$  -- for the conjugated variables $\bar a_s$   in the following sense.

We regard a  real vertex $c_j^{(p)}$ of the diagram $\gD$ as a  position to put there a component $a^{(p)}_{\xi_j}(t)$
 of the random vector $a^{(p)}(t)$,  and a real vertex $\bar c_j^{(p)}$ -- as a position for   a component $\bar a^{(p)}_{\s_j}(t')$
 of the conjugate vector,
 with  suitable $t = t(\bf l)$, $t' = t'(\bf l)$.
  More precisely, for $ p\geq 0$ we view the real vertices $c_j^{(p)}$, $\bar c_j^{(p)}$  as random functions of the variables $\bf l, \xi, \s$, defined by
\be\label{values}
c_j^{(p)}({\bf l},\xi,\s)=a_{\xi_j}^{(p)} (t), \qquad  
\bar c_j^{(p)}({\bf l},\xi,\s)=\bar a_{\s_j}^{(p)} (t'),
\ee
where $t$ and $t'$ are components of the time-vector $\bf l$ in \eqref{time}, assigned to the vertices $c_j^{(p)}$ and $\bar c_j^{(p)}$. 
%  assigned to the vertices $c_j^{(p)}$ and $\bar c_j^{(p)}$, respectively. 
We say that the vertices $c_j^{(p)}$ and $\bar c_j^{(p)}$ take values $a_{\xi_j}^{(p)} (t)$ and $\bar a_{\s_j}^{(p)} (t')$.
For  virtual vertices $w_j$ and $\bar w_j$ we say that they take values $w_{\xi_j}$ and $\bar w_{\s_j}$.
% here $w$ is just a letter, what we need is to keep trace of the indices $\xi_j$ and $\s_j$.  
 %Below, abusing notation, we sometimes write the values of vertices instead of the vertices.
%E.g., this is the case in fig.~\ref{f:F1}(a) and in fig.~\ref{f:F5} from Appendix~\ref{s:lem_diag}.

\subsubsection{Restrictions on  the multi-indices $(\xi,\s)$} \label{s_3.1.5}

To calculate the value of a diagram $\gD\in\gD_m\times\ov\gD_n$ we will make substitutions \eqref{values}, using 
 multi-indices  $(\xi, \sigma)\in \Z_L^{(4N+2)d}$. 
At some stages of our constructions it will be necessary to put restrictions on the 
$(\xi, \sigma)$ and substitute in the diagram only the corresponding  multi-indices
 which we call {\it admissible}. The set of all admissible multi-indices   $(\xi, \sigma)$  is denoted
 $$\cA_s(\gD)\subset \R^{(4N+2)d},$$
 where we  take $s,\xi_i,\s_j$ in $\R^d$ rather than  in $\Z^d_L$  to be able to extend  the correlation from Theorem~\ref{t:main''} to a function 
 on $\R^d =\{ s\}$.
The set $\cA_s(\gD)$ is defined by the following three relations:

$\bullet$ We have
\be\lbl{p1}
\xi_0=\s_0=s.
\ee

$\bullet$  For all $1\leq i\leq N$, 
\be\lbl{de'==1}
\de'^{\xi_{2i-1}\xi_{2i}}_{\s_{2i-1}\s_{2i}}=1.
\ee

$\bullet$
If the vertices $c_i$ and $\bar c_j$ are adjacent, the corresponding indices are equal:
\be\lbl{p4}
c_i\volna \bar c_j \Rightarrow \xi_i=\s_j.
\ee
We will also use the corresponding discrete subset 
\be\lbl{adm_entire}
\cA_s^L(\gD)=\cA_s(\gD)\cap \Z^{(4N+2)d}_L,
\ee
which is exactly the set of multi-indices in which we sum in \eqref{i:J}.
In  view of condition \eqref{p1} it may be not empty only if $s\in\Z^d_L$.

\subsubsection{More notation} 
\lbl{s:more_not}
Let $\gD\in\gD_m\times\ov\gD_n$.

$\bullet$ We denote by $E(\gD)$ the set of all edges of a diagram $\gD$, 
$
E(\gD)=\{(c_i,\bar c_j)\}.
$

$\bullet$ Let $(\xi,\s)\in\cA_s(\gD)$ be an admissible index--vector. Then $\xi_i=\s_j$  if $c_i\sim \bar c_j$. 
Denoting by $\phi$ an edge $\phi=(c_i,\bar c_j)$, we set 
\be\lbl{ga-phi}
\ga^\phi(\xi,\s):=\ga_{\xi_i}=\ga_{\s_j}
\ee
($\gamma_s$'s are defined in \eqref{diss_op}). 

\indent $\bullet$ Recall that in each pair of adjacent vertices in $\gD$  always one vertex is  real while another  is virtual.
Then, by $l^{\phi}_r$ and $l^{\phi}_w$ we denote the times assigned to the real and virtual vertices, 
 connected by the edge $\phi$. So each $l^\phi_{r,w}$ is a  mapping from $E(\gD)$ to the set $\{\tau_1, \tau_2, l_1, \dots, l_N\}$. 

$\bullet$ By $ l^{\hat c_i}$  we denote the time, assigned to a vertex  $\hat c_i$.

$\bullet$ By 
$$L(\gD), \bar L(\gD)\subset\{0,\ldots 2N\}$$ 
we denote the sets such that 
$\{c_i\}_{i\in L(\gD)}$ and $\{\bar c_i\}_{i\in \bar L(\gD)}$
form the sets of all non-conjugated leaves and all conjugated leaves, correspondingly.
In other words, abusing notation and identifying the vertices with their values, we have
$\{c_i\}_{i\in L(\gD)}=\{\az{\xi_i}\}$ and $\{\bar c_i\}_{i\in \bar L(\gD)}=\{\bz{\s_i}\}$.
E.g., for the diagram from fig.~\ref{f:F4}(b) we have
$L(\gD) = \{1,2,5,6\}$ and $\bar L(\gD) = \{1,3,4,5\}$.

\subsection{Formula for the product 
$a^{(m)}_s \bar a^{(n)}_s $}

Let us denote                                                                          
\be\label{tau}
\tau:=(\tau_1,\tau_2),
\ee
where we recall that $\tau_1,\tau_2$ are the times for the roots $c_0, \bar c_0$. Recall also  
that the sets $E(\gD)$, $L(\gD)$, $\bar L(\gD)$, the functions $\ga^\phi=\ga^\phi(\xi,\s)$  
and the times $l^{\phi}_{r,w}$ and $l^{c_j,\bar c_j}$ are defined in Section~\ref{s:more_not};
in difference with the functions $\ga^\phi$,
the  time variables $l^{\phi}_r, l^{\phi}_w, l^{c_j}, l^{\bar c_j}$ are independent from $(\xi,\s)$.
To each diagram $\gD\in\gD_m\times\ov\gD_n$ we associate the (random) function
\begin{align}
\non
G_\gD(\tau,l,\xi,\s):=c_\gD
&\prod_{\phi\in E(\gD)} e^{-\ga^{\phi}(\xi,\s)\,(l^{\phi}_r - l^{\phi}_w)}\mI_{\{-T\leq l^{\phi}_w \leq l^{\phi}_r\}} (\tau,l)
\\ \lbl{G_D}
&\prod_{j\in L(\gD)}\az{\xi_{j}}(l^{c_j})\prod_{j\in \bar L(\gD)}\bz{\s_{j}}(l^{\bar c_j}),
\end{align}  
which is defined on the set of admissible multi-indices $ (\xi,\s)\in \cA_s^L(\gD)$ (see \eqref{adm_entire}).
See below for a discussion of this formula.
Here $\mI_{\{-T\leq l^{\phi}_w \leq l^{\phi}_r\}}$ denotes the indicator function of the set in which $-T\leq l^{\phi}_w \leq l^{\phi}_r$ and $c_\gD$ is the constant 
\be\lbl{c_D-def}
c_\gD=(-1)^{\#w_\gD}i^N,
\ee
where 
$\#w_\gD$  is the number of non-conjugated virtual vertices of the diagram $\gD$,
$N=m+n$ and $i=\sqrt{-1}$ is the imaginary unit.
Set
\be\lbl{Omega_def}
\om_j(\xi,\s)=\om^{\xi_{2j-1}\xi_{2j}}_{\s_{2j-1}\s_{2j}} 
\qnd
\om=(\om_1,\ldots,\om_N),
%\qnd \Om(l,\xi,\s)= \sum_{j=1}^N l_j \,\om_j(\xi,\s),
\ee
 we recall that the quadratic form $\om^{s_1s_2}_{s_3s}$ is defined in \eqref{omega}.
For vectors $\vec x=(x_1,\ldots,x_N)$ and $\vec t=(t_1,\ldots,t_N)$ set also
\be\lbl{Theta-def}
\Theta(\vec x,\vec t)=\prod_{j=1}^N\theta(x_j,t_j),
\ee
where  $\theta$ is as in \eqref{theta}.

 %we recall that $\theta:\R^2\mapsto\C$ is a measurable bounded function (see Remark~\ref{r:theta}).

%See a proof of the following lemma for a discussion of the formula \eqref{G_D}.
\begin{lemma}
\lbl{l:a^ma^n}
For any integers $m,n\geq 0$ satisfying $N:=m+n\geq 1$, $s\in\Z^d_L$ and 
$\tau_1,\tau_2\geq -T$, we have
\be\lbl{a^ma^n}
a^{(m)}_s(\tau_1) \bar a^{(n)}_s(\tau_2)
=\sum_{\gD\in\gD_m\times\ov\gD_n} I_s(\gD),
\ee
where $I_s(\gD)=I_s(\tau;\gD)$,
\be
\lbl{a^ma^n-I}
I_s(\tau;\gD)=\int_{\R^N}
dl\, 
L^{-Nd}\sum_{(\xi,\s)\in
	\cA_s^L(\gD)} \,
G_\gD(\tau,l,\xi,\s)\,
\Theta(\om(\xi,\s),\nu^{-1}l). 
\ee
\end{lemma}

Components of the 
formula \eqref{a^ma^n-I} are  quite simple. 
Indeed, the factor   $\prod_{j\in L(\gD)}\az{\xi_{j}}(l^{c_j})\prod_{j\in \bar L(\gD)}\bz{\s_{j}}(l^{\bar c_j})$, 
entering $G_\gD(\tau,l,\xi,\s)$ (see 
 \eqref{G_D}), is just the product of values of all leaves of the diagram $\gD$. The
term $l^{\phi}_r - l^{\phi}_w$ 
 is a difference of the times assigned to vertices, coupled by an edge $\phi$, and the corresponding exponent in $G_\gD(\tau,l,\xi,\s)$ 
  comes from the iteration of  the Duhamel formula \eqref{an} that generates  $\phi$. It appears together with the indicator function $\mI_{\{-T\leq l^{\phi}_w \leq l^{\phi}_r\}}$, which is there 
   due to the fact that in \eqref{an} the integration is performed not over $\R$ but only over the interval $[-T,\tau]$. 
 The factor $\theta(\om_j,\nu^{-1}l_j)$, entering the function $\Theta(\om,\nu^{-1}l)$, comes from the iteration of the Duhamel formula, generating
 the $j$-th block of~$\gD$. 

The assertion  of Lemma~\ref{l:a^ma^n} easily (but clumsily) follows by iterating the %Duhamel
  formula \eqref{an}.
Instead of giving a 
complete proof we restrict ourselves to considering  an example below.
 
 \subsubsection{Example}
  Here we establish formula \eqref{a^ma^n} for the case $m=1$ and $n=0$. 
  %To this end
 %we multiply by $\bar a_s^{(0)}(\tau_2)$ the both sides of \eqref{a1.1} with $\tau=\tau_1$.
 % There we write $l_1$ instead of $l$ and denote by $\xi_1$, $\xi_2$ the indices $s_1,s_2$ assigned to the non-conjugated variables 
 %$a^{(0)}_{s_1}$, $a^{(0)}_{s_2}$, and by $\s_1,\s_2$~-- the indices $s_3,s_4$, assigned to $\bar a^{(0)}_{s_3}$ and $\de^s_{s_4}$. 
 We get from  \eqref{a1.1} that 
 \begin{align}
 \non
 a^{(1)}_s(\tau_1)\bar a^{(0)}_s(\tau_2) = i \int_{-T}^{\tau_1} dl_1\;L^{-d}&\sum_{\xi_1,\xi_2,\s_1,\s_2} \de^{\prime\xi_1\xi_2}_{\s_1\s_2} \de^s_{\s_2}\,e^{-\ga_{\s_2}(\tau_1-l_1)}
  \\
 \lbl{a1s''}
 &\big(a^{(0)}_{\xi_1} a^{(0)}_{\xi_2}{\bar a_{\s_1}}^{(0)}\big)(l_1) \bar a^{(0)}_{s}(\tau_2) \,
\theta(\om_1,\nu^{-1}l_1),
 \end{align}
 where we use notation \eqref{Omega_def}.
 The set $\gD_1\times\ov\gD_0$ consists of a unique diagram, given in fig.~\ref{f:F3}(a), which we denote by $\gD^{1,0}$.
 Now the total number of vertices is  $4N+2=6$, and the 
  corresponding set of admissible multi-indices $\cA_s^L(\gD^{1,0})$ has the form 
 $$
 \cA_s^L(\gD^{1,0})=\big\{(\xi_0, \xi_1, \xi_2, \sigma_0, \sigma_1, \sigma_2)
% \xi,\s) 
 \in\Z^{6d}_L:\, \xi_0=\s_0=s, \;\de^{\prime \xi_1\xi_2}_{\s_1\s_2}=1,\;\xi_0=\s_2\big\}.
 $$
 Then we rewrite the r.h.s. of  \eqref{a1s''} as
\[
%a^{(1)}_s(\tau_1)\bar a^{(0)}_s(\tau_2) =
 i \int_{-T}^{\tau_1} dl_1\;L^{-d}\!\!
 \sum_{(\xi,\s)\in\cA_s^L(\gD^{1,0})}  e^{-\ga_{\s_2}(\tau_1-l_1)}
\big(a^{(0)}_{\xi_1} a^{(0)}_{\xi_2}{\bar a_{\s_1}}^{(0)}\big)(l_1) \bar a^{(0)}_{\s_0}(\tau_2) \,
\theta(\om_1,\nu^{-1}l_1).
\]
 The unique edge of the diagram $\gD^{1,0}$ is $\phi:=(c_0^{(1)},\bar w_2)$, so using the introduced in Section~\ref{s:more_not} notation we write
 $e^{-\ga_{\s_2}(\tau_1-l_1)}=e^{-\ga^\phi(l_r^\phi-l_w^\phi)}$ and
 	$\int_{-T}^{\tau_1}\,dl_1=\int_\R  \mI_{\{-T\leq l^{\phi}_w \leq l^{\phi}_r\}}\, dl_1$.
 Then, noting that 	the term $a^{(0)}_{\xi_1} a^{(0)}_{\xi_2}{\bar a_{\s_1}}^{(0)} \bar a^{(0)}_{\s_0} $ is a product of values of all leaves in the diagram $\gD^{1,0}$, 
 we see that \eqref{a1s''} takes the form
 $a^{(1)}_s(\tau_1)\bar a^{(0)}_s(\tau_2)=I_s(\gD^{1,0})$,
 where the sum $I_s(\gD^{1,0})$ is defined in \eqref{a^ma^n-I}.
 Thus, formula \eqref{a^ma^n} holds for $m=1, n=0$.

\section{Feynman diagrams and formula for expectations $\EE a^{(m)}_{s} \bar a^{(n)}_{s}$} 
\lbl{sec:FD}

In this section we compute the expectation $\EE a^{(m)}_{s}(\tau_1) \bar a^{(n)}_{s}(\tau_2)$.
Due to \eqref{a^ma^n}, it suffices to find $\EE I_s(\gD)$ for every diagram $\gD\in\gD_m\times\ov\gD_n$.
Note that the randomness enters $I_s(\gD)$ only via the product 
of values of leaves $a_{\xi_i}^{(0)}$, $\bar a_{\s_j}^{(0)}$ in \eqref{G_D}, which are Gaussian random variables.
Since their correlations are given by \eqref{corr_a_in_time},
due to the Wick theorem $\EE I_s(\gD)$ is given by a sum over all Wick pairings of values $a_{\xi_i}^{(0)}(l^{c_i})$ of non-conjugated leaves in $\gD$ with values $\bar a_{\s_j}^{(0)}(l^{\bar c_j})$ of conjugated leaves. 
We parametrize this sum by {\it Feynman diagrams} $\gF$ (see \cite{Jan} for the notion), which are obtained from the diagram $\gD$ by connecting the Wick paired leaves by edges, see fig.~\ref{f:F6}. 
\begin{figure}[t]
a)\qu	\parbox{5cm}{ 
		\begin{tikzpicture}[]
		\node at (-1,0) (c0) {$c_0^{(2)}$};
		\node at (1,0) (bc0) {$\bar c_{0}^{(0)}$};
		
		\node at (-2.1, -1.2) (c1) {$c_1^{(1)}$};
		\node at (-1.4, -1.2) (c2) {$c_2^{(0)}$};
		\node at (-0.7, -1.2) ( bc1) {$\bar c_1^{(0)}$};
		\node at (0, -1.3) (bw2)  {$\bar w_{2}$};  
		
		\node at (-2.1, -2.5) (c3) {$c_3^{(0)}$};
		\node at (-1.4, -2.5) (c4) {$c_4^{(0)}$};
		\node at (-0.7, -2.5) (bc3) {$\bar c_3^{(0)}$};
		\node at (0, -2.6) (bw4) {$\bar w_{4}$}; 
		
		\draw [line width=0.25mm] (c0.south)--(bw2.north);
		\draw [line width=0.25mm](c1.south)--(bw4.north);
		\draw [line width=0.25mm] (c3.south) -- (-2.1,-3) --(1,-3) --  (bc0.south);
		\draw [line width=0.25mm](c2.south)--(-0.9, -2.2);
		%c4-bc1:
		\draw [line width=0.25mm](c4.north)--(-0.9, -1.5);
		\end{tikzpicture}
	}
\hfill
b)\qu\parbox{5cm}{ 
\begin{tikzpicture}
\node at (-1,0) (c0) {$c_0^{(2)}$};
\node at (1,0) (bc0) {$\bar c_{0}^{(0)}$};

\node at (-2.5, -1.2) (c1) {$c_1^{(1)}$};
\node at (-1.6, -1.2) (c2) {$c_2^{(0)}$};
\node at (-0.5, -1.2) ( bc1) {$\bar c_1^{(0)}$};
\node at (0.4, -1.3) (bw2)  {$\bar w_{2}$};  

\node at (-2.5, -2.5) (c3) {$c_3^{(0)}$};
\node at (-1.6, -2.5) (c4) {$c_4^{(0)}$};
\node at (-0.5, -2.5) (bc3) {$\bar c_3^{(0)}$};
\node at (0.4, -2.6) (bw4) {$\bar w_{4}$}; 

%c0-bw2:
\draw [<-,line width=0.25mm] (-0.8,-0.3)-- node[above, near end]{{\scriptsize $f_\gF$}} (0.35, -1) ;
%c1-bw4
\draw [<-,line width=0.25mm](c1.south)--node[above, near end]{{\scriptsize $f_\gF$}}(bw4.north);
%c3-bc0:
\draw [<-,line width=0.25mm] (c3.south) -- (-2.5,-3.2) -- node[below]{{\scriptsize $f_\gF$}}  (1,-3.2) --  node[right]{{\scriptsize $f_\gF$}} (bc0.south);
%c2-bc3:
\draw [<-,line width=0.25mm](c2.south)--(-0.9, -2.2);
%c4-bc1:
\draw [<-,line width=0.25mm](c4.north)--(-0.8, -1.5);

\end{tikzpicture}
}
	\hfill
	\caption{a) A Feynman diagram $\gF$ obtained from the diagram $\gD$ from fig.~\ref{f:F4}(a). \qu b) Orientation of the Feynman diagram from item (a) and associated permutation~$f_\gF$.}
	\lbl{f:F6}
\end{figure}

Next we give a detailed definition of the Feynman diagrams and discuss related notions. 

\subsection{Feynman diagrams}  
\lbl{sub:FD}

Let us fix $m,n\geq 0$ satisfying $N=m+n\geq 1$, and consider a diagram $\gD\in\gD_m\times\ov\gD_n$. %corresponding to the product $a^{(m)}_{s} \bar a^{(n)}_{s}$. 
Its 
 set of leaves  consists of $N+1$ conjugated leaves $\bar c_i^{(0)}$ and $N+1$ non-conjugated leaves 
 $c_j^{(0)}$. Consider any partition of this set to $N+1$ non-intersecting pairs of the form
$(c_i^{(0)},\bar c_j^{(0)})$, where the leaves $c_i^{(0)}$ and $\bar c_j^{(0)}$
{\it do not belong to the same block} of $\gD$ (if such partition exists).
We denote by $\gF$ a diagram obtained from  $\gD$ 
by connecting the leaves in each pair by an edge
(see fig.~\ref{f:F6}(a)) and call it a Feynman diagram. 
We never couple leaves  belonging to the same block since 
  for any $j$-th block the variables $a_{\xi_\al}^{(0)}$, $\bar a_{\s_\beta}^{(0)}$,
assigned to the vertices in this block as in \eqref{values}, satisfy 
$
\EE a^{(0)}_{\xi_\alpha} \bar a^{(0)}_{\sigma_\beta} =0
$
for $\alpha, \beta\in \{2j-1, 2j\}$ and any admissible mutli-index $(\xi,\s)\in\cA_s^L(\gD)$, since $\xi_\al\ne\s_\beta$ in view of \eqref{notethat} and \eqref{de'==1}.

%in this case the Wick pairing 
%$\EE a_{\xi_i}^{(0)}\bar a_{\s_j}^{(0)}=0$ for every $(\xi,\s)\in\cA_s(\gD)$, because $\xi_i\neq \s_j$ by \eqref{de'==1} 
% (we recall the relation $\{s_1,s_2\}\ne\{s_3,s\}$ from \eqref{def_del}). 

Let $\gF(\gD)$ be the set of all Feynman diagrams,  which may be 
obtained from $\gD$ using different  partitions of the set of leaves to pairs. 
We define
\be\lbl{gF-def}
\gF_{m,n}=\bigcup_{\gD\in\gD_m\times\ov\gD_n} \gF(\gD).
\ee
This is the set of all Feynman diagrams corresponding to the product $a^{(m)}_s \bar a^{(n)}_s$.
By construction,  the set of vertices of any Feynman diagram $\gF\in \gF_{m,n}$ is partitioned into pairs of 
adjacent vertices $(c_i,\bar c_j)$. Each pair has either the form 
$(c^{(p)}_i,\bar w_j)$ or $(w_i,\bar c^{(p)}_j)$, where $p\geq 1$, or the form 
$(c_i^{(0)},\bar c_j^{(0)})$.
Moreover, adjacent vertices  never belong to the same block. To any $\gD\in\gD_m\times\ov\gD_n$ the construction above 
corresponds many diagrams $\gF\in \gF_{m,n}$, but the inverse mapping
\be\lbl{D_FF}
\gF_{m,n} \to \gD_m\times\ov\gD_n, \quad \gF \mapsto \gD_\gF,
\ee
is well defined: the diagram $\gD_\gF$ is obtained from $\gF$ by erasing the edges that couple its leaves.

 In fact, by demanding  that in Feynman diagrams  coupled leaves never 
   belong  to the same block, we have excluded only a part of vanishing Wick pairings.
To exclude the remaining part, in Section~\ref{s:ch_of_c} we will replace the set $\gF_{m,n}$ of Feynman diagrams by a smaller set $\gF^{\,true}_{m,n} \subset \gF_{m,n}$. For the Wick pairings corresponding to  diagrams $\gF\notin\gF^{\,true}_{m,n}$
 the assumption  $\{s_1,s_2\}\ne\{s_3,s\}$, following from  \eqref{def_del},   again is violated due to
  \eqref{de'==1}, but in a less trivial way (see Proposition~\ref{l:Az}).

\subsubsection{Set of admissible multi-indices $\cA_s'(\gF)$} 
\lbl{s:adm_ind}

Let us orient edges of a Feynman diagram $\gF\in\gF_{m,n}$ in the direction from conjugated vertices to non-conjugated, as in fig.~\ref{f:F6}(b),
and consider a permutation $f_\gF$ of the set 
$\{0, \dots, 2N\}$ such that
\be\label{pi_F}
f_\gF(j)= i \; \text{ if $\bar c_j\sim c_i$, i.e. an edge of $\gF$ goes from $\bar c_j$ to $c_i$,}
\ee
see fig.~\ref{f:F6}(b).
For a vector $\xi\in\R^{(2N+1)d}$ we denote 
\be\lbl{xi_f}
\xi_{f_\gF}=(\xi_{f_\gF(0)},\ldots,\xi_{f_\gF(2N)}).
\ee
Let $\cA'_s(\gF)$ be a set of all multi-indices   $\xi\in\R^{(2N+1)d}$ satisfying the following relations: for $\s=\xi_{f_\gF}$,
\smallskip

\begin{itemize}
	\item We have
	\be\lbl{p'1}
\xi_0=\s_0=s.
\ee
\item For any $1\leq i\leq N$,
\be\lbl{p'2} \de'^{\xi_{2i-1} \xi_{2i}}_{\s_{2i-1}\s_{2i}}=1.
\ee
\end{itemize}
\smallskip
 
\noindent 
In other words, we require \eqref{p'1} and \eqref{p'2} to be satisfied by the multi-indices  $\xi,\s$, where $\s$ is obtained from $\xi$ accordingly to the 
edges of  diagram $\gF$: $\s_j=\xi_i$ if a vertex $\bar c_j$ is coupled with the vertex $c_i$.
In particular, if $\xi\in\cA_s^{\prime}(\gF)$ then $(\xi,\xi_{f_\gF})\in \cA_s(\gD_\gF)$, where $ \cA_s(\gD)$ is defined in 
Section~\ref{s_3.1.5} and  the diagram $\gD_\gF$ -- in \eqref{D_FF}. 
Indeed, \eqref{p4} holds by the definition \eqref{pi_F} of the permutation $f_\gF$ while \eqref{p1} and \eqref{de'==1} are obvious.

We also consider the discrete set $\cA_s^{\prime L}(\gF)=\cA'_s(\gF)\cap \Z^{(2N+1)d}_L$.

\smallskip

Our motivation behind these definitions is as follows.  
As it is explained in the beginning of the section, each term $\EE I(\gD)$ can be written as a sum over the Feynman diagrams $\gF$, 
$$\EE I_s(\gD)=\sum_{\gF\in\gF(\gD)} J_s(\gF).$$ 
Each summand $J_s(\gF)$ has the form \eqref{a^ma^n-I} where $G_\gD$ is replaced by a function obtained by the corresponding to $\gF$ Wick pairing of the 
values  $a_{\xi_i}^{(0)}$, $\bar a_{\s_j}^{(0)}$ of leaves. In this formula for $J_s(\gF)$ we should take summation only over those
 multi-indices  $(\xi,\s)\in\cA_s^L(\gD)$ for which $\xi_i=\s_j$ once the vertices $c_i$ and $\bar c_j$ are adjacent in $\gF$, that is over vectors of the form $(\xi,\xi_{f_\gF})$, where $\xi\in\cA_s^{\prime L}(\gF)$.

\subsection{Formula for the expectation $\EE a^{(m)}_s \bar a^{(n)}_s$} 
\lbl{s:F_not}

Below we introduce some more notation. Let $\gF\in\gF_{m,n}$ be a Feynman diagram.

 $\bullet$ We denote by $E(\gF)$ the set of edges of the diagram $\gF$, then 
  $E(\gD_\gF)\subset E(\gF)$ (see \eqref{D_FF}). We also set 
  $$
   E_D(\gF):=E(\gD_\gF), \quad E_{L}(\gF) := E(\gF) \setminus  E_D(\gF)\quad\text{so} \quad E(\gF)=E_D(\gF)\cup E_L(\gF).
$$

$\bullet$ For  an edge $\vartheta \in E(\gF)$ which couples some vertices $c_i$ and $\bar c_j$, we set 
$c_\vartheta:=c_i$, $\bar c_\vartheta:=\bar c_j$, and denote 
 by $l^{c_\vartheta}$ and $l^{\bar c_\vartheta}$  the times assigned to the vertices $c_\vartheta$ and $\bar c_\vartheta$. 
 We also set
 \be\non
 \ga^{\vartheta}(\xi):=\ga_{\xi_i}
 \qnd
 b^\vartheta(\xi):=b(\xi_i) \quad \text{if \ $\vartheta $ couples $c_i$ and $c_j$},
 \ee
 where $\gamma_s$ is defined in \eqref{diss_op} and $b(s)$ enters the definition of the random force $\eta^\om$. 

$\bullet$ If $\phi\in E_D(\gF)$, we define the times $l_r^\phi$ and $l_w^\phi$ as in Section~\ref{s:more_not}.
 
\smallskip
 
The main result of this section is Lemma \ref{l:Ea^ma^n},  stated below. 
Recalling \eqref{tau} and \eqref{G_D}, \eqref{corr_a_in_time}, to each Feynman diagram $\gF$ we associate the density 
 function
$c_\gF F^\gF(\tau, l,\xi)$, where $c_\gF = c_{\gD_\gF}$ (see \eqref{c_D-def}) and $F^\gF(\tau, l,\xi)$ is the real function 
\begin{align}\non
F^\gF(\tau, l,\xi)=&\prod_{\phi\in E_D(\gF)} e^{-\ga^{\phi}(l^{\phi}_r - l^{\phi}_w)}\mI_{\{-T\leq l^{\phi}_w \leq l^{\phi}_r\}} (\tau,l)\\ \lbl{F=}
&\prod_{\psi\in E_L(\gF)} 
\Big(e^{-\ga^{\psi}|l^{c_\psi}-l^{\bar c_\psi}|}- e^{-\ga^{\psi}(l^{c_\psi}+l^{\bar c_\psi}+2T)}\Big)
\frac{(b^{\psi})^2}{\ga^{\psi}},
\end{align}
where
$\ga^{\phi}=\ga^{\phi}(\xi)$,  $\ga^{\psi}=\ga^{\psi}(\xi)$,
$b^\psi=b^\psi(\xi)$.
We also associate to  $\gF$  the quadratic forms 
\be\lbl{Omeggga}
 %\Om^\gF(l,\xi):=\Om(l,\xi,\xi_{f_\gF})=\sum_{j=1}^N l_j \om_j^\gF(\xi), \quad\text{where}\quad 
\om_j^\gF(\xi):=\om_j(\xi,\xi_{f_\gF})= \omega^{\xi_{2j-1} \xi_{2j}}_{\xi_{f_{\gF}(2j-1)} \xi_{f_{\gF}(2j)}}
 \qu\mbox {and set}\qu \om^\gF=(\om_1^\gF,\ldots,\om_N^\gF),
\ee
see \eqref{Omega_def} and  \eqref{xi_f}.
\begin{lemma}
	\lbl{l:Ea^ma^n}
	For any integers $m,n\geq 0$ satisfying $N=m+n\geq 1$, for $s\in\Z^d_L$ and 
	$\tau_1,\tau_2\geq -T$, we have
	\be\lbl{formula_Ea^ma^n}
	\EE a^{(m)}_s(\tau_1) \bar a^{(n)}_s(\tau_2)
	=   \sum_{\gF\in\gF_{m,n}} c_\gF J_s(\gF),
	\ee
	where $J_s(\gF)=J_s(\tau;\gF)$
	\be\lbl{Ea^ma^n-J}
	J_s(\tau;\gF)=
	\int_{\R^N}
	dl\, 
	L^{-Nd}\sum_{\xi\in\cA^{\prime L}_s(\gF)}\, F^\gF\,(\tau, l, \xi)\, 
	\Theta(\om^\gF(\xi),\nu^{-1}l).
	\ee	
	 The density $F^\gF$, given by \eqref{F=}, is real, smooth in $\xi \in\R^{(2N+1)d}$, piecewise smooth in $l\in \R^N$ and 
	  such that  for any vector $\ka \in\Z_+^{d(2N+1)}$ we have:
	\be\lbl{F-prop}
	|\p_{\xi}^{\ka}F^\gF(\tau,l,\xi)|\leq C_{\ka}^{\#}(\xi)\,e^{- \de \big(\sum_{i=1}^m|\tau_1-l_i|
	+\sum_{i=m+1}^N|\tau_2-l_i|\big)},
	\ee
	for any $\xi\in\cA_s'(\gF)$,  $l\in\R^N$ and $\tau_1,\tau_2\geq -T$, where $\de=\de_N>0$. 
%	Each sum $J_s(\gF)$ converges absolutely. 
	\end{lemma}

Proof of \eqref{formula_Ea^ma^n} can be obtained by taking the expectation of both sides of \eqref{a^ma^n} and is  essentially explained in previous discussion of this section.  
Estimate \eqref{F-prop} follows from the explicit form \eqref{F=} of the density $F^\gF$. 
Nevertheless, we give an accurate proof  in Appendix~\ref{s:Ea^ma^n}.
Below we do not use the explicit form of $F^\gF$ but only the estimate \eqref{F-prop}.

Let us note that 
\be\lbl{N=1-empty}
\text{
if $\;m+n=1, \;$ then $\;\EE a^{(m)}_s \bar a^{(n)}_s=0$.}
\ee
 Indeed, in this case  $\gF_{m,n}=\emptyset$, since every partition of the set of leaves into pairs couples vertices from the same block. Accordingly, we will often assume that $m+n\ge 2$. 
 %So below we usually assume $m+n\geq 2$.

\section{Change of coordinates and final formula for expectations $\EE a^{(m)}_s\bar a^{(n)}_s$}
\lbl{s:ch}

Lemma~\ref{l:Ea^ma^n} gives an explicit formula for the expectation
$\EE a^{(m)}_s \bar a^{(n)}_s$,
however, the structure of the set of multi-indices 
$\cA_s^{\prime L}(\gF)$, over which we take a summation in \eqref{Ea^ma^n-J}, is too complicated for further analysis.
In this section we find new coordinates in which this set  and the quadratic forms \eqref{Omeggga} 
 take convenient form.
% appropriate for study of the asymptotic behaviour as $\nu\to 0$ and $L \to \infty$ of the sum $J_s(\gF)$. 

By \eqref{p'2} the vectors $\xi\in\cA_s^{\prime L}$ satisfy the system of equations
\be\lbl{lin_syst_c}
\xi_{2k-1}+\xi_{2k}=\s_{2k-1} + \s_{2k}, \qquad \s:=\xi_{f_\gF}, \qquad 1\leq k\leq N. 
\ee
We start with constructing coordinates $x=(x_0,\dots,x_{2N})$, where $x_0=\xi_0$ and for every $1\leq k \leq N$ we set either 
$x_{2k-1}=\xi_{2k-1}-\s_{2k-1}$ and $x_{2k}=\xi_{2k}-\s_{2k}$ or
$x_{2k-1}=\xi_{2k-1}-\s_{2k}$ and $x_{2k}=\xi_{2k}-\s_{2k-1}$,
so that equations \eqref{lin_syst_c} take trivial form $x_{2k}=-x_{2k-1}$, $\forall k$.
We will make a choice (different for different $k$ and dependent on the diagram $\gF$) between the two possibilities above  in such a way that the transformation $\xi\mapsto x$ is invertible. Then, components $z_k:=x_{2k-1}$, $1\leq k\leq N$, form coordinates on the set $\cA_s^{\prime L}$ while components $x_{2k}$ are functions of $z=(z_k)_{1\leq k\leq N}$. 

To construct this transformation $\xi\mapsto x$ we add to the diagram $\gF$ dashed edges which couple vertices inside each block in such a way that the obtained diagram becomes a Hamilton cycle, see fig.~\ref{f:F7}(b) (we show that it is always possible to do). 
 For every $k$ we choose
between the two possibilities above in accordance with the dashed edges. 
It turns out that in the coordinates $z$  the function $\Theta(\om^\gF,\nu^{-1}l)$ with $\theta(x,t)=e^{ixt}$ takes a simple form \eqref{Theta-exp}-\eqref{Om_xy}. This leads to an  explicit formula for the sum  $J_s(\gF)$ and next in 
Sections~\ref{s:est},\,\ref{est-imroved} allows  to analyse its asymptotic behaviour as $\nu\to 0$, $L\to\infty$.
Next we explain construction of the transformation $\xi\mapsto x$ in detail.

\subsection{Cycles $\gC_\gF, \hat{\gC}_\gF$ and permutation $\pi_\gF$}
\lbl{s:cycles}

Let us take a Feynman diagram $\gF\in\gF_{m,n}$ with $N=m+n\geq 2$. 
By construction, if two vertices of $\gF$ belong to the same block then they are not adjacent. 
Now inside each block 
$\hat B_k$,  $k\geq 1$, (see \eqref{B_kkk}-\eqref{barB_kkk}) we join  by four dashed edges conjugated vertices with non-conjugated in all possible ways, see fig.~\ref{f:F7}(a).
We also  join by a dashed edge the two roots $c_0$, $\bar c_0$ and denote the resulting graph by $\mathfrak{U}_\gF$.

\begin{figure}[t]
	\qu a)\qu\parbox{5cm}{ 
		\begin{tikzpicture}
		\node at (-1,0) (c0) {$c_0^{(2)}$};
		\node at (1,0) (bc0) {$\bar c_{0}^{(0)}$};
		
		\node at (-2.5, -1.2) (c1) {$c_1^{(1)}$};
		\node at (-1.6, -1.2) (c2) {$c_2^{(0)}$};
		\node at (-0.5, -1.2) ( bc1) {$\bar c_1^{(0)}$};
		\node at (0.4, -1.3) (bw2)  {$\bar w_{2}$};  
		
		\node at (-2.5, -2.5) (c3) {$c_3^{(0)}$};
		\node at (-1.6, -2.5) (c4) {$c_4^{(0)}$};
		\node at (-0.5, -2.5) (bc3) {$\bar c_3^{(0)}$};
		\node at (0.4, -2.6) (bw4) {$\bar w_{4}$}; 
		
		\draw [line width=0.25mm] (-0.8,-0.3)-- (0.35, -1);
		%c1-bw4:
		\draw [line width=0.25mm](c1.south)--(bw4.north);
		%c3-c0:
		\draw [line width=0.25mm] (c3.south) -- (-2.5,-3.4) --(1,-3.4) --  (bc0.south);
		%c2-bc3:
		\draw [line width=0.25mm](-1.6, -1.5)--(-0.9, -2.2);
		%c4-bc1:
		\draw [line width=0.25mm](-1.6, -2.2)--(-0.7, -1.5);
		
		%c0-bc0:
		\draw [dashed,line width=0.25mm](-0.8,-0.1)--(0.6,-0.1);
		%c2-bc1
		\draw [dashed,line width=0.25mm](-1.5, -1.3)--(-0.9, -1.3);
		%c2-bw2
	%	\draw [dashed,line width=0.25mm](-1.5, -1.3)..controls (-0.6, -1.8) ..(0.2, -1.5);
		\draw [dashed,line width=0.25mm](-1.5, -1.5) .. controls (-0.8,-1.7) .. (0.2, -1.5);
		%c1-bw2:
		\draw [dashed,line width=0.25mm] (-2.2, -0.9) .. controls (-1.1,-0.4) .. (0.2, -1.1);
		%c1-bc1:
	%	\draw [dashed,line width=0.25mm] (-2.2, -0.9) .. controls (-1.5,-0.7) .. (-0.7, -1.1);
		\draw [dashed,line width=0.25mm](-2.4, -1.5) .. controls (-1.7,-1.7) .. (-0.7, -1.5);
		%c3-bc3:
		\draw [dashed,line width=0.25mm] (-2.4, -2.8) .. controls (-1.7,-3) .. (-0.7, -2.8);
		%c4-bw4:
		\draw [dashed, line width=0.25mm] (-1.5, -2.8) .. controls (-0.8,-3) .. (0.2, -2.8);
		%c4-bc3:
		\draw [dashed, line width=0.25mm] (-1.5, -2.6) -- (-0.7, -2.6);
		%c3-bw4:
		\draw [dashed,line width=0.25mm] (-2.4, -2.85) .. controls (-1.1,-3.4) .. (0.2, -2.85);
		\end{tikzpicture}
	}
	\hfill
	b)\qu\parbox{5cm}{ 
		\begin{tikzpicture}
		\node at (-1,0) (c0) {$c_0^{(2)}$};
		\node at (1,0) (bc0) {$\bar c_{0}^{(0)}$};
		
		\node at (-2.5, -1.2) (c1) {$c_1^{(1)}$};
		\node at (-1.6, -1.2) (c2) {$c_2^{(0)}$};
		\node at (-0.5, -1.2) ( bc1) {$\bar c_1^{(0)}$};
		\node at (0.4, -1.3) (bw2)  {$\bar w_{2}$};  
		
		\node at (-2.5, -2.5) (c3) {$c_3^{(0)}$};
		\node at (-1.6, -2.5) (c4) {$c_4^{(0)}$};
		\node at (-0.5, -2.5) (bc3) {$\bar c_3^{(0)}$};
		\node at (0.4, -2.6) (bw4) {$\bar w_{4}$}; 
		
		%c0-bw2:
		\draw [<-,line width=0.25mm] (-0.8,-0.3)-- 
		%node[above, near end]{{\scriptsize $f_\gF$}} 
		(0.35, -1) ;
		%c1-bw4
		\draw [<-,line width=0.25mm](c1.south)--
		%node[above, near end]{{\scriptsize $f_\gF$}}
		(bw4.north);
		%c3-bc0:
		\draw [<-,line width=0.25mm] (c3.south) -- (-2.5,-3.2) -- 
		%node[below]{{\scriptsize $f_\gF$}}  
		(1,-3.2) --  
		%node[right]{{\scriptsize $f_\gF$}} 
		(bc0.south);
		%c2-bc3:
		\draw [<-,line width=0.25mm](-1.6, -1.5)--(-0.9, -2.2);
		%c4-bc1:
		\draw [<-,line width=0.25mm](-1.6, -2.2)--(-0.8, -1.5);
		
		%c0-bc0:
		\draw [dashed,->,line width=0.25mm](-0.8,-0.1)--
		%node[above]{{\scriptsize $g_\gF$}}
		(0.6,-0.1);
		%c2-bc1
		\draw [dashed,->,line width=0.25mm](-1.5, -1.3)--
		%node[above, near end]{{\scriptsize $g_\gF$}}
		(-0.9, -1.3);
		%c1-bw2:
		\draw [dashed,->,line width=0.25mm] (-2.2, -0.9) .. controls (-1.1,-0.4) .. %node[above, very near start]{{\scriptsize $g_\gF$}}
		(0.2, -1.1);
		%c3-bc3:
		\draw [dashed,->,line width=0.25mm] (-2.4, -2.8) .. controls (-1.7,-3) .. (-0.7, -2.8);
		%c4-bw4:
		\draw [dashed,->,line width=0.25mm] (-1.5, -2.8) .. controls (-0.8,-3) .. (0.2, -2.8);
		\end{tikzpicture}
	}
	\caption{ a)  The graph $\mathfrak{U}_\gF$ obtained from the Feynman diagram $\gF$ from fig.~\ref{f:F6}. b) A Hamilton cycle $\gC_\gF$ in the graph $\mathfrak{U}_\gF$ from item (a). 
		%together with the permutation $f_\gF$.
	}
	\lbl{f:F7}
\end{figure}

	\begin{lemma}
	\lbl{l:cycle}
	For each Feynman diagram $\gF\in\gF_{m,n}$ the graph $\mathfrak{U}_\gF$ has a Hamilton cycle
	in which solid and dashed edges alternate.
\end{lemma} 
Proof of Lemma~\ref{l:cycle}, given in Appendix~\ref{s:cycle}, 
follows from the fact that the diagram, made by the blocks of $\gF$,  is connected. 
In general, the Hamilton cycle in Lemma~\ref{l:cycle} is not unique, so we fix one for each diagram $\mathfrak U_\gF$ and denote it as $\gC_\gF$. Examples of the Hamilton cycles $\gC_\gF$ are given in fig.~\ref{f:F7}(b) and fig.~\ref{f:ex}. 
It is straightforward to see that, by construction of the diagrams $\mathfrak{U}_\gF$, the cycles $\gC_\gF$ contain all solid edges of $\mathfrak{U}_\gF$ (which are the edges of $\gF$), and a half of the dashed edges: for each block exactly two of four dashed edges that couple vertices inside the block enter the cycle $\gC_\gF$. On the cycle $\gC_\gF$  the dashes and solid edges 
alternate. 

We orient the dashed edges of $\gC_\gF$ in the direction from non-conjugated vertices to conjugated and recall that the solid edges are oriented in the opposite direction, from conjugated vertices to non-conjugated. Then, $\gC_\gF$ becomes an oriented cycle, see fig.~\ref{f:F7}(b).

\begin{comment}
We orient the dashed edges of the diagram $\gC_\gF$ in the direction from the non-conjugated vertices to conjugated, and orient its solid edges  
(i.e. the edges of $\gC_\gF$ which are also edges of $\gF$) in the direction from the conjugated vertices to non-conjugated, see fig.~\ref{f:F7}(b). Now we define a permutation $f_\gF$ of the set 
$\{0, \dots, 2N\}$ by the relation 
\be\label{pi_F}
f_\gF(j)= i \; \text{ if a solid edge of $\gC_\gF$ goes from $\bar c_j$ to $c_i$.
}
\ee
In  terms of this permutation assumption \eqref{p4} may be reformulated as 
\be\label{wt_pi}
\sigma_j =\xi_{f_\gF(j)} \quad \forall\, 0\le j\le 2N, \quad \forall\,(\xi, \sigma)\in \cA_s(\gF). 
\ee
\end{comment}

\begin{comment}
To construct the desired transformation $\xi\mapsto x$ we define the permutation  
\be\non
\pi_\gF:=f_\gF\circ g_\gF
\ee
of the set $\{0, \dots, 2N\}$, where we recall that the permutation $f_\gF$ is defined in \eqref{pi_F}.
It is convenient to represent $\pi_\gF$  graphically. To this end we define the reduced  cycle graph $\hat \gC_\gF$, obtained from $\gC_\gF$ by the following procedure: for every vertex $c_i$ of 
$\gF$ we merge the dashed edge  $(i,g_\gF(i))$ and the two vertices  $c_i, \bar c_{g_\gF(i)}$ to the new vertex 
 $c_i$, see fig.~\ref{f:F8}. The cycle $\hat \gC_\gF$ is an oriented graph with the vertices 
$c_0, \dots, c_{2N}$, so permutation $\pi_\gF$  takes the form
\end{comment}
Let us denote by $\hat\gC_\gF$ the reduced oriented cycle $\gC_\gF$, that is the cycle $\gC_\gF$ in which we keep only non-conjugated vertices and "forget" all conjugated ones, see  fig.~\ref{f:F8}.
The cycle $\hat\gC_\gF$ defines a cyclic permutation $\pi_\gF$ of the set of parameters for the non-conjugated vertices 
$\{0,\dots,2N\}$ as
$$
\pi_\gF(j)=i \qu \text{if an edge of $\hat \gC_\gF$ goes from $c_j$ to $c_i$. }
$$
\begin{figure}[t]
\parbox{3cm}{ 
		\begin{tikzpicture}
		\node at (0,0) (c0) {$c_0$};
		\node at (1,-1) (c3) {$c_3$};
		\node at (1,-2) (c2) {$c_2$};
		\node at (-1,-2) (c4) {$c_4$};
		\node at (-1,-1) (c1) {$c_1$};
		
		\draw[->, line width=0.25mm] (c0) --node[below,near start]{{\footnotesize $\pi_\gF$}}(c3);
		\draw[->, line width=0.25mm] (c3) --(c2);
		\draw[->, line width=0.25mm] (c2) -- node[above]{{\footnotesize $\pi_\gF$}}(c4);
		\draw[->, line width=0.25mm] (c4) --(c1);
		\draw[->, line width=0.25mm] (c1) --(c0);
		
		\draw[dashed] (c1.north) .. controls (0,2) and (2,-1) .. node[above]{{\small $A_1$}}(c2);
		\draw[dashed] (c3) .. controls (1.5,-1.5) and (2,-3) .. node[below,near end]{{\small $A_2$}}(c4.south);
		\end{tikzpicture}
	}
\begin{comment}
\hfill
b)\fbox{\parbox{4.4cm}{ 
	\begin{tikzpicture}
	\node at (0,0) (c0) {$c_0$};
	\node at (1,-1) (c2) {$c_2$};
	\node at (2,-2) (c4) {$c_4$};
	\node at (1,-3) (c1) {$c_1$};
	\node at (-1,-3) (c5) {$c_5$};
	\node at (-2,-2) (c3) {$c_3$};
	\node at (-1,-1) (c6) {$c_6$};
	
	\draw[->, line width=0.25mm] (c0) --(c2);
	\draw[->, line width=0.25mm] (c2) --(c4);
	\draw[->, line width=0.25mm] (c4) --(c1);
	\draw[->, line width=0.25mm] (c1) --(c5);
	\draw[->, line width=0.25mm] (c5) --(c3);
	\draw[->, line width=0.25mm] (c3) --(c6);
	\draw[->, line width=0.25mm] (c6) --(c0);
	
	\draw[dashed, line width=0.25mm] (c5) .. controls (-3.5,-2) .. (c0.west);
	\end{tikzpicture}
}}
\end{comment}
\caption{Cycle $\hat \gC_\gF$ corresponding to the cycle $\gC_\gF$ in fig.~\ref{f:F7}(b), the associated permutation $\pi_\gF$ and 
	 the arcs $A_1:=[c_1,c_2]$,
	 $A_2:=[c_3,c_4]$. Here we have $\al_{12}^\gF=-\al^\gF_{21}=1$.}
\lbl{f:F8}
\end{figure}
 The graph $\mathfrak{U}_\gF$ and the Hamilton cycle $\gC_\gF$ play a central role in our
  construction. Nevertheless, below we use only the "derivative" objects: the reduced cycle $\hat\gC_\gF$ and the permutation $\pi_\gF$. 

\subsection{The change of coordinates}
\lbl{s:ch_of_c}
For a fixed Feynman diagram $\gF\in\gF_{m,n}$ with $N=m+n\geq 1$, we set  
\be\lbl{x,y-xi,x}
x_0:=\xi_0,\qu x_j:=\xi_j-\xi_{\pi_{\gF}(j)}\quad \forall 1\leq j\leq 2N,
\ee
and define the $(2N+1)$--vector 
 $x=(x_j)_{0\leq j\leq 2N}$, $x_j\in\R^d$. 
 Since  $\pi_\gF$ is a cyclic permutation of the set $\{0,\ldots, 2N\}$, then the transformation $\xi\mapsto x$ is a bijection of $(\R^{d})^{2N+1}$. Indeed, it is invertible since iterating \eqref{x,y-xi,x} we get
 \be\non
 \xi_j=x_j+\xi_{\pi_{\gF}(j)} =x_j+x_{\pi_{\gF}(j)} + \xi_{\pi^2_{\gF}(j)}
 =\ldots= \sum_{i=0}^k x_{\pi^i_{\gF}(j)} +\xi_0,
 \ee
 where $k=k(j)$ denotes minimal positive integer satisfying $\pi^{k+1}_{\gF}(j)=0$.
 In other words, since $\xi_0=x_0$,
  \be\lbl{xi(x)}
  \xi_j=\sum_{i:\,c_i\in [c_j,c_0]} x_i, \qquad 0\leq j\leq 2N,
  \ee
where  by 
  $[c_a,c_b]$, $a\ne b$, we denote a set (an {\it arc}) of vertices  of the cycle $\hat\gC_\gF$ that are situated between the vertices $c_a$ and $c_b$ according to the orientation of $\hat\gC_\gF$, including $c_a$ and $c_b$. E.g., in fig.~\ref{f:F8} $[c_4,c_3]=\{c_4,c_1,c_0,c_3\}.$
  For $a=b$ we set $[c_a,c_b]=\{c_a\}$.
  
  Next we write the set of admissible multi-indices $\cA_s'(\gF)\subset \R^{d(2N-1)} $ in the coordinates $x$.
  An important role below is played  by the $N\times N$ {\it incidence matrix}  $\al^\gF=(\al_{ij}^\gF)_{1\leq i,j\leq N}$ with components
  \footnote{Formula \eqref{al_ij} admits the following interpretation. An element $\al_{ij}^\gF$ equals $\pm 1$ if the arcs $[c_{2i-1},c_{2i}]$ and $[c_{2j-1},c_{2j}]$ are linked, and equals $0$ otherwise. The sign "$+$" or "$-$" is determined by the order of linking; see fig.~\ref{f:F8}.  So the incidence matrix $\al^\gF$ describes the linking of the arcs  $[c_{2j-1},c_{2j}]$ on 
   the reduced cycle $\hat \gC_\gF$, or equivalently on the Hamiltonian cycle $\gC_\gF$ in the diagram  $\gF$.  
   }
  \be\lbl{al_ij}
  \al_{ij}^\gF=\left\{
  \begin{array}{cl}
  	1&\qmb{if}\qu c_{2j-1}\in [c_{2i-1},c_{2i}], \;\;  c_{2j}\notin [c_{2i-1},c_{2i}],\\
  	-1&\qmb{if}\qu c_{2j-1}\notin [c_{2i-1},c_{2i}], \;\;  c_{2j}\in [c_{2i-1},c_{2i}],\\
  	0&\qmb{otherwise}.
  \end{array}
  \right.
  \ee
  It is straightforward to see that the matrix $\al^\gF$ is skew-symmetric,  so it is S-regular  in the following sense:

  \begin{definition}\label{Sreg}
  	A square matrix $\al=\{\al_{ij}\}$ is called S-regular if it is skew-symmetric and all its elements $\al_{ij}$ are such that
  	$
  	\al_{ij}\in \{-1, 0, 1\}.
  	$
  	If in addition  all  rows and columns of $\al$  are non-zero vectors, then it is called SS-regular. 
  \end{definition}

  Given below  Propositions \ref{l:change} and 
  \ref{l:Om-change}  are in the hart of the proof of the main result of this work -- 
  Theorem~\ref{t:mainpr}. Previous constructions in Sections~\ref{s:D}-\ref{s:cycles} essentially were made in order to create 
  the notation,   needed to state and  prove these results.

\begin{proposition}
\lbl{l:change}
For any diagram $\gF\in\gF_{m,n}$ and any $s\in\R^d$,
the set $\cA_s'(\gF)$ consists of all vectors $\xi=\xi(x)$, where $\xi(x)$ is given by \eqref{xi(x)} and $x\in(\R^{d})^{2N+1}$ satisfies
 the following relations, for every  $1\leq k \leq N$:
\begin{align}\lbl{zk1} 
&(a) \
x_0=s, \qquad\qu
(b) \
x_{2k}=-x_{2k-1}, \\
\lbl{zk2}
& (a) \ x_{2k-1}\neq 0 \qquad\qu
(b) \  \sum_{i=1}^N\al_{ki}^\gF x_{2i-1}\neq 0.
\end{align}  
Moreover, for vectors $x$ satisfying \eqref{zk1}(b) we have
\be\lbl{differ-xi}
\xi_{2k-1}{ (x)}- \xi_{\pi_\gF(2k)}(x) =  \sum_{i=1}^N \al_{ki}^\gF x_{2i-1}.
\ee
\end{proposition}
{\it Proof.} 
	In this proof we omit the lower index $\gF$ in the notation $\pi_\gF$ and $f_\gF$. We start by noting that for every $k\geq 1$ 
	\be\lbl{pi=f}
	\{\pi(2k-1),\pi(2k)\}=\{f(2k-1),f(2k)\}.
	\ee 
	Indeed, the permutation $\pi$ encodes the change of indices of non-conjugated vertices of the cycle $\gC_\gF$ under the double shift, the first along dashed edges and the  second along solid edges. The first shift preserves the set of indices $\{2k-1,2k\}$ while the second is encoded by the permutation $f$.
 
Now let us recall that the set $\cA_s'(\gF)$ consists of all multi-indices 
 $\xi$ satisfying \eqref{p'1} and \eqref{p'2} with $\s=\xi_f$.
In view of what is told above,  we have $\{\xi_{f(2k-1)},\xi_{f(2k)}\}=\{\xi_{\pi(2k-1)},\xi_{\pi(2k)}\}$, so 
$\de'^{\xi_{2k-1}\xi_{2k}}_{\xi_{f(2k-1)}\xi_{f(2k)}}=\de'^{\xi_{2k-1}\xi_{2k}}_{\xi_{\pi(2k-1)}\xi_{\pi(2k)}}$.
Then equation \eqref{p'2} is equivalent to 
$\de'^{\xi_{2k-1}\xi_{2k}}_{\xi_{\pi(2k-1)}\xi_{\pi(2k)}}=1$, that is to the equation
\be\lbl{de'1}
\xi_{2k-1}+\xi_{2k}=\xi_{\pi(2k-1)} + \xi_{\pi(2k)}
\ee
jointly  with 
\be\lbl{intersec=empty}
\{\xi_{2k-1},\xi_{2k}\}\cap \{\xi_{\pi(2k-1)},\xi_{\pi(2k)}\}=\emptyset
\ee
(see \eqref{notethat}). 
Equation \eqref{de'1} is equivalent to \eqref{zk1}(b).  
The equality $\xi_0=s$ from \eqref{p'1} is equivalent to \eqref{zk1}(a). The equality $\s_0=s$, in view of \eqref{xi(x)}, takes the form
\be\lbl{s=s}
s= \sigma_0 =
\xi_{f(0)}=\sum_{i:\,c_i\in [c_{f(0)},c_0]} x_i.
\ee
Since $f(0)=\pi(0)$, then $[c_{f(0)},c_0]=[c_{\pi(0)},c_0]$, and this interval contains all vertices of the cycle $\hat\gC_\gF$. 
Then, using \eqref{zk1}(b), we see that the  r.h.s.  of \eqref{s=s} equals to $x_0$,
so \eqref{s=s} again takes the form \eqref{zk1}(a).

Now it remains to show that \eqref{intersec=empty} is equivalent to \eqref{zk2}.
In view of \eqref{de'1}, relation \eqref{intersec=empty} is equivalent to 
\be\lbl{intersec=empty''}
\xi_{2k-1}\neq \xi_{\pi(2k-1)} \qnd \xi_{2k-1}\neq \xi_{\pi(2k)}. 
\ee
The first inequality above is equivalent to  \eqref{zk2}(a),  while the second is equivalent to  \eqref{zk2}(b) if
\eqref{differ-xi} is established. So it remains to verify \eqref{differ-xi}.
%it suffices to show that 
%\be\lbl{differ-xi}
%\xi_{2k-1}- \xi_{\pi(2k)} =  \sum_{i=1}^N \al_{ki}^\gF x_{2i-1},
%\ee
%for every $1\leq k\leq N$.
  Due to \eqref{xi(x)}, 
$$
\xi_{2k-1}- \xi_{\pi(2k)}= \sum_{i:\,c_i\in [c_{2k-1},c_0]} x_i - \sum_{i:\,c_i\in [c_{\pi(2k)},c_0]} x_i. 
$$
Let us first assume that $c_{\pi(2k)} \in [c_{2k-1},c_0]$ and $c_{\pi(2k)}\ne c_{2k-1}$.
Then the identity above takes the form
\be\non
\xi_{2k-1}- \xi_{\pi(2k)} =\sum_{i:\,c_i\in [c_{2k-1},c_{2k}]} x_i,
\ee
where we have used that the vertex situating in $\hat\gC_\gF$ before $c_{\pi(2k)}$ is $c_{2k}$. In view of the assumption above, $c_0\notin  [c_{2k-1},c_{2k}]$. 
Then, due to cancellations provided by \eqref{zk1}(b), we get \eqref{differ-xi}.

The case $c_{\pi(2k)} \notin [c_{2k-1},c_0]$ can be considered similarly. 
If $c_{\pi(2k)}= c_{2k-1}$, that is $\pi(2k)= 2k-1$, then the  both sides of equation \eqref{differ-xi} vanish. Indeed,  $\al^\gF_{ki}=0$ $\forall i$ since the arc $[c_{2k-1},c_{2k}]$ contains all vertices $c_j\in\hat\gC_\gF$. 
\qed
\begin{comment}
Let us simplify the last sum above, using \eqref{zk1}(b). 
The terms $x_{2k-1}$ and $x_{2k}$ cancel each other, so the sum can be written as
$\sum_{i:\,c_i\in A_k} x_i$, where we recall that $A_k=(c_{2k-1},c_{2k})$. 
Now we fix some $1\leq i\leq 2N$ and assume that we have neither $A_i\rightarrow A_k$ nor $A_i\leftarrow A_k$.
Then the vertices $c_{2i-1}$ and $c_{2i}$  simultaneously 
belong or  belong not to the arc $A_k$, and 
by \eqref{zk1}(b)
this implies that $x_{2i-1}$ and $x_{2i}$ do not give input to the sum. 
Then, using again \eqref{zk1}(b) and the fact that $c_0\notin A_k$, we arrive at \eqref{differ-xi}.
The relation $c_0\notin A_k$ follows from our assumption $(c_{2k},c_0]=[c_{\pi(2k)},c_0] \subset [c_{2k-1},c_0]$. 
\end{comment}

\medskip

Let us denote 
\be\non
 z_j:=x_{2j-1}\in\R^d \qu\qnd\qu z=(z_j)_{1\leq j\leq N} \in \R_L^{Nd}. 
\ee
Due to Proposition~\ref{l:change}, vector $z$ forms coordinates on the set $\cA_s'(\gF)$ and, by \eqref{zk1}(b), $x_i=(-1)^{i+1} z_{\lceil i/2\rceil}$. 
Then, using \eqref{zk1}(a), we see that for $\xi\in \cA_s'(\gF)$ relation \eqref{xi(x)} takes the form
\be\lbl{xi,s(z)}
\xi_j(z)=s+\sum_{i:\,c_i\in [c_j,c_0)}(-1)^{i+1} z_{\lceil i/2\rceil},\quad 1\leq j\leq 2N,
\ee
where $[c_j,c_0):=[c_j,c_0]\sm\{c_0\}$,  and $\xi_0=s$. 
Here we emphasize that the linear mapping $z\mapsto \xi$ depends on $\gF$ (through the interval $[c_j,c_0)$) and
 that $\xi$ is an affine vector-function of the parameter  $s\in\R^d$. 
Thus, in the $z$-coordinates
\be\lbl{A_s-A^z}
\cA_s'(\gF)=\{ \xi(z):\; z\in \cZ(\gF)\},
\ee
where, by \eqref{zk2},
\be\lbl{ZZZ}
\cZ(\gF)=\{z\in \R^{dN}:\qu z_j\neq 0 \qnd \sum_{i=1}^N \al_{ji}^\gF z_i \neq 0 \quad \forall 1\leq j\leq N\}.
\ee

 \begin{remark}
	Since the choice of the Hamilton cycle $\gC_\gF$ in general is not unique, the obtained parametrization $z\mapsto\xi$ is not unique as well. However, one can show that if $z'\mapsto\xi$ is another parametrization,  obtained by the procedure above, and $\al^{\gF\,\prime}$ is the associated incidence matrix, then for each $j$ we have either $z_j'(\xi)=z_j(\xi)$ or $z_j'(\xi)=\sum_{i=1}^N \al_{ji}^\gF z_i$. In the latter case we also have the symmetric relation $z_j(\xi)=\sum_{i=1}^N \al_{ji}^{\gF\,\prime} z'_i$.
\end{remark}

Next we write the quadratic forms $\om_j^\gF(\xi)$, 
% and $\Om^\gF(l,\xi)$
defined in \eqref{Omeggga}, in the $z$-coordinates.
\begin{proposition}
\lbl{l:Om-change}
Functions $\om_j^\gF(\xi)$, restricted to the set $\cA_s'(\gF)\ni\xi$ and written in the $z$-coordinates, take the form
\be\lbl{om_j^F(z)}
\om_j^\gF(z)=2z_j\cdot \sum_{i=1}^N \al_{ji}^\gF z_i,
\ee
where the  S-regular 
 incidence matrix $\al^\gF=(\al_{ij}^\gF)_{1\leq i,j\leq N}$ is defined in~\eqref{al_ij}. 
\end{proposition}
{\it Proof.} In this proof we again skip the lower index $\gF$ in the notation $\pi_\gF$.
Due to \eqref{pi=f} and their definition \eqref{Omega_def},\eqref{Omeggga}, the functions $\om_j^\gF(\xi)$ have the form 
$$
\om_j^\gF(\xi)=|\xi_{2j-1}|^2 -|\xi_{\pi(2j-1)}|^2 +|\xi_{2j}|^2 -|\xi_{\pi(2j)}|^2.
$$
Using \eqref{zk1}(b), we obtain 
\be\non
\begin{split}
\om_j^\gF(\xi)&=x_{2j-1}\cdot(\xi_{2j-1}+\xi_{\pi(2j-1)}) + x_{2j}\cdot(\xi_{2j}+\xi_{\pi(2j)}) \\
&=x_{2j-1}\cdot(\xi_{2j-1}+\xi_{\pi(2j-1)}-\xi_{2j}-\xi_{\pi(2j)} )\\
&=x_{2j-1}\cdot(2\xi_{2j-1}-x_{2j-1} - 2\xi_{\pi(2j)}-x_{2j} ) \\
&=2x_{2j-1}\cdot(\xi_{2j-1}-\xi_{\pi(2j)}).
\end{split}
\ee
Thus, by \eqref{differ-xi}, 
$
\om_j^\gF(z)=2z_j\cdot \sum_{i=1}^N \al_{ji}^\gF z_i.
$
\qed

\smallskip

 Finally, we note that instead of considering the set of Feynman diagrams $\gF_{m,n}$ it suffices to study only its subset. Denote by 
$$\gF^{\,true}_{m,n}\subset \gF_{m,n}$$  the subset of diagrams $\gF$ for which the corresponding incidence matrix $\al^\gF$  
 is not only S-regular, but
is SS-regular  (see Definition~\ref{Sreg}).
% does not have zero lines  (and zero columns since it is skew-symmetric).
In view of \eqref{ZZZ} we have:
\begin{proposition}\lbl{l:Az}
	If $\gF\notin\gF^{\,true}_{m,n}$, then $\cZ(\gF) =\emptyset$. If $\gF\in\gF^{\,true}_{m,n}$, then $\cZ(\gF)$ is 
	  the complement to the union of a finite system of linear subspaces of positive 
		codimension. In particular, it is  is an open
		dense subset of $ \R^{dN}$.
\end{proposition}
Due to \eqref{A_s-A^z}, Proposition~\ref{l:Az} implies that  the integrals $J_s(\gF)$ from 
\eqref{Ea^ma^n-J} with $\gF\notin\gF^{\,true}_{m,n}$ vanish.

\subsection{Final formula for expectations $\EE a^{(m)}_s\bar a^{(n)}_s$}
\lbl{s:ff-exp}

Let $F_s^\gF(\tau,l,z)$ be the function $F^\gF(\tau,l,\xi)$, defined in \eqref{F=}, restricted to the set $\cA_s'(\gF)$ and written in the $z$-coordinates,
\be\lbl{F=F(z)}
F_s^\gF(\tau,l,z)=F^\gF\big(\tau,l,\xi(z)\big).
\ee
The function $F^\gF_s$ depends on $s\in\R^d$ through the transformation $z\mapsto \xi$, see \eqref{xi,s(z)}.
Changing  in the sums $J_s(\gF)$ the coordinates $\xi$ to $z$ and using \eqref{useme} 
we see that  Lemma~\ref{l:Ea^ma^n} joined with \eqref{A_s-A^z}
 and Propositions~\ref{l:Az}, \ref{l:Om-change} implies the following result
  (we recall \eqref{theta}):
  %Here, as before, 
 %$\theta$ is arbitrary complex valued measurable bounded function (see Remark~\ref{r:theta}). 
 \begin{theorem}
 	\lbl{p:aman}
 	For any integers $m,n\geq 0$ satisfying $N=m+n\geq 1$, $s\in\Z^d_L$ and $\tau_1,\tau_2\geq -T$, we have
\be\lbl{final_Ea^ma^n}
	\EE a^{(m)}_s(\tau_1) \bar a^{(n)}_s(\tau_2)
	= \sum_{\gF\in\gF^{\,true}_{m,n}} c_\gF  J_s(\gF),
	\ee
		where $\gF^{\,true}_{m,n}$ is a finite set of diagrams, defined in Section~\ref{s:ch_of_c}, the constants $c_\gF\in\{\pm 1,\pm i\}$ are defined in Section~\ref{s:F_not} and 
	\be
	\lbl{J(F)-z}
	J_s(\tau;\gF)=
\int_{\R^N}
dl\, 
L^{-Nd}\sum_{z\in\cZ(\gF)\cap\Z^{Nd}_L}\,F_s^\gF(\tau,l,z)
 \,\Theta(\om^\gF(z),\nu^{-1}l). 
\ee
Here $\Theta(\om^\gF,\nu^{-1}l)=\prod_{j=1}^N \theta(\om_j^\gF,\nu^{-1}l_j)$, while components $\om_j^\gF$ of the vector $\om^\gF$ are given by \eqref{om_j^F(z)}. The set $\cZ(\gF)$ is defined in \eqref{ZZZ} and
the incidence matrix $\al^\gF$ from \eqref{ZZZ}, \eqref{om_j^F(z)} is SS-regular. 
%skew-symmetric, does not have zero lines and columns and its components satisfy $\al_{ij}^{\gF}\in\{0,\pm 1\}$. 
 The density function  $F^\gF_s$ is real and satisfies  
\be\lbl{F-prop1}
|\p_{s}^\mu\p_{z}^{\ka}F_s^\gF(\tau,l,z)|\leq C_{\mu,\ka}^{\#}(s)C_{\mu,\ka}^{\#}(z)\,e^{-\de \big(\sum_{i=1}^m|\tau_1-l_i|
	+\sum_{i=m+1}^N|\tau_2-l_i|\big)}
\ee
with a suitable $\delta=\de_N>0$, for any vectors $\mu\in\Z^d_+$, $\ka\in\Z_+^{dN}$, and any $s\in\R^d$, $z\in\R^{Nd}$.
\end{theorem} 

Since the density $F_s^\gF$ in \eqref{J(F)-z} is  an affine function of  
$s\in\R^d$ and satisfies \eqref{F-prop1}, the sum $J_s(\gF)$ is a Schwartz function of $s\in\R^d$.
Thus, the correlations $\EE a^{(m)}_s(\tau_1) \bar a^{(n)}_s(\tau_2)$ extends to a Schwartz function of $s\in\R^d$ via the equality 
\eqref{final_Ea^ma^n}, as is affirmed  in Theorem~\ref{t:main''}.
In view of this result, below we always assume that $s\in\R^d$.

	Subsequent analysis of the sums $J_s(\gF)$ is based on the lemma below, where the function 
	$\Theta(\om^\gF(z),\nu^{-1}l)$ is defined as in \eqref{c_D-def} with  $\theta(y,t)=e^{ity}$ (cf.  \eqref{theta-def}).
 
	\begin{lemma}\lbl{l:ph_func} 
If $\theta(y,t)=e^{ity}$, then 
\be\lbl{Theta-exp}
\Theta(\om^\gF(z),\nu^{-1}l)=e^{i\nu^{-1}\Om^\gF(l,z)},
\ee
where the phase function $\Om^\gF$ is given by
\be\lbl{Om_xy}
\Om^\gF(l,z)
= \sum_{1\leq i,j\leq N} \al_{ij}^\gF\, (l_i-l_j)\,z_i\cdot z_j.
\ee
\end{lemma}
{\it Proof.}
Since $\theta(x,t)=e^{itx}$, the function $\Theta$ is given by \eqref{Theta-exp} with  
$
\Om^\gF(l,z)=\sum_{j=1}^N l_j \om^\gF_j(z).
$
From \eqref{om_j^F(z)} it follows that
 $\Om^\gF=2 \sum_{1\leq i,j\leq N} l_j\al_{ji}^\gF\, z_j\cdot z_i$. 
 Now \eqref{Om_xy} follows from the skew symmetry  of the matrix  $\al^\gF$.
 \qed

\subsection{ Continuous approximation}
	
 From now on  till Section \ref{s:app_eq_proofs}
 we always assume that  
 	$$\theta(y,t)=e^{ity}.$$  

Using that  by Proposition~\ref{l:Az} 
 \ $\cZ(\gF)$ with $\gF\in\gF^{\,true}_{m,n}$ is an open dense subset of 
$\R^{dN}$,  we approximate the sum $L^{-Nd}\sum_{z\in\cZ(\gF)\cap \Z^{Nd}_L}$ from \eqref{J(F)-z} by an integral over $\R^{dN}$ by
applying Theorem 3.1 from \cite{DK}. 
For $d\geq 2$ we get
$$
\Big|J_s(\gF)-\int_{\R^N}
dl\, 
\int_{\R^{dN}}dz\,F_s^\gF(\tau,l,z)
e^{i\nu^{-1}\Om^\gF(l,z)} \Big| \leq C^{\#}(s)L^{-2}\nu^{-2},
$$
where we used \eqref{Theta-exp}. When
 $d=1$ the r.h.s. of the inequality above should be modified by adding the term $C^{\#}(s) L^{-1}$, 
see Remark 3.2 in~\cite{DK}.

Together with \eqref{final_Ea^ma^n} this estimate implies  the main result of the section:
\begin{theorem}\lbl{th:Ea^ma^n-cont}
	Assume that $d\geq 2$ and the function $\theta$ is as above. 
Then for any integers $m,n\geq 0$ satisfying $N=m+n\geq 1$, any $s\in\R^d$ and 
$\tau_1,\tau_2\geq -T$, we have
	\be\lbl{int_Ea^ma^n}
	\Big|\EE a^{(m)}_s(\tau_1) \bar a^{(n)}_s(\tau_2)
	-  \sum_{\gF\in\gF^{\,true}_{m,n}} c_\gF\wt J_s(\gF) 
	\Big| \leq C^{\#}(s)L^{-2}\nu^{-2},
	\ee
	where
	\be\lbl{J_c(F)}
	\wt J_s(\tau;\gF)=\int_{\R^N}
	dl\, 
	\int_{\R^{dN}}dz\,F_s^\gF(\tau,l,z)
	e^{i\nu^{-1}\Om^\gF(l,z)}.
	\ee
	The constant $c_\gF = c_{\gD_\gF}$ is defined in \eqref{c_D-def} and the diagram $\gD_\gF$~-- in \eqref{D_FF}. The real valued density $F_s^\gF$ is given by \eqref{F=F(z)} together with \eqref{F=} and satisfies  the estimate \eqref{F-prop1}. 
	The phase function $\Om^\gF$ is given by  \eqref{Om_xy} where the incidence matrix $\al^\gF$  is SS-regular. 
	%skew-symmetric, does not have zero lines and columns and its components satisfy $\al_{ij}^\gF\in\{0,\pm 1\}$.

	If $d=1$ then the r.h.s.  of \eqref{int_Ea^ma^n} should be modified by adding to it the term $C^{\#}(s) L^{-1}$.
\end{theorem}
\begin{comment}
Theorem~\ref{th:Ea^ma^n-cont} allows to get an explicit formula for the correlation 
$\EE a_s^{(m)}(\tau_1)\bar a_s^{(n)}(\tau_2)$ for arbitrary $m$ and $n$, 
up to an error term of the size $O(L^{-2}\nu^{-2})$ 
(however, the cardinality of the set $\gF^{\,true}_{m,n}$ growth very fast with $m, n$, so number of integrals $\wt J_s(\gF)$ associated with the correlation 
$\EE a_s^{(m)}\bar a_s^{(n)}$  also does).
In Section~\ref{s:ex_int} we demonstrate this in the case $m=n=2$, which is relevant for study of the quasisolution \eqref{quasisol}. 
\end{comment}
Theorem~\ref{th:Ea^ma^n-cont} provides  an explicit formula for a correlation 
$\EE a_s^{(m)}\bar a_s^{(n)}$ with  arbitrary $m$ and $n$, 
up to an error term of the size $O(L^{-2}\nu^{-2})$.
In  Section~\ref{s:ex_int}  we consider an example when $m=n=2$.

Since the functions $F^\gF$ and $\Om^\gF$ are given by explicit formulas \eqref{F=F(z)} and \eqref{Om_xy}, the integral over the variable $l$ in \eqref{J_c(F)} can be found explicitly. 
In Section~\ref{s:time out} we give the result of this computation,
 relevant for further study of the energy spectrum of the quasisolutions.      

\begin{comment}
Proof of Theorem~\ref{t:main''} consists of two independent parts. 
In the first one, which is now completed, we established Theorem~\ref{th:Ea^ma^n-cont},
while in the second part, to which is devoted the rest of the paper, we estimate the integrals 
\eqref{J_c(F)}. Before passing to the second part, let us briefly summarize the road we passed to conclude the first:
\be\lbl{scheme}
\begin{split}
&\fbox{\parbox{3.15cm}{
Diagrams $\gD$ and formula \eqref{a^ma^n}}
}
\Rightarrow
\fbox{\parbox{3.15cm}{
	Feynman diagrams $\gF$ and \eqref{formula_Ea^ma^n}}
}
\Rightarrow
\fbox{\parbox{3.15cm}{
	Graph $\mathfrak{U}_\gF$ and Hamilton cycle $\gC_\gF$}
}
\\
&
\Rightarrow
\fbox{\parbox{4.7cm}{
		Transformation $\xi\rightarrow z$, matrix $\al^\gF$ and \eqref{Om_xy}-\eqref{J(F)-z} }
}
\Rightarrow
\fbox{\parbox{3.15cm}{
		Approximation of sums by integrals}
}
\end{split}
\ee
Here we emphasize the matrix $\al^\gF$ since the explicit form of the phase function \eqref{Om_xy} plays an important role in what follows, while for the function $F_s^\gF$ we use only the estimate \eqref{F-prop1} but not its explicit form. 
\end{comment}

\section{Main estimate}
\lbl{s:est}

\subsection{Estimation of integrals $\wt J_s(\gF)$}\label{s_6.1}

 Next we estimate the integrals $\wt J_s(\gF)$ from Theorem~\ref{th:Ea^ma^n-cont}. 
 
 \begin{theorem}\lbl{cor:final}
 	For any integers $m,n\geq 0$, satisfying $N=m+n\geq 2$, any $\gF\in\gF^{\,true}_{m,n}$ and $s\in\R^d$, 
 	\be\lbl{est wt_J}
 	|\wt J_s(\gF)|\leq C^\#(s)
 	\nu^{\min(\lceil N/2\rceil,d)}\chi_d^{N}(\nu), 
 	\ee
 	uniformly in $\tau_1,\tau_2\geq -T$, wherethe function $\chi_d^N$ is defined in \eqref{chi^N_d}.
 \end{theorem}   

Together with Theorem~\ref{th:Ea^ma^n-cont}, estimate \eqref{est wt_J}
implies the desired inequality  \eqref{mainn} and concludes the proof of Theorems~\ref{t:main''} and \ref{t:mainpr}.
\footnote{We recall that $\gF_{m,n}^{\,true}=\emptyset$ for $N=1$, see \eqref{N=1-empty}, so we study only the case $N\ge2$.}

We deduce \eqref{est wt_J} from an abstract theorem below where we estimate integrals of the following more general form. Let 
\be\lbl{J===}
J_s=\int_{\R^K}
dl\, 
\int_{\R^{dM}}dz\,F_s(l,z)
e^{i\nu^{-1}Q(l,z)},
\ee
where $M\geq 2$, $K\geq 1$, $d\geq 1$,
$l=(l_1,\ldots, l_K)$ and 
$z$ is the polyvector 
$(z_1,\ldots,z_M)$ with  $z_i=(z_i^1,\ldots,z_i^d) \in\R^d$.
The phase function $Q(l,z)$ is assumed to be linear in $l$ and quadratic in $z$, 
$$
Q(l,z)=\big( Q(l)z\big)\cdot z=\sum_{i,j=1}^M ( q_{ij}\cdot l)\, (z_i\cdot z_j)= \sum_{i,j=1}^M  q_{ij}\cdot l \,\sum_{m=1}^d z_i^mz_j^m.
$$
Here
 $q_{ij}=q_{ji}=(q_{ij}^k)_{1\leq k\leq K}$ are $K$-vectors,
  so for $l\in \R^K$, $Q(l)=\big(q_{ij}\cdot l\big)_{1\leq i,j\leq M}$ is a real  symmetric 
$M\times M$--matrix. 
Denoting by $Q_k$, $1\le k \le K$, the real symmetric matrices
$$
Q_k =(q^k_{ij})_{1\le i, j\le M}\,,
$$
we write $Q(l)$ as $\sum_k Q_k l_k$.  We denote by $R\le K$ the rank of the system of vectors 
$
q_{ij} \in \R^K$, $1\le i, j\le M.
$
It equals to the rank of the system of matrices $\{ Q_k, 1\le k\le K\} $ in the space of $M\times M$ matrices. Indeed, consider the row
$$
(Q_1, \dots, Q_K),
$$
interpreting it as a matrix with columns $Q_j$ (regarded as  vectors in $\R^{M^2}$). Then the two ranks we talk about are the column\,- and
raw-ranks of this matrix, so they coincide. 

Note that the phase function $\Om^\gF$ in \eqref{Om_xy} may be written as $Q(l, z)$.

The complex  density function $F_s(l,z)$  in \eqref{J===} is assumed to be  bounded and measurable in $l\in \R^K, z\in\R^{dN}$,
 smooth in $z$,  depend on the parameter $s\in\R^d$,  and satisfy
\be\lbl{F_in_z}
|\p_{z}^{\ka}F_s(l,z)|\leq C_\ka^{\#}(s)C_\ka^{\#}(z)h_\ka(|l-l_0|),
\ee
for any  $\ka\in \Z_+^{dM}$ and some $l_0\in\R^K$. Each 
function $h_\ka$ is  assumed to be monotone decreasing, bounded and such that the function  $h_\ka(|l|)$ is integrable over $\R^K$. 
Then
\be\lbl{F_in_l}
0 \leq h_\ka \leq C_{\ka},\qquad\qquad \int_0^\infty h_\ka(r)r^{K-1}\,dr\leq C_\ka,
\ee
for appropriate constants $C_\ka>0$. 
For $k\in\N$ we introduce the functions 
\be\lbl{psi_d^k}
\psi_{d}^k(\nu)=
\left\{\begin{array}{cl}
	-\ln\nu &\qmb{if}\qu k=d, \\
	1 &\qmb{otherwise}.
\end{array}
\right.
\ee

\begin{theorem}\lbl{t:est wt_J}
	Assume that 
	$
	\operatorname{tr} Q_k =0
	$
for each $k$. Then, under assumption \eqref{F_in_z} 
\be\lbl{st_ph est}
|J_s|\leq C^{\#}(s) \nu^{\min(R,d)}\psi_d^R(\nu).
\ee
The function $C^\#$ depends on $F_s$ only through 
 the functions $C_\ka^\#$ and the constants $C_\ka$ in \eqref{F_in_z}, \eqref{F_in_l}. It 
 also depends on the tensor $(q^k_{ij})$.
%The function $C^\#$   also depends on the tensor $(q^k_{ij})$, but if the numbers $q^k_{ij}$ are integer, it may be chosen to depend only on 
%$q_{max} = \max_{i,j,k} |q^k_{ij}|.$

\end{theorem} 
 Theorem~\ref{t:est wt_J} is proven in the next subsection.
In Appendix~\ref{s:est int quadr} we note that it  implies an asymptotic estimate for parameter-depending 
 integrals of some quotients with asymptotically  degenerating  divisors, 
including those which are obtained from the integrals $\wt J_s(\gF)$ by explicit integration over $\R_l^N$. 
That result  is not needed for the present paper, however we believe that it is of independent interest. It is related to \cite{K}.

In view of \eqref{F-prop1}, 
the integrals $\wt J_s(\gF)$ satisfy the assumptions of Theorem~\ref{t:est wt_J}
with the functions $C^\#_\ka$ and the constants $C_\ka$ independent from $\tau_1,\tau_2$.
Here, in view of \eqref{Om_xy}, we have $q_{ij} \cdot l=\al_{ij}^\gF(l_i-l_j)$ and it can be shown that 
the rank $R$ satisfies $R\geq \lceil N/2 \rceil$ (this follows solely from the fact that the matrix $\al^\gF$ 
 is skew-symmetric and does not have zero rows and columns since it is SS-regular).  
%skew-symmetric and does not have zero lines and columns). 
So Theorem~\ref{t:est wt_J} implies  estimate
\eqref{est wt_J}, where the function $\chi_d^N$ is replaced by $\psi_d^R$.
In the case $R=d=\lceil N/2 \rceil$ (which holds true for certain matrices $\al^\gF$) this estimate is slightly weaker than \eqref{est wt_J}. Indeed, it 
gives
\be\non
|\wt J_s(\gF)|\leq C^{\#}(s) \nu^{\lceil N/2 \rceil} \ln\nu^{-1},
\ee
so  compare to  \eqref{est wt_J} we get an additional factor 
$\ln\nu^{-1}$, unless $N=2,3$. 
This is insufficient for the purposes of \cite{DK}. 
To overcome this difficulty, in Theorem~\ref{t:est wt_J'} we prove that if the tensor $(q_{ij}^k)$ has a block-diagonal form,  the  estimate \eqref{st_ph est} improves and implies \eqref{est wt_J}. 
This is mostly a technical issue, and  ideologically \eqref{est wt_J} follows from \eqref{st_ph est}. 
So we postpone Theorem~\ref{t:est wt_J'} till the next Section~\ref{est-imroved} and first establish 
estimate \eqref{st_ph est}.

\subsection{Proof of Theorem~\ref{t:est wt_J}}
\lbl{s:est wt_J}

In this proof the positive constants $C, C_1$, etc. and the functions $C^\#$ depend on $F_s$
as  in Theorem~\ref{t:est wt_J}.  To simplify presentation  we  assume that in \eqref{F_in_z} \, $l_0=0$,
 but emphasize that the estimates we obtain are uniform in $l_0$.
\smallskip
 
Let $\la_1(l), \ldots, \la_M(l)$ be (real) continuous in $l$  eigenvalues of the symmetric matrix $Q(l)$, 
enumerated in  decreasing order: 
\be\lbl{la_order}
|\la_1(l)|\geq |\la_2(l)|\geq \ldots\geq |\la_M(l)| \qmb{for any}\qu l\in\R^K.
\ee
The corresponding normalized eigenvectors are vector--functions of $l$,  forming an orthonormal basis in $\R^M$ for 
every $l$, and analytic outside an
 analytic subset of $\R^K_l$ of positive codimension (while as functions of $l\in\R^k$ they even are not continuous). 
Denoting by $w(l,z)=w=(w_1,\ldots,w_M)$, $w_i\in\R^d$, the polyvector $z=(z_1,\ldots,z_M)$, written in  this basis, we write 
%\footnote{   we use the same basis in $\R^d$ to construct  vectors $w_j$ from the vectors $z_j$, $1\le j\le M$. }
%for every component $z_i,w_j$ of the vectors $z$ and $w$),  
 function $Q$ as 
$$
Q(l,z)=\sum_{i=1}^M \la_i (l)|w_i|^2\,. %, \qmb{where}\qu w=w(l,z).
$$
Since the transformation $z\mapsto w$ is orthogonal for every $l$, then $dz=dw$ and 
\be\label{integral}
J_s=\int_{\R^K}
dl\, 
\int_{\R^{dM}}dw\,H_s(l,w)
e^{i\nu^{-1}\sum_{i=1}^M \la_i (l)|w_i|^2}.
\ee
Here 
$H_s(l,w)=F_s(l,z(l,w))$, so that  the function $H_s(l,w)$ again satisfies the estimate \eqref{F_in_z}.
The following observation plays the key role in the proof.
\begin{lemma}\lbl{l:H-S}
	Let $A=(a_{ij})$ be a real symmetric 
	$M\times M$-matrix, $M\geq 2$, such that  $\operatorname{tr} A=0$, and let $\la_1,\ldots,\la_M$ be its eigenvalues. 
	Then 
	\be\lbl{HiSh}
	\sum_{i,j}a_{ij}^2= -2\sum_{i<j} \la_i\la_j.
	\ee
\end{lemma} 
{\it Proof.} 
Let $\|A\|_{HS}$ be the Hilbert-Schmidt norm of the matrix $A$. 
From one hand, $\|A\|_{HS}^2$ equals to the l.h.s. of \eqref{HiSh}.
From another hand, 
$$
\|A\|_{HS}^2=\sum_{i=1}^M \la_i^2=\Big(\sum_{i=1}^M \la_i\Big)^2 - 2\sum_{i<j} \la_i\la_j.
$$
It remains to note that $\Big(\sum_{i=1}^M \la_i\Big)^2=(\operatorname{tr} A)^2=0$.
\qed

\smallskip 
 
Applying Lemma~\ref{l:H-S} to the matrix $Q(l)$, we find
\be\non
\frac12\sum_{i,j} \big(q_{ij} \cdot l \big)^2=
-\sum_{i<j} \la_i(l)\la_j(l) =:\cK(l) \ge0. 
\ee
Let us consider the sets 
\be\non
\cN_\nu:=\{l\in\R^K:\,\cK(l)\geq \nu^2 \} \qnd \cN_\nu^c:=\R^K\setminus \cN_\nu.
\ee
Below for a set $V\subset \R^K_l$  we denote by $\lan J_s,V\ran$ the integral \eqref{integral} 
with the domain of integration $\R_l^K$  replaced by  $V$.
Recall that $R$ is the rank of the system of $K$--vectors $(q_{ij})$. 
\begin{lemma}\lbl{l:over_ov_N} 
 Under assumption \eqref{F_in_z} we have
$
|\lan J_s,\cN_\nu^c\ran|\leq C^{\#}(s)\nu^{R}.
$
\end{lemma}
Proof of Lemma \ref{l:over_ov_N}
 is given at the end of this section. 
 
\noindent It remains to establish  estimate \eqref{st_ph est} for the integral $\lan J_s,\cN_\nu\ran$. 
In view of~\eqref{la_order}, 
\be\lbl{good_est}
|\la_1\la_2|\geq 
 M^{-2} \sum_{i<j} \big|\la_i\la_j\big| \ge  M^{-2}\cK.
\ee
So for $l\in\cN_\nu$ we have $|\la_1(l)\la_2(l)|\geq M^{-2}\nu^2$.
It is known (see \cite{Hor, DS}) 
 that the (generalized) Fourier transform of the function 
 $$
 \R^{2d} \ni (w_1, w_2)\mapsto 
 e^{- i\nu^{-1}(\la_1|w_1|^2+ \la_2|w_2|^2)}
 $$
 is  
 $$
 \zeta (\pi  \nu)^d (|\la_1 \la_2|)^{-d/2}e^{i\nu(\la_1^{-1}|\xi_1|^2+ \la_2^{-1}|\xi_2|^2 )  /4},
 $$
 where $\zeta$ is a  complex constant of unit norm; here we define the Fourier transform of a regular function $u(x)$ as 
 $
 \int e^{-ix\cdot \xi} u(x)\,dx. 
 $
Inspired by the stationary phase method \cite{Hor,DS}, we apply the  Parseval's identity to the partial 
 integral over $w_1, w_2$ and get
\begin{align*}
&\int_{\R^{2d}} H_s(l,w)e^{i\nu^{-1}(\la_1|w_1|^2 + \la_2|w_2|^2)}\,dw_1dw_2 \\
\non
&=\zeta \frac{(\nu/4\pi)^d}{|\la_1\la_2|^{d/2}} 
\int_{\R^{2d}} \hat H_s^{1,2}(l,\xi_1,\xi_2,w_{\geq 3})e^{-i\frac{\nu}{4}(\la_1^{-1}|\xi_1|^2 + \la_2^{-1}|\xi_2|^2)}\,d\xi_1d\xi_2,
\end{align*}
for any 
 $w_{\geq 3}=(w_3,\ldots,w_M)$, where  $\hat H_s^{1,2}$ stands for 
  the Fourier transform of  function $H_s$ with respect to the variables $w_1,w_2$.\,\footnote{
  See Section~12.2 in \cite{DK} for justification of this formula without using the generalized Fourier transform.}
  Since $H_s$ satisfies assumption \eqref{F_in_z} (where $l_0=0$), 
  the r.h.s. above is bounded in absolute value by 
  $$
   C^{\#}(s)  C^{\#}( w_{\geq 3})
  \frac{\nu^d}{| \la_1\la_2|^{d/2}} h(| l|),
  $$
  where the function $h\geq 0$ satisfies \eqref{F_in_l}. So 
\begin{align*}
|\lan J_s,\cN_\nu\ran|\le
  \int_{\cN_\nu}  \frac{C^{\#}(s)  \nu^{d}h(| l|) }{|\la_1(l)\la_2(l)|^{d/2}}\,dl \,
  &\int_{\R^{(M-2)d}}  C^{\#}(w_{\geq 3}) \,dw_{\geq 3} 
  \\&\le % Ê\le %\;\big|\hat H_s^{1,2}(l,\xi_1,\xi_2,w_{\geq 3})\big|\\
 C_1^{\#}(s) \nu^{d} \int_{\cN_\nu}\frac{h(|l|)  }{|\la_1(l)\la_2(l)|^{d/2}}\,dl.
\end{align*}
Then, due to \eqref{good_est},  we have
\be\non
|\lan J_s,\cN_\nu\ran|\leq C^{\#}(s) \nu^{d} 
\int_{\cN_\nu}\frac{h(|l|)}{(\cK(l))^{d/2}}\,dl
=: C^{\#}(s)I^\nu.
\ee
\begin{lemma}\lbl{l:est-for-I-nu}
	Assume that the function $h$ satisfies assumption \eqref{F_in_l}. 
	Then $I^\nu\leq C\nu^{\min(R,d)}\psi^{R}_d(\nu)$.
\end{lemma}
Lemma~\ref{l:est-for-I-nu} together with Lemma~\ref{l:over_ov_N} concludes the proof of the theorem. The lemmas are established below. 
\qed

\smallskip

{\it Proof of Lemma \ref{l:over_ov_N}.}
Since   function $H_s$ satisfies \eqref{F_in_z} with $l_0=0$,  then 
 %and bounding the complex exponent by one we get 
$$
| \lan J_s, \cN_\nu^c\ran | \leq C^\#(s)\int_{\cN_\nu^c} h_0(|l|)\,dl.
$$
Let us denote by $v_1,\ldots, v_R$ a maximal collection of linearly independent vectors from the set
$(q_{ij})_{1\leq i,j\leq M}$. In the integral $ \lan J_s, \cN_\nu^c\ran $ we 
 make a linear (non-degenerate) change of variable from $l$ to $t=(t_1,\ldots,t_K)$ in such a way that $t_k= v_k\cdot l$ for $k\leq R$.
 %Then $dl= A dt$, where the matrix $A$ depends on the tensor $(q^k_{ij})$ and is bounded by $C=C(q_{max})$ if
 %the numbers $q^k_{ij}$ are integer. 
 Let 
$t^<=(t_1,\ldots,t_R)$ and $t^>=(t_{R+1},\ldots,t_K)$.
Due to the identity $\cK(l)=\frac12\sum \big(q_{ij}\cdot l \big)^2$, 
\be\lbl{K-t equiv}
\frac{|t^<|^2}{2}\leq \cK(l(t))\leq C|t^<|^2.
\ee
%where $C=C(q)$ and $C=C(q_{max})$ if $q$ is an integer tensor. 
So denoting by $\cN_{\nu, t}^c$ the set $\cN_\nu^c$ written in the $t$-coordinates,  we obtain
$
\cN_{\nu, t}^c \subset\{t:\, |t^<|< \sqrt{2}\nu\}.
$
%Set $t_0=t(l_0)$ and $t^>_{0}=t^>(l_0)$. 
Since the function $h_0$ is decreasing, 
 inequalities $|l|\geq C|t|$ and 
$|t|\geq|t^>|$ imply that 
$h_0(|l|)\leq h_0(C|t|)\leq h_0(C|t^>|)$.
Then
\begin{align*}
|\lan J_s, \cN_\nu^c\ran| &\leq C^\#(s)\int_{t^<\in\R^R:\,|t^<|\,< \,  \sqrt 2\nu}  dt^<\, \int_{\R^{K-R}}
h_0(C|t^>|)  \, dt^>\\
&\leq  C_1^\#(s) \nu^R \int_{\R^{K-R}}  h_0(C|t^>|)\,dt^>.
\end{align*}
The last integral is bounded by $C\int_0^\infty h_0(r)r^{K-R-1}\,dr<C_1$, in view of assumption \eqref{F_in_l}.
\qed

\smallskip
 
{\it Proof of Lemma \ref{l:est-for-I-nu}.}
We make the change of coordinates $l\mapsto t$ as in the proof of Lemma~\ref{l:over_ov_N} and use the notation $t^<, \,t^>,$ introduced there.
Denoting by $\cN_{\nu, t}$ the set $\cN_\nu$ written in the $t$-coordinates, in view of \eqref{K-t equiv} we find
$\cN_{\nu, t}\subset\{t:\,|t^<|\geq C'\nu\}$.
Then, using that $h(|l|)\leq h(C_1|t|)$, by \eqref{K-t equiv} we get
$$
I^\nu \leq 2^{d/2}\,\nu^{d}\int_{t:\,|t^<|\geq C\nu}\frac{h(C_1|t|)}{|t^<|^d}\,dt. 
$$
Since the function $h(|t|)$ is integrable, the r.h.s.   above is bounded by $C\nu^d$, if we take the integral only over the set where $|t^<|\geq 1 $. Then, it remains to get the desired estimate for the integral over the set where $C\nu \leq |t^<|\leq 1$. 
Since the function $h$ is decreasing, we have $h(C_1|t|)\leq h(C_1|t^>|)$, so this integral is bounded by
$$
2^{d/2}\,\nu^{d}\int\limits_{t\in\R^K:\, C\nu \leq |t^<|\leq 1}\frac{h(C_1|t^>|)}{|t^<|^d}\,dt \leq 
C_2\nu^{d}\int\limits_{t^<\in\R^R:\, C\nu \leq |t^<|\leq 1}\frac{1}{|t^<|^d}\,dt^<,
$$
where we have integrated out the variable $t^>$.
Passing to the spherical coordinates, we see that the latter integral is bounded by
$$
C_3\nu^{d}\int_{C\nu}^1\frac{r^{R-1}}{r^{d}}\,dr\leq 
\left\{
\begin{array}{cl}
C_4\nu^{d} &\qmb{if}\qu d\leq R-1 \\
C_4\nu^d \ln\nu^{-1} &\qmb{if}\qu d=R \\
C_4\nu^R &\qmb{if}\qu d>R.
\end{array}
\right.
$$
The r.h.s.  of the last inequality can be rewritten as $C_4\nu^{\min(R,d)}\psi_d^R(\nu).$
\qed

\section{Refinement of estimate \eqref{st_ph est} and proof of Theorem \ref{cor:final}}
\lbl{est-imroved}

In this section we establish Theorem~\ref{t:est wt_J'}, in which we improve estimate \eqref{st_ph est}, and deduce from it Theorem~\ref{cor:final}.  

\subsection{Refinement of \eqref{st_ph est} }

We consider integral \eqref{J===}, where to simplify the formulation of result (and since it suffices for our purpose) we assume $M=K$ and denote $N:=M=K\geq 2$. We consider a decomposition of the set  $\{1,\ldots N\}$ 
into $p\geq 1$ non-empty disjoint subsets
\be\lbl{0-M part}
\{1,\ldots N\}=\I_1\cup\ldots\cup\I_p, 
\ee
and denote by $l^k$ the vector 
$l^{k}=(l_i)_{i\in\I_k}$. 
%When convenient, we will identify $l^k$ with a vector in $\R^K$ with zero components $l_r$, $r\notin \I_k$.

We assume that the phase function $Q$ admits a block-decomposition
$$
Q(l,z)=\sum_{k=1}^p { U}_k(l,z), \qquad U_k(l,z)=\sum_{i,j\in\I_k}  { (q_{ij}\cdot l^k)} \, (z_i\cdot z_j),
$$ 
where $q_{ij}=q_{ji}\in\R^{|\I_k|}$  for $i,j\in\I_k$.
Concerning  the density $F_s$ we assume that it satisfies the estimate
\be\lbl{F_in_z'}
|\p_{z}^{\ka}F_s(l,z)|\leq C_\ka^{\#}(s)C_\ka^{\#}(z)\prod_{i=1}^p h^i_\ka(|l^i-l^i_0|),
\ee
for any $\ka$ and some vectors $l^i_0\in\R^{|\I_i|}$.
The functions $h^i_\ka$ are assumed to satisfy the same assumption that we imposed on the functions $h_\kappa$ in \eqref{F_in_z}, where in \eqref{F_in_l} we replace $K$ by $|\I_i|$.
Denote by $R_k$ the rank of the system of $|\I_k|$-vectors $(q_{ij})_{i,j\in\I_k}$.
\begin{theorem}\lbl{t:est wt_J'}
	Assume that $\sum_{i\in\I_k}q_{ii} =0\in\R^{|\I_k|}$ for any $1\leq k\leq p$. Then, under assumption \eqref{F_in_z'} we have
	\be\non
	|J_s|\leq C^{\#}(s)\prod_{k=1}^p\nu^{\min(R_k,d)}\psi^{R_k}_d(\nu),
	\ee
	where the function $C^\#$ depends on $F_s$  as in Theorem~\ref{t:est wt_J}. 
\end{theorem} 
{\it Proof.} 
We argue by induction. In the case $p=1$ the assertion follows from Theorem~\ref{t:est wt_J}. Assume that theorem is proven for the case of $p-1$ subsets $\I_k$ in \eqref{0-M part} and let us establish it for the case of $p$ subsets. 
We recall the notation $l^p=(l_i)_{i\in\I_p}$ and set $l^<=(l_i)_{i\notin\I_p}$. Similarly, we denote $z^p=(z_i)_{i\in\I_p}$ and $z^<=(z_i)_{i\notin\I_p}$.
Let $N_p:=|\I_p|$ and
$$
I_s(l^<,z^<)=\int_{\R^{N_p}}dl^p\int_{\R^{dN_p}}dz^p\,F_s(l,z)e^{i\nu^{-1}U_p(l,z)},
$$
where we recall that the function $U_p(l,z)$ depends only on $l^p$ and $z^p$.
Then, 
$$
\p_{z^<}^\kappa \, I_s(l^<,z^<)=\int_{\R^{N_p}}dl^p\int_{\R^{dN_p}}dz^p\;\big(\p_{z^<}^\kappa \, F_s(l,z)\big) \,e^{i\nu^{-1}U_p(l,z)},
$$
for any $\kappa$. Literally repeating the proof of Theorem~\ref{t:est wt_J}, we get
$$
|\p_{z^<}^\kappa I_s(l^<,z^<)|\leq C^\#(s)C^\#(z^<)\prod_{i=1}^{p-1}h^i(|l^i-l_0^i|)g(\nu),
$$
where $g(\nu)=\nu^{\min(R_p,d)}\psi^{R_p}_d(\nu)$ and the functions $h^i$ satisfy the same assumptions that the functions $h_\kappa^i$ from \eqref{F_in_z'}.
Thus, we see that the function $G_s(l^<,z^<)=(g(\nu))^{-1} I_s(l^<,z^<)$ satisfies assumption \eqref{F_in_z'} with $p=p-1$.
Note that 
$$
J_s=g(\nu)\int_{\R^{N-N_p}}dl^<\int_{\R^{d(N-N_p)}}dz^<\;G_s(l^<,z^<)e^{i\nu^{-1}\sum_{k=1}^{p-1}U_k(l,z)},
$$
where $U_k(l,z)$ depend only on $l^<,z^<$.
Then, by the induction hypothesis,
$$
|J_s|\leq C^{\#}(s) g(\nu)  \prod_{k=1}^{p-1}\nu^{\min(R_k,d)}\psi^{R_k}_d(\nu)=
C^{\#}(s)\prod_{k=1}^{p}\nu^{\min(R_k,d)}\psi^{R_k}_d(\nu).
$$
\qed

\subsection{Proof of Theorem \ref{cor:final}}

Recall that the phase function $\Om^\gF$ from the definition \eqref{J_c(F)} of the integral $\wt J_s(\gF)$ is given by \eqref{Om_xy} and that for $\gF\in\gF^{\,true}_{m,n}$ the  matrix $\al^\gF$ is  SS-regular  (see Denition~\ref{Sreg}). 
%does not have zero lines and columns. 
Consider a partition \eqref{0-M part} with $p\geq 1$ such that 
the matrices 
$\al_k=(\al_{ij}^\gF)_{i,j\in\I_k}$ are irreducible and $\al_{ij}^\gF=0$ once $i\in\I_k$ and $j\in\I_r$ with $k\neq r$.
% . Then 
%\be\non
%N_k\geq 2\; \forall k \qu\qnd\qu \sum_{k=1}^p N_k=N,
%\ee
%where in the first inequality we have used that .
Then the integral $\wt J_s(\gF)$ satisfies assumptions of Theorem~\ref{t:est wt_J'}
with $q_{ij}\cdot l=\al_{ij}^\gF(l_i-l_j)$.
\begin{lemma}\lbl{l:rank}
	Let $A=(a_{ij})$ be an irreducible $n\times n$-matrix, $n\geq 2$, and the vectors $r_{ij}\in\R^n$ be given by $r_{ij}=a_{ij}(e_i-e_j)$, where $(e_i)_{1\leq i\leq n}$ is a basis in $\R^n$.
	Then rank $R$ of the system of $n$-vectors $(r_{ij})_{1\leq i,j\leq n}$ equals  $n-1$. 
\end{lemma}
Proof of Lemma~\ref{l:rank} is a simple exercise from linear algebra and we postpone it to Appendix~\ref{s:l_rank}.  
Set $N_k=|\I_k|$. 
Since the $N_k\times N_k$-matrices $\al_k$ are skew-symmetric and non-zero,  we have $N_k\geq 2$.
Applying Lemma~\ref{l:rank} to the system of vectors $(q_{ij})_{i,j\in \I_k}$, we see that its rank $R_k$ equals to $N_k-1$.
Then by Theorem~\ref{t:est wt_J'} we get the estimate
\be\lbl{fin_est}
|\wt J_s(\gF)|\leq C^{\#}(s) \prod_{i=1}^p\nu^{\min(N_i-1,d)}\psi^{N_i-1}_d(\nu).
\ee 
Let us show that \eqref{fin_est} implies the desired estimate \eqref{est wt_J}.
Assume first that $N_i-1< d$ for every $1\leq i\leq p$. Then, in view of \eqref{fin_est} and the identity $\sum_{i=1}^p N_i=N$, we have
$$
|\wt J_s(\gF)|\leq C^{\#}(s)\nu^{N-p}.
$$ 
Since $N_i\geq 2$ for any $i$, the number of blocks $p$ satisfies $p\leq \lfloor N/2\rfloor$, where $\lfloor \cdot \rfloor$ denotes the (lower) entire part.
Then $N-p\geq \lceil N/2\rceil $, so we get $\nu^{N-p}\leq \nu^{\lceil N/2 \rceil}$ which implies \eqref{est wt_J}.

Now we assume that
\be\lbl{N_j-1<d}
N_j-1\geq d \qmb{for some}\qu 1\leq j\leq p,
\ee
so $\nu^{\min(N_j-1,d)}=\nu^d$.
Since for every $i$ we have $\nu^{\min(N_i-1,d)}\leq \nu$,
by  \eqref{fin_est} we find
$$
|\wt J_s(\gF)|\leq C^{\#}(s) \nu^{d+p-1}  \prod_{i=1}^p\psi^{N_i-1}_d(\nu)
\leq C_1^\#(s)\nu^{d+p-2}.
$$
In the case $p\geq 2$ this implies \eqref{est wt_J}. 
Now it remains to consider the case when \eqref{N_j-1<d} holds with   $p=1$.  Then 
$N-1\geq d$. If $N-1>d$ then \eqref{fin_est} with $p=1$ implies   \eqref{est wt_J}.
Otherwise $N-1=d$, so
\be\lbl{J<nu^N_1-1}
|\wt J_s(\gF)|\leq C^{\#}(s) \nu^{N-1} \ln\nu^{-1}. 
\ee
If $N-1>\lceil N/2 \rceil$ then 
$\nu^{N-1} \ln\nu^{-1}< C \nu^{\lceil N/2 \rceil}$ and we are done.
The opposite situation 
is possible only if $N\leq 3$. Since $d=N-1$, it takes place if  $d=2$, $N=3$ or $d=1,$ $N=2$.  
In these cases we find $N-1=\lceil N/2 \rceil=d$, which is 2 
or $1$ correspondingly, so again 
\eqref{J<nu^N_1-1} implies \eqref{est wt_J}.
\qed

\section{Addenda}
\label{s_add}

\subsection{Optimality of the estimate in Theorem~\ref{t:main''} }
\lbl{app:3}

Throughout this section we assume that $N=m+n\geq 2$ and $d\geq 2$. 
Accordingly to Theorem~\ref{th:Ea^ma^n-cont}, we have
\be\lbl{EE-opt}
\EE a^{(m)}_s(\tau_1) \bar a^{(n)}_s(\tau_2)
=\sum_{\gF\in\gF^{\,true}_{m,n}} \wt J_s(\gF) 
+ O\big( C^{\#}(s) L^{-2}\nu^{-2}\big),
\ee
where we recall that the integrals $\wt J_s(\gF)$ are given by \eqref{J_c(F)} and $\gF^{\,true}_{m,n}$ is a certain subset of the set of Feynman diagrams $\gF$  (which are defined in Section~\ref{sec:FD}), associated with the product $a^{(m)}_s \bar a^{(n)}_s$.
Each integral $\wt J_s(\gF)$ satisfies the estimate \eqref{est wt_J}, and this implies the estimate in 
 Theorem~\ref{t:main''}. 
 Since $\nu\ln \nu^{-1} \leq C$ these estimates admit the following slightly weaker forms:
 \begin{align}\lbl{weaker est1}
 &|\wt J_s(\gF)|\leq C^\#(s)
 \nu^{\min(N/2,d)}\\
 \non
 &\abs{\EE a_s^{(m)}(\tau_1) \bar a_s^{(n)}(\tau_2)}
 \leq C^{\#}(s)\big(\nu^{-2}L^{-2} + \nu^{\min( N/2,d)}\big).
 \end{align}
In this appendix we discuss the question if it is possible to get rid of the minimum with  $d$ in the estimates above, that is, if it is true that
\be\lbl{est improved}
|\wt J_s(\gF)|\leq C^\#(s)
\nu^{N/2},\qquad
\abs{\EE a_s^{(m)}(\tau_1) \bar a_s^{(n)}(\tau_2)}
\leq C^{\#}(s)\big(\nu^{-2}L^{-2} + \nu^{N/2}\big),
\ee
 for any $N\geq 2$. We discussed the motivation for this question in Section~\ref{s:meaning}. 
 
 Let $\gF^2_{m,n}\subset \gF^{\,true}_{m,n}$ be a subset of Feynman diagrams $\gF$ for which 
 there is $q=q(\gF)$ such that  the matrix $\al^\gF$ from \eqref{Om_xy} satisfies $\al_{ij}^\gF=0$ if both $i\neq q$ and $j\neq q$ (so, all elements of the matrix $\al^\gF$ outside the $q$-th line and column  are zero).  
It can be shown that $\gF^2_{m,n}\neq\emptyset$ if $m,n\geq 1$.  
\begin{proposition}\lbl{l:large_int}
	 If $m,n\geq 1$ and $N> 2d$, then for any Feynman diagram $\gF\in\gF^2_{m,n}$ we have $ C_1^\#(s)\nu^d \leq |\wt J_s(\gF)| \leq  C_2^\#(s)\nu^d$.
\end{proposition}
Proposition~\ref{l:large_int} shows that the first estimate from \eqref{est improved} is false for $\gF\in\gF^2_{m,n}$ when $N>2d$, so in general we have only \eqref{weaker est1}.
Then, if the large integrals $\wt J_s(\gF)$ of the size  $\nu^d$ did not cancel each other in the sum from \eqref{EE-opt}, the second estimate from \eqref{est improved} would also fail.
Moreover, if they did not cancel in the sums 
from \eqref{spec_form} that give $n_s^k$,
then as we have discussed in Section~\ref{s:meaning}, this would imply that 
the energy spectrum corresponding to a truncation $\sum_{i=0}^{m}\rho^iv_s^{(i)}$ with $m> d$ of the decomposition \eqref{decomp} cannot be approximated by a solution of the WKE \eqref{WKE}.
However, we find some strong cancellations in the sum from~\eqref{EE-opt}:
\begin{proposition}\lbl{l:cancellation} 
	We have 
	$
	\big|\sum_{\gF\in\gF^2_{m,n}}\wt J_s(\gF)\big| \leq C^{\#}(s)\nu^{N-1}
	$
	for $N\geq 2$.
\end{proposition}
Since $\nu^{N-1}\leq \nu^{N/2}$, 
this estimate does not prevent the second estimate from \eqref{est improved} to be true. 
We expect that for $N\gg 2d$ the integrals $\wt J_s(\gF)$ with $\gF\in\gF^2_{m,n}$ are the biggest (see below) but still there are many other diagrams $\gF\notin\gF^2_{m,n}$ for which the corresponding 
 integrals  also are large (bigger than $\nu^{N/2}$).
Nevertheless, we expect that similar cancellations take place for them as well,  but do not know if they go till the order $\nu^{N/2}$. So, we state the following
\begin{problem}\lbl{c:conj}
	Prove that for any $d\geq 2$ and $m,n$ satisfying $N=m+n\geq 2$, 
	\be\lbl{conj-est}
	\Big|\sum_{\gF\in\gF^{\,true}_{m,n}} \wt J_s(\gF) \Big| \leq C^\#(s) \nu^{N/2 }.
	\ee
\end{problem}
If \eqref{conj-est} is true, we get the second estimate from \eqref{est improved}. In particular, under the "kinetic" scaling $\rho=\sqrt\eps\nu^{-1/2}$ the energy spectrum corresponding to a truncation $\sum_{i=0}^m \rho^i  v_s^{(i)}$ of series \eqref{decomp} with any $m\geq 2$ can be approximated by a solution of the WKE \eqref{WKE} at least with the accuracy $\eps^2$,
if $L^{-2}\nu^{-2}\leq \nu^{m}$.
\smallskip

{\it Discussion of  proofs of Propositions~\ref{l:large_int}, \ref{l:cancellation}  and of Problem~\ref{c:conj}}. 
Below we do not give proofs but present only their rough ideas.
Let us consider the matrix $\al^{\gF}_l=\big(\al_{ij}^\gF(l_i-l_j)\big)_{1\leq i,j\leq N}$, which depends on $l\in\R^N$ (see \eqref{Om_xy}). 
Since it is symmetric, it has $N$ real eigenvalues, which we enumerate in the decreasing order 
$
|\la_1(l)|\geq |\la_2(l)|\geq \ldots\geq |\la_N(l)|\geq 0
$,
for any $l\in\R^N$. 
It is possible to show that the set $\gF^2_{m,n}$ can be equivalently defined as follows. For $k\geq 1$ we denote by $\gF^k_{m,n}$ a set of Feynman diagrams $\gF$ for which the eigenvalues 
$\la_1(l), \ldots, \la_k(l)$ are not identically zero while 
$\la_i(l)\equiv 0$ for any $i\geq k+1$. 
Since the matrix $\al_l^\gF$ is non-zero but has zero trace, we have $\gF^1_{m,n}=\emptyset$, so $\gF^{\,true}_{m,n}=\cup_{k=2}^N \gF^k_{m,n}$. 

Let us take a diagram $\gF\in\gF^2_{m,n}$ and consider the integral over $dz$ from the definition \eqref{J_c(F)} of $\wt J_s(\gF)$. 
Applying to it the Parseval identity in the eigendirections corresponding to the eigenvalues $\la_1$ and $\la_2$, and using that for $\gF\in\gF^2_{m,n}$ we have $\la_3=\ldots=\la_N=0$, we find that 
$\displaystyle{\wt J_s(\gF)\sim \nu^d \int_{\R^N} \frac{h_s^\gF(l)\, dl}{|\la_1(l)\la_2(l)|^{d/2}}}$ if $N>2d$, for an appropriate integrable function $h_s^\gF$, which is fast decaying in $s$. 
Due to Lemma~\ref{l:H-S}, we have  $\la_1\la_2=-\frac12 \sum\limits_{1\leq i,j\leq N} (\al_{ij}^\gF)^2(l_i-l_j)^2$.
Using this formula, we show that the latter integral converges, which proves Proposition~\ref{l:large_int}.

 Below by a graph of a Feynman diagram we mean the diagram, where we do not specify which vertices are virtual, which have positive degree and which are leaves, so the graph is just a collection of vertices coupled by edges. To establish Proposition~\ref{l:cancellation}, we note that, up to a renumbering of vertices, all diagrams $\gF\in\gF^2_{m,n}$ have the same graph which we denote by $\mathfrak{G}^2$. To recover the set of diagrams $\gF^2_{m,n}$ from this graph, it suffices to look over all possible placings of virtual vertices (once positions of the virtual vertices are known, degrees of the other vertices are recovered uniquely).

The graph $\mathfrak{G}^2$ is such that  in a $k$-th block with any $k\neq 1,\, m+1$, the virtual vertex can be placed in two different positions, $c_{2k}$ or $\bar c_{2k}$.
Now, using the stationary phase method \cite{DS,Hor}, we develop each integral $\wt J_s(\gF)$ in $\nu$. 
Change of position of a virtual vertex inside of a block, roughly speaking, changes only the sign in first $p$ orders of this decomposition. 
This is where the cancellations come from. 
The number $p$ here equals to the number of blocks in which position of the virtual vertex is not determined uniquely, that is $p=N-2$.
After some work, we deduce from this Proposition~\ref{l:cancellation}.

Let us now discuss Problem~\ref{c:conj}. 
Take a diagram $\gF_0\notin\gF^2_{m,n}$ and consider its graph $\mathfrak{G}_{\gF_0}$. We expect that in the sum
\be\lbl{sum_cancel}
\sum_{\gF\in\gF^{\,true}_{m,n}:\,\mathfrak{G}_\gF=\mathfrak{G}_{\gF_0}}
	\wt J_s(\gF)
\ee
there are similar cancellations.
However, the graph $\mathfrak{G}^2$ is "very symmetric" (we are able to draw it explicitly), while for a general diagram $\gF_0$ there are less symmetries in the graph $\mathfrak{G}_{\gF_0}$.
That is why we expect that in the sum \eqref{sum_cancel} the number $p$ of orders, in which the cancellations take place, is smaller than in the sum from Proposition~\ref{l:cancellation},
so we expect to get less "additional degrees" of $\nu$ than in Proposition~\ref{l:cancellation}. 
Then the estimate \eqref{conj-est} can be true only if
the integrals $\wt J_s(\gF)$ are smaller than those from 
Proposition~\ref{l:cancellation}, once $N>2d$. Namely, if
 there exists $r>d$ sufficiently large such that for the diagrams $\gF$ satisfying $\mathfrak{G}_\gF=\mathfrak{G}_{\gF_0}$ we have
$
 |\wt J_s(\gF)|\leq C^\#(s)\nu^r.
$
 
It is plausible that   this is  indeed  the case and below we explain the reason.
  Let $k>2$ be such that $\gF_0\in\gF^k_{m,n}$. Then it is possible to show that $\gF\in\gF^k_{m,n}$ as well for any  diagram $\gF$ satisfying $\mathfrak{G}_\gF=\mathfrak{G}_{\gF_0}$.
 Then, arguing as in the proof of Proposition~\ref{l:large_int} and applying the stationary phase method in the eigendirections corresponding to the eigenvalues $\la_1,\ldots,\la_k$, we get
  $\displaystyle{\wt J_s(\gF)\sim \nu^{kd/2} \int_{\R^N} \frac{h_s^\gF(l)\, dl}{|\la_1(l)\cdots\la_k(l)|^{d/2}}}$, if the latter integral converges. Perhaps, the factor $\nu^{kd/2} $ could lead to the desired estimate. 
  However, we are not able to establish the convergence since we do not know a good estimate from below for the product  $\la_1(l)\cdots\la_k(l)$ in the case $k>2$, in difference with the situation when $k=2$ considered in Proposition~\ref{l:large_int}.

\subsection{Proof of Lemma \ref{l:Ea^ma^n}}
\label{s:Ea^ma^n}

We compute the expectation of the both sides of \eqref{a^ma^n}.
Let us take any diagram $\gD\in\gD_m\times\ov\gD_n$ and fix $(\xi,\s)\in\cA_s^L(\gD)$.
Recall that 
$\EE \az{s'} \az{s''}=\EE \bz{s'} \bz{s''}=0$, 
for any $s',s''$.
Then, by the Wick theorem we find
\be\lbl{Wick-1}
\EE \prod_{i\in L(\gD)}\az{\xi_{i}}(l^{c_i})
\prod_{j\in \bar L(\gD)}\bz{\s_{j}}(l^{\bar c_j}) 
=\sum_{f} \prod_{j\in \bar L(\gD)} 
\EE\az{\xi_{f(j)}}(l^{c_{f(j)}})\bz{\s_j}(l^{\bar c_j}),
\ee
where the summation is taken over all bijections $f:\, \bar L(\gD)\mapsto L(\gD)$ of the set $\bar L(\gD)$ indexing conjugated leaves to the set $L(\gD)$ indexing non-conjugated leaves. 
Let us fix any such bijection $f$  and couple by an edge in the diagram $\gD$ the leaf $c_{f(j)}$ with the leaf $\bar c_j$ for every $j\in \bar L(\gD)$ (that is, the leaves which take values $\az{\xi_{f(j)}}$and $\bz{\s_j}$); denote the obtained diagram by $\gF_f$.
The summation over the bijections $f$ in \eqref{Wick-1} can be replaced by the summation over the diagrams $\gF_f$. 

The only difference between the set $\gF(\gD)$ of Feynman diagrams obtained from the diagram $\gD$ with the set of diagrams $\gF_f$ is that, by definition, in any $\gF\in\gF(\gD)$ there are no leaves from the same block which are adjacent. 
Thus, for any $\gF_f\notin\gF(\gD)$ there are adjacent leaves $c_{f(j)}^{(0)}\sim \bar c_j^{(0)}$ that belong to the same block.
Since by \eqref{de'==1} we have 
$\xi_{f(j)}\neq \s_j$ for $(\xi,\s)\in\cA_s^L(\gD)$, then $\EE\az{\xi_{f(j)}}\bz{\s_j}=0$ by \eqref{corr_a_in_time}, so the product from the r.h.s. of \eqref{Wick-1}, corresponding to the bijection $f$, vanishes. Therefore, it suffices to take the summation only over the set of Feynman diagrams $\gF(\gD)$.  
Then, denoting $\xi_\psi:=\xi_i$ and $\s_\psi:=\s_j$ if vertices $c_i$ and $\bar c_j$ are  coupled by an edge $\psi$, 
we rewrite \eqref{Wick-1} as
\begin{align}\non
\EE \prod_{i\in L(\gD)}\az{\xi_{i}}(l^{c_i})
&\prod_{j\in \bar L(\gD)}\bz{\s_{j}}(l^{\bar c_j}) 
=\sum_{\gF\in\gF(\gD)} \prod_{\psi\in E_L(\gF)} 
\EE\az{\xi_\psi}(l^{c_\psi})\bz{\s_\psi}(l^{\bar c_\psi})
\\
%\lbl{exp_prod-est}
\non
=&\sum_{\gF\in\gF(\gD)} \prod_{\psi\in E_L(\gF)} 
\Big(e^{-\ga^{\psi}|l^{c_\psi}-l^{\bar c_\psi}|}- e^{-\ga^{\psi}(l^{c_\psi}+l^{\bar c_\psi}+2T)}\Big)
\frac{(b^{\psi}(\xi))^2}{\ga^{\psi}(\xi)}\de_{\xi_\psi}^{\s_\psi},
\end{align}
by \eqref{corr_a_in_time}, where we recall that the set $E_L(\gF)$, the times $l^{c_\psi}, l^{\bar c_\psi}$ and the functions $b^\psi$, $\ga^\psi$ are defined in Section \ref{s:F_not}.
Since $E(\gD)= E_D(\gF)$ for $\gF\in\gF(\gD)$,  we get
\be\non
\EE G_\gD(\tau,l,\xi,\s) = \sum_{\gF\in\gF(\gD)} c_\gD F^\gF(\tau,l,\xi)\,\prod_{\psi\in E_L(\gF)} \de_{\xi_\psi}^{\s_\psi}, \qu\forall \, \tau,l \mbox{ and } (\xi,\s)\in\cA_s^L(\gD).
\ee
Then, taking the expectation of the both sides of \eqref{a^ma^n-I} and using that, by definition of $c_\gF$, we have $c_\gF=c_\gD$ for $\gF\in\gF(\gD)$, 
we find
\be
\non
\begin{split}
\EE I_s(\gD)=\sum_{\gF\in\gF(\gD)}&c_\gF
\int_{\R^N}
dl\\ 
&L^{-Nd}\sum_{(\xi,\s)\in\cA_s^{L}(\gD)}\,
F^\gF(\tau,l,\xi) \prod_{\psi\in E_L(\gF)} \de_{\xi_\psi}^{\s_\psi} \, \Theta(\om(\xi,\s),\nu^{-1}l).
\end{split}
\ee
Next we note that $(\xi,\s)\in\cA_s^L(\gD)$ and $\de_{\xi_\psi}^{\s_\psi}=1$ for every edge $\psi\in E_L(\gF)$ if and only if $\xi\in\cA_s^{\prime L}(\gF)$ and $\s=\xi_{f_\gF}$, where we recall that $\xi_{f_\gF}$ is defined in \eqref{xi_f}.
Then in the sum above we can replace the summation $\sum_{(\xi,\s)\in\cA_s^{L}(\gD)}$ by 
$\sum_{\xi\in\cA_s^{\prime L}(\gF)}$ and substitute $\s=\xi_{f_\gF}$. Then 
$\om(\xi,\s)$ replaces by $\om^\gF(\xi)$ (see \eqref{Omeggga}) and we get
$\EE I_s(\gD)=\sum_{\gF\in\gF(\gD)} c_\gF J_s(\gF)$. Now the desired formula \eqref{formula_Ea^ma^n} follows from the identity
$
\sum_{\gD\in\gD_m\times\ov\gD_n}\sum_{\gF\in\gF(\gD)}=\sum_{\gF\in\gF_{m,n}},
$
which is due to \eqref{gF-def}.

\smallskip

It remains to derive the estimate \eqref{F-prop}.
Recall that $\ga_s\geq 1$ for any $s\in\R^d$.
Let $\wt F^\gF$ be a function defined as $F^\gF$, where in the product over $\phi\in E_D(\gF)$
the exponent $\ga^\phi$ is replaced by $\ga^\phi-1/2$, that is
$$
F^\gF(\tau,l,\xi)=\wt F^\gF(\tau,l,\xi)\prod_{\phi\in E_D(\gF)} e^{-\frac12 (l_r^\phi-l_w^\phi) }
\mI_{\{-T\leq l^{\phi}_w \leq l^{\phi}_r\}}.
$$
Since the product over $\phi$ above does not depend on $\xi$, it suffices to show that
\be\lbl{chp-1}
|\p_{\xi^\ka} \wt F^\gF(\tau,l,\xi)|\leq C_{\ka}^\#(\xi) 
\ee
for $\xi\in\cA^{\prime}_s(\gF)$ and 
\be\lbl{chp-2}
\prod_{\phi\in E_D(\gF)} e^{-\frac12 (l_r^\phi-l_w^\phi) }
\mI_{\{-T\leq l^{\phi}_w \leq l^{\phi}_r\}} \leq e^{-\delta \big(\sum_{i=1}^m|\tau_1-l_i| + \sum_{i=m+1}^N|\tau_2-l_i|\big)},
\ee
where $\de=\de_N>0$.

We start with the estimate \eqref{chp-1}.
Since $b$ is a Schwartz function and, clearly, 
$\prod_{\psi\in E_L(\gF)} (b^\psi(\xi))^2 = \prod_{i\in L(\gF)}(b(\xi_i))^2$, we find
\be\non
|\p_{\xi^\ka} \wt F^\gF(\tau,l,\xi)|\leq C_{\ka}^\#\big((\xi_i)_{i\in L(\gF)}\big) P_\ka(\xi),
\ee
where $P_\ka$ is a polynomial. Thus, it remains to show that 
$C_{\ka}^\#\big((\xi_i)_{i\in L(\gF)}\big) \leq C_{1\ka}^\#(\xi)$
for any $\xi\in\cA_s^{\prime}(\gF)$, or equivalently 
\be\lbl{xi-leaf-xi}
\big|(\xi_i)_{i\in L(\gF)}\big|\geq C^{-1}|\xi_j|, \qquad C>0, 
\ee
for any $j\notin L(\gF)$. Inequality \eqref{xi-leaf-xi} easily follows from assumption \eqref{p'2}. 
Indeed, since for  $j\notin L(\gF)$ a vertex $c^{(p)}_{j}$ has positive degree $p\geq 1$, it is coupled in $\gF$ with a virtual vertex $\bar w_{2k}$.
Then, for $\xi\in\cA_s'(\gF)$ we have $f_\gF(2k)=j$, so  \eqref{p'2} implies  
$\xi_{2k-1}+\xi_{2k}=\xi_{f_\gF(2k-1)}+\xi_{j}$. Then, among the indices $\xi_{2k-1}, \xi_{2k}$ and $\xi_{f_\gF(2k-1)}$ there is at least one, say $\xi_{2k-1}$, satisfying $|\xi_{2k-1}|\geq |\xi_j|/3$.
If $2k-1\in L(\gF)$ then we get  \eqref{xi-leaf-xi}, so we are done.
Otherwise, the vertex $c^{(p_1)}_{2k-1}$ is not a leaf and we continue the procedure in the same way until we arrive at a leaf. If it is $\xi_{2k}$ or $\xi_{f_\gF(2k-1)}$ who dominates $|\xi_j|/3$ then we argue in the same way.

Now we establish the estimate \eqref{chp-2}.
 We recall that $E_D(\gF)$ is the set of edges of the diagram $\gF$ which couple its virtual vertices with vertices of positive degree. 
Fix any $1\leq k\leq m$. There exist edges $\phi_1,\ldots,\phi_p\in E_D(\gF)$ which form a path from the block $\hat B_k$ of the diagram $\gF$ to the root $a^{(m)}_s$.
That is, the edge $\phi_1$ couples the virtual vertex of the block $\hat B_k$ with a vertex of a block $\hat B_{j_1}$, the edge $\phi_2$ couples the virtual vertex of the block $\hat B_{j_1}$ with a vertex from a block $\hat B_{j_2}$, etc. Finally, the edge $\phi_p$ couples the virtual vertex of the block $\hat B_{j_{p-1}}$ with the root $c^{(m)}$ (here, essentially, $\hat B_{j_{p-1}} = B_1$). 
Accordingly to the indicator function in \eqref{chp-2}, the times $l_k,l_{j_1},\ldots, l_{j_{p-1}}$ assigned to the blocks $\hat B_k,\hat B_{j_1}, \ldots, \hat B_{j_{p-1}}$  satisfy
$-T\leq l_k\leq l_{j_1}\leq \ldots \leq l_{j_{p-1}}\leq \tau_1$.
Then, bounding by one those exponents from \eqref{chp-2} which correspond to edges $\phi\notin\{\phi_1,\ldots, \phi_p\}$, we get 
\be\lbl{est-times}
\begin{split}
\prod_{\phi\in E_D(\gF)}e^{-\frac12 (l_r^\phi-l_w^\phi) }
\mI_{\{-T\leq l^{\phi}_w \leq l^{\phi}_r\}} &\leq 
e^{-\frac12\Big((l_{j_1}-l_{k})+(l_{j_2}-l_{j_1}) + \ldots+ (\tau_1-l_{j_{p-1}}) \Big)}
\mI_{-T\leq l_{k}\leq l_{j_1}\leq\ldots\leq \tau_1}
\\
&=e^{-\frac12 (\tau_1-l_k)}\mI_{-T\leq l_{k}\leq l_{j_1}\leq\ldots\leq \tau_1}
\leq e^{-\frac12 |\tau_1-l_k|}.
\end{split}
\ee
Since \eqref{est-times} holds for any $1\leq k\leq m$, we get that its l.h.s.  is bounded by $e^{-\frac1{2m} \sum_{i=1}^m|\tau_1-l_i|}$.

Now we fix any  $m+1\leq k\leq N$ and find a path from the block $\hat B_k$ to the root $\bar c^{(n)}$. Arguing as above we get
\be\non
\prod_{\phi\in E_D(\gF)}e^{-\frac12 (l_r^\phi-l_w^\phi) }
\mI_{\{-T\leq l^{\phi}_w \leq l^{\phi}_r\}} \leq 
e^{-\frac1{2n} \sum_{i=m+1}^N|\tau_2-l_i|}.
\ee
Combining the two obtained estimates we get that \eqref{chp-2} holds with $\de=\frac{1}{4N}$.

\smallskip

We conclude the proof of the lemma by noting that the absolute convergence of the sum $J_s(\gF)$ follows from the estimate \eqref{F-prop}.
\qed

\subsection{Proof of Lemma \ref{l:cycle}}
\lbl{s:cycle}
\begin{figure}[t]
	a) \parbox{5cm}{
		\begin{tikzpicture}[]
		\node at (0,0) (c2r-1) {$c_{2r-1}$};
		\node at (0,-1.5) (bc2r-1) {$\bar c_{2r-1}$};
		\node at (-1.5,-1.5) (cj) {$\bullet$};
		\node at (-1.5,0) (ci) {$\bullet$};
		
		\node at (1.5,0) (bc2r) {$\bar c_{2r}$};
		\node at (1.5,-1.5) (c2r) {$ c_{2r}$};
		\node at (3,-1.5) (cjj) {$\bullet$};
		\node at (3,0) (cii) {$\bullet$};
		
		\node at (-0.7,-0.7) {$\gC_1$};
		\node at (2.3,-0.7) {$\gC_2$};

		\draw[dashed, line width=0.25mm] (c2r-1) -- (bc2r-1);
		\draw[line width=0.25mm] (bc2r-1) -- (cj);
		\draw[dotted, line width=0.25mm] (cj) -- (ci);
		\draw[line width=0.25mm] (ci) -- (c2r-1);
		
		\draw[dashed, line width=0.25mm] (c2r) -- (bc2r);
		\draw[line width=0.25mm] (bc2r) -- (cii);
		\draw[dotted, line width=0.25mm] (cjj) -- (cii);
		\draw[line width=0.25mm] (cjj) -- (c2r);
		\end{tikzpicture}
	}
	\hfill
	b) \parbox{5cm}{
		\begin{tikzpicture}[]
		\node at (0,0) (c2r-1) {$c_{2r-1}$};
		\node at (0,-1.5) (bc2r-1) {$\bar c_{2r-1}$};
		\node at (-1.5,-1.5) (cj) {$\bullet$};
		\node at (-1.5,0) (ci) {$\bullet$};
		
		\node at (1.5,0) (bc2r) {$\bar c_{2r}$};
		\node at (1.5,-1.5) (c2r) {$ c_{2r}$};
		\node at (3,-1.5) (cjj) {$\bullet$};
		\node at (3,0) (cii) {$\bullet$};
		
		\node at (0.7,-0.7) {$\gC_{12}$};

		\draw[line width=0.25mm] (bc2r-1) -- (cj);
		\draw[dotted, line width=0.25mm] (cj) -- (ci);
		\draw[line width=0.25mm] (ci) -- (c2r-1);
		
		\draw[line width=0.25mm] (bc2r) -- (cii);
		\draw[dotted, line width=0.25mm] (cjj) -- (cii);
		\draw[line width=0.25mm] (cjj) -- (c2r);
		
		\draw[dashed, line width=0.25mm] (c2r-1) -- (bc2r);
		\draw[dashed, line width=0.25mm] (c2r) -- (bc2r-1);
		\end{tikzpicture}
	}
	\caption{Construction of the cycle $\gC_{12}$ from the cycles $\gC_1$ and $\gC_2$. In this figure we do not make difference between solid and dashed edges.}
	\lbl{f:F10}
\end{figure}
Let us consider a graph $\gC'_\gF$ which is obtained from $\mathfrak{U}_\gF$ by erasing in each block two (of four) dashed edges. We choose the edges we preserve arbitrarily, but in such a way that in the graph  $\gC'_\gF$ each vertex is incident to exactly two  edges, one of which is solid  and another one is dashed.
 We will modify the choice of the preserved dashed edges in certain blocks of  $\gC'_\gF$ in such a way that $\gC'_\gF$ will become a cycle.
 Since in the graph  $\gC'_\gF$ every vertex is incident to two edges,  $\gC'_\gF$  decouples to a union of $k\geq 1$ disjoint cycles $\gC_i$, $1\leq i \leq k$, in which solid and dashed edges alternate. If $k=1$ then we are done. Assume that $k> 1.$
We claim that in this case there exists a block $\hat B_r=\{c_{2r-1},c_{2r}, \bar c_{2r-1}, \bar c_{2r}\}$ such that some but not all of its vertices belongs to cycle $\gC_1$.

Indeed, assume that the claim is not true, so that each block either entirely belongs to the cycle $\gC_1$, or its intersection with $\gC_1$ is empty. 
Take a block $\hat B_p$ belonging to $\gC_1$.
Then any vertex $\hat c_j\notin \hat B_p$ which is adjacent with a vertex from the block $\hat B_p$ (by construction of  $\gC'_\gF$ such vertex exists), belongs to the cycle $\gC_1$ and, consequently, all other vertices from the block to which belongs $\hat c_j$ belong to $\gC_1$ as well.
Arguing in this way, by induction we obtain that the cycle $\gC_1$ contains all blocks of the diagram  $\gC'_\gF$, as well as the vertices $c_0$ and $\bar c_0$.
Thus, $\gC_1= \gC'_\gF$ which contradicts to the assumption $k> 1$, so the claim is true.

Let $\hat B_r$ be a block as in the claim. 
By construction of the diagram  $\gC'_\gF$, its dashed edges either provide the couplings
$ c_{2r-1}\volna \bar c_{2r-1}$ and $ c_{2r}\volna \bar c_{2r}$
or the couplings $c_{2r-1}\volna \bar c_{2r}$ and $c_{2r}\volna \bar c_{2r-1}$.
For definiteness, we assume that $c_{2r-1} \volna \bar c_{2r-1}\in\gC_1$ and $c_{2r}\volna \bar c_{2r}\in \gC_2$, as in fig.~\ref{f:F10}(a). 
Let us erase the (dashed) edges that couple the vertices $c_{2r-1}$ with $\bar c_{2r-1}$ and 
$c_{2r}$ with $\bar c_{2r}$, and
instead couple the vertices 
$c_{2r-1}$ with $\bar c_{2r}$ and $c_{2r}$ with $\bar c_{2r-1}$.
Then from the diagram  $\gC'_\gF$ we obtain another diagram
 $\gC''_\gF$, constructed from $\mathfrak{U}_\gF$ in the same way as the diagram $\gC'_\gF$, in which there is one cycle $\gC_{12}$ instead of the two disjoint cycles $\gC_1$ and $\gC_2$ in  $\gC'_\gF$, see fig.~\ref{f:F10}(b). So the total number of cycles in the diagram  $\gC''_\gF$ is $k-1$,
where we recall that by our assumption in  $\gC'_\gF$ there are $k>1$ cycles.  
Then, arguing by induction, we arrive at a diagram $\gC_\gF$ which is a cycle itself.  

\subsection{Example: computation of $\EE a_s^{(2)}\bar a_s^{(2)}$ }
\lbl{s:ex_int}

In this section  we apply techniques, developed in Sections~\ref{s:D}-\ref{s:ch}, to analyse the expectation $\EE a_s^{(2)}\bar a_s^{(2)}$. 
Namely, we fix one of Feynman diagrams $\gF\in\gF^{\,true}_{2,2}$,~-- say, that from fig.~\ref{f:ex},~-- and find explicitly the corresponding integral $\wt J_s(\gF)$ (in total  there are about 25 different Feynman diagrams in the set $\gF^{\,true}_{2,2}$). 

\begin{figure}[t]
	\begin{tikzpicture}
	\node at (-1.2, 0) (c0) {$c_0^{(2)}$};
	
	\node at (-0.2, -1.3) (b_w2) {$\bar w_{2}$};
	\node at (-0.9, -1.2) (b_c1) {$\bar c_1^{(0)}$};
	\node at (-1.6, -1.2) (c2) {$c_2^{(0)}$};
	\node at (-2.3, -1.2) (c1) {$c_1^{(1)}$};
	
	\node at (-0.2, -2.6) (b_w4) {$\bar w_4$};
	\node at (-0.9, -2.5) (b_c3) {$\bar c_3^{(0)}$};
	\node at (-1.6, -2.5) (c4) {$c_4^{(0)}$};
	\node at (-2.3, -2.5) (c3) {$c_3^{(0)}$};
	
	%c0-b_w2
	\draw [line width=0.25mm] (c0.south)--(b_w2.north);
	%c1-b_w4
	\draw [line width=0.25mm] (c1.south)--(b_w4.north);
	%c2-b_c3
	\draw [line width=0.25mm] (c2.south)--(-1.1, -2.3);

	\node at (2,0) (b_c0) {$\bar c_0^{(2)}$};
	
	\node at (1, -1.2)  (c5) {$c_5^{(0)}$};
	\node at (1.7, -1.3) (w6) {$w_6$};
	\node at (2.4, -1.2) (b_c5) {$\bar c_5^{(1)}$};
	\node at (3.1, -1.2) (b_c6) {$\bar c_{6}^{(0)}$};  
	
	\node at (1, -2.5) (c7) {$c_7^{(0)}$};
	\node at (1.7, -2.5) (w8) {$w_8^{(0)}$};
	\node at (2.4, -2.5) (b_c7) {$\bar c_7^{(0)}$};
	\node at (3.1, -2.5) (b_c8) {$\bar c_{8}^{(0)}$}; 
	%b_c0-w4
	\draw [line width=0.25mm] (b_c0)--(w6.north);
	%b_c5-w8
	\draw [line width=0.25mm](b_c5)--(1.5, -2.3);
	%c5--b_c8
	\draw [line width=0.25mm](c5.south)--(2.9, -2.3);
	%b_c1-c7
	\draw [line width=0.25mm](b_c1.south)--(0.8, -2.4);
	%c3-b_c7
	\draw [line width=0.25mm](-2.4, -2.8)--(-2.4, -3.1)--(2.3, -3.1)--(2.3, -2.8);
	%c4-b_c6
	\draw [line width=0.25mm](-1.7, -2.8)--(-1.7,-3.3)--(3.5,-3.3)--(3.5,-1.3)--(3.3,-1.3);
	
	%c0-bc_0
	\draw [dashed, line width=0.25mm] (-1, -0.1) -- (1.6,-0.1);
	%c1-b_c1
	\draw [dashed, line width=0.25mm] (-2.05,-0.95) ..controls (-1.6,-0.7) .. (-1.05, -1);
	%c3-b_c3
	\draw [dashed, line width=0.25mm] (-2.2, -2.8) .. controls (-1.5, -3) .. (-1.1,-2.8);
	%c4-b_c4
	\draw [dashed, line width=0.25mm] (-1.5, -2.8) .. controls (-0.8, -3) .. (-0.4,-2.8);
	%c2-b_w2
	\draw [dashed, line width=0.25mm] (-1.4,-0.9) ..controls (-1,-0.7) .. (-0.3, -1.1);
	%c5-b_c5
	\draw [dashed, line width=0.25mm] (1.25, -0.95) ..controls (1.6,-0.7) .. (2.25, -1);
	%w6-b_c6
	\draw [dashed, line width=0.25mm] (1.85, -1.05) ..controls (2.3,-0.7) .. (2.95, -1);
	%c7-b_c8
	\draw [dashed, line width=0.25mm](1.1, -2.8)..controls (2,-3) .. (2.9, -2.8);
	%w8-b_c7 
	\draw [dashed, line width=0.25mm](1.8, -2.6)-- (2.1, -2.6);
	
	\end{tikzpicture}
	\caption{A diagram $\gF\in\gF^{\,true}_{2,2}$ and the corresponding cycle $\gC_\gF$.}
	\lbl{f:ex}
\end{figure}

The cycle $\gC_\gF$ for the considered diagram $\gF$ can be chosen as in fig~\ref{f:ex}, so that the associated cycle $\hat \gC_\gF$ has the form
\be\non
c_0\mapsto c_6 \mapsto c_4 \mapsto c_1 \mapsto c_7\mapsto c_5 \mapsto c_8\mapsto c_3 \mapsto c_2 \mapsto c_0.
\ee
Accordingly to \eqref{Om_xy}, the phase function $\Om^\gF$ has the form
$$
\Om^\gF(l,z)=2\big((l_1-l_2)z_1\cdot z_2 + (l_1-l_3)z_1\cdot z_3 + (l_3-l_2)z_2\cdot z_3 + (l_4-l_3)z_3\cdot z_4\big).
$$
Assuming for simplicity $T=\infty$, by \eqref{F=} the function $F^\gF$ reads 
\be\non
\begin{split}
	F^\gF(\tau&,l,\xi)= 
	\mI_{l_2\leq l_1\leq \tau_1}(\tau,l)  \mI_{l_4\leq l_3\leq \tau_2}(\tau,l)  \\
	&\exp \big(-\ga_{\xi_0}(\tau_1-l_1)-\ga_{\xi_1}(l_1-l_2)-\ga_{\xi_6}(\tau_2-l_3)-\ga_{\xi_8}(l_3-l_4)\big) \\
	&\exp \big(\ga_{\xi_2}(l_1-l_2) -\ga_{\xi_3}|l_2-l_4|-\ga_{\xi_4}|l_2-l_3|-\ga_{\xi_5}(l_3-l_4) -\ga_{\xi_7}|l_1-l_4|\big)\\
	&B(\xi_2)B(\xi_3)B(\xi_4)B(\xi_5)B(\xi_7),
\end{split}
\ee
where we denote $B(s)=(b(s))^2/\ga_s$. 
Here the first two lines correspond to the product over the set $E_D(\gF)$ in the formula \eqref{F=} and the last to lines ~-- to the product over $E_L(\gF)$.
Accordingly to \eqref{F=F(z)}, to write down the function $F_s^\gF$ from \eqref{J_c(F)} it remains to write the functions $\xi_i(z)$, which are given by~\eqref{xi,s(z)}: 
\be\non\begin{split}
	&\xi_0=s, \qu \xi_1=s+z_2+z_3 \\
	&\xi_2=s-z_1,\qu \xi_3=s-z_1+z_2, \qu
	\xi_4=s+z_3, \qu \xi_5=s-z_1+z_2+z_3-z_4, \\
	&\xi_6=s,\qu
	\xi_7=s-z_1+z_2+z_3,\qu
	\xi_8=s-z_1+z_2-z_4.
\end{split}	
\ee 
In Appendix~\ref{s:time out}, assuming for simplicity $\tau_1=\tau_2$ and $T=\infty$,  we explicitly compute the integral over the time variable $dl$ in \eqref{J_c(F)} and discuss the motivation for this. Below we present the result of this computation for the example we are considering
(here we do not restrict ourselves to the case $\tau_1=\tau_2$ but assume $T=\infty$).

We have
$\om_1^\gF(z)=2z_1\cdot(z_2+z_3)$, 
$\om_2^\gF(z)=2z_2\cdot(-z_1-z_3),$
$\om_3^\gF(z)=2z_3\cdot(-z_1+z_2-z_4)$
and 
$\om_4^\gF(z)=2z_4\cdot z_3$.
Computing the integral over  $dl$ we get
\be\non
\wt J_s(\gF)=\frac{1}{2\ga_s}\sum_{j=1}^6 e^{\pm\ga_s(\tau_1-\tau_2)}\int_{\R^{dN}} \frac{B(\xi_2)B(\xi_3)B(\xi_4)B(\xi_5)B(\xi_7)}{K^j(z)}\,dz,
\ee
where for $j=1,2,3$ we take "$+$" in the term $\pm\ga_s$, for $j=4,5,6$ we take "$-$", $\xi=\xi(z)$ and
\be\non
K^j(z)=\big(\Ga_1^j(z) + i\nu^{-1} D_1^j(z)\big)
\big(\Ga_2^j(z) + i\nu^{-1} D_2^j(z)\big)
\big(\Ga_3^j(z) + i\nu^{-1} D_3^j(z)\big).
\ee
Here the quadratic forms $D^j_k$ are given by sums of $\om^\gF_k$ while the functions $\Ga_k^j\geq 1$~-- by sums of $\ga_{\xi_i}$.
Namely, omitting the upper index $\gF$,
$$
D_1^1=\om_4,\qu D_2^1=\om_3+\om_4,\qu D_3^1=\om_2+\om_3+\om_4=-\om_1,
$$  
since $\sum_{j=1}^4\om_j=0$.
Similarly,
\begin{align*}
&D_1^2=\om_4,\qu D_2^2=\om_2+\om_4,\qu D_3^2=-\om_1, \\
&D_1^3=\om_2,\qu D_2^3=\om_2+\om_4,\qu D_3^3=-\om_1,\\
&D_1^4=\om_4,\qu D_2^4=\om_2+\om_4,\qu D_3^4=-\om_3,\\
&D_1^5=\om_2,\qu D_2^3=\om_2+\om_4,\qu D_3^3=-\om_3,\\
&D_1^6=\om_2,\qu D_2^6=\om_1+\om_2,\qu D_3^6=-\om_3.
\end{align*}
The functions $\Ga_k^j$ read
\begin{align*}
&\Ga_1^1=\ga_{\xi_3}+\ga_{\xi_5}+ \ga_{\xi_7} + \ga_{\xi_8},\;
\Ga_2^1=\ga_{s}+\ga_{\xi_3}+ \ga_{\xi_4} + \ga_{\xi_7},\;
\Ga_3^1=\ga_s+\ga_{\xi_1}+\ga_{\xi_2}+ \ga_{\xi_7},\\
&\Ga_1^2=\Ga_1^1,\qu
\Ga_2^2=\ga_{\xi_1}+\ga_{\xi_2}+ \ga_{\xi_4} +\ga_{\xi_5}+ \ga_{\xi_7}+\ga_{\xi_8},\qu
\Ga_3^2=\Ga_3^1,\\
&\Ga_1^3=\ga_{\xi_1}+\ga_{\xi_2}+ \ga_{\xi_3} + \ga_{\xi_4},\qu
\Ga_2^3=\Ga_2^2,\qu
\Ga_3^3=\Ga_3^1,\\
&\Ga_1^4=\Ga_1^1,\qu
\Ga_2^4=\Ga_2^2,\qu
\Ga_3^4=\ga_s+\ga_{\xi_4}+\ga_{\xi_5}+ \ga_{\xi_8},\\
&\Ga_1^5=\Ga_1^3,\qu
\Ga_2^5=\Ga_2^2,\qu
\Ga_3^5=\Ga_3^4,\\
&\Ga_1^6=\Ga_1^3,\qu
\Ga_2^6=\Ga_2^1,\qu
\Ga_3^6=\Ga_3^4.
\end{align*}

\subsection{Proof of Lemma~\ref{l:rank}} 
\lbl{s:l_rank}
Since the matrix $A$ is irreducible, there exists $k$ and indices $i_1,\ldots,i_k$ satisfying $i_j\neq i_{j+1}$ $\forall j$, such that 
$a_{i_1i_2}a_{i_2i_3}\cdots a_{i_{k-1}i_k}\neq 0$ and the sequence $i_1,\ldots,i_k$ contains every number from $1$ to $n$.
We construct a set $B$ of linearly independent vectors $r_{ij}$ by the following inductive procedure. For the base of induction we consider the indices $i_1,i_2$ and set 
$B=\{r_{i_1i_2}\}$. For the inductive step, we assume that the indices $i_1,\ldots, i_{m-1}$ are already considered, and we consider the index $i_{m}$. If the set $B$ does not contain any vector $r_{uv}$ for which $u=i_m$ or $v=i_m$ then we add the vector $r_{i_{m-1}i_m}$ to $B$ and pass to the next step. Otherwise we just  pass to the next step. We conclude the procedure when all indices $i_1,\dots,i_k$ are considered. 

Since the sequence  $i_1,\ldots,i_k$ contains every $1\leq j\leq n$, after the end of the procedure above we have $|B|=n-1$. Due to the form of the vectors $r_{ij}$, vectors from the set $B$ are linearly independent, so the rank $R$ satisfies $R\geq n-1$. Thus, it remains to show that $R<n$. If $R=n$ then the basis vectors $e_m$ can be uniquely expressed through the vectors $r_{ij}$. However, this is not possible since the vectors $r_{ij}$ are invariant with respect to translations of the set $(e_i)$: if we replace $e_i$ by $e_i+e$ for every $i$ and a vector $e$ then the vectors $r_{ij}$ remain unchanged. 
\qed

\subsection{Approximate equation}
\lbl{s:app_eq_proofs}

In this section we adopt the method  of Feynman diagrams, developed for equation \eqref{ku4}, to  equation \eqref{ku33}, and prove
  Proposition~\ref{l:approx_eq}. In fact,  for the proof 
   we do not need to develop the theory in so detailed way (cf.  proof of Corollary 2.2 in \cite{DK}), but we also use the results of this section 
    in our  work \cite{DKMV}. 

\subsubsection{Extended Feynman diagrams and correlations $\EE\fa_s^{(m)}\bar \fa_s^{(n)}$}
Similarly to Section~\ref{s:int_repr} we write equation \eqref{ku33} in the interaction representation 
$
\fa_s(\tau)=v_s(\tau) e^{-i\nu^{-1}\tau |s|^2}\,:
$
\begin{equation}\non
\dot \fa_s + \gamma_s \fa_s
=  i\rho L^{-d} {\sum}_{s_1,s_2,s_3}  (\dess-\de^{=12}_{\ 3s}) \fa_{s_1} \fa_{s_2} \bar \fa_{s_3}
e^{  i \nu^{-1}\tau \omega^{12}_{3s}}
+b(s) \dot\beta_s  \,,\quad s\in{{\mathbb Z}}^d_L\,,
\end{equation}
where $\de^{=12}_{\ 3s}=1$ if $s_1=s_2=s_3=s$ and $\de^{=12}_{\ 3s}=0$ otherwise.
Note that 
\be\label{mutual}
\dess \cdot  \de^{=12}_{\ 3s} \equiv  0.
\ee
%the conditions $\de'^{12}_{3s}=1$ and $\de^{=12}_{\ 3s}=1$ are mutually exclusive.
Expansion \eqref{decomp}, written in the $\fa$-coordinates, takes the form 
$$
\fa_s(\tau)=\sum_{i=0}^\infty\rho^i \fa_s^{(i)}, \quad \fa_s^{(i)}(-T)=0,
$$ 
where $\fa_s^{(0)}=a_s^{(0)}$ and, similarly to \eqref{an}, 
\be\label{fan}
\begin{split}
	\fa^{(m)}_s&(\tau) 
	= \sum_{m_1+m_2+m_3=m-1} i \int_{-T}^\tau   dl \\
	&\times L^{-d}\sum_{s_1, \ldots,s_4}
	(\de'^{12}_{34}-\de^{=12}_{\ 34})\,e^{-\ga_{s_4}(\tau-l)} \,
	\de^s_{s_4}\big( \fa_{s_1}^{(m_1)} \fa_{s_2}^{(m_2)} {\bar \fa_{s_3}}^{(m_3)}\big)(l)\;
	\theta(\om^{12}_{34},\nu^{-1}l).
\end{split}
\ee   
Here and below in this section   $\theta$ is as in \eqref{theta}. 

%as arbitrary measurable bounded function, see Remark~\ref{r:theta}.

Literally repeating the proof of Lemma~\ref{l:a^ma^n}, we see that the product 
$\fa_s^{(m)}(\tau_1)\bar \fa_s^{(n)}(\tau_2)$ is given by formula \eqref{a^ma^n} where in the definition \eqref{a^ma^n-I} of the sum $I_s(\gD)$ the set $\cA_s^L(\gD)$ is replaced by the set $\mathcal{B}_s^L(\gD)$, defined by relations \eqref{p1}, \eqref{p4} and 
\be\lbl{de'=}
\de'^{\xi_{2i-1}\xi_{2i}}_{\s_{2i-1} \s_{2i}}
+\de^{=\xi_{2i-1}\xi_{2i}}_{\ \s_{2i-1} \s_{2i}}=1
\ee 
(we recall \eqref{mutual}), cf. \eqref{de'==1}.
Moreover, the function $G_\gD$ from \eqref{a^ma^n-I} is replaced by the function
$
(-1)^{r(\xi,\s)} G_\gD(\tau,l,\xi,\s),
$
where $r(\xi,\s)$ is the number of indices $1\leq i \leq N$ for which $\de^{=\xi_{2i-1}\xi_{2i}}_{\ \s_{2i-1} \s_{2i}}=1$.

\begin{comment}
Let $\de\cA_s(\gD):=\wt\cA_s^L(\gD)\sm\cA_s^L(\gD)$ (we skip the upper index $L$ for simplicity of notation). Then, we find
\be\non
\Del_s^{m,n}(\tau_1,\tau_2):=\fa^{(m)}(\tau_1)\bar \fa^{(n)}(\tau_2)- a^{(m)}(\tau_1)\bar a^{(n)}(\tau_2) = \sum_{\gD\in\gD_m\times\ov\gD_n}\de I_s(\gD),
\ee
where 
\be\non
\de I_s(\gD;\tau)=
\int_{\R^N}
dl\, 
L^{-Nd}\sum_{(\xi,\s)\in
\de\cA_s(\gD)} \,
\wt G_\gD(\tau,l,\xi,\s)\,
\Theta(\om(\xi,\s),\nu^{-1}l)\,,
\ee
$\tau=(\tau_1,\tau_2)$ and the function 
$\wt G_\gD(\tau,l,\xi,\s)$, defined for 
$(\xi,\s)\in\de\cA_s(\gD)$, is given by the formula \eqref{G_D} multiplied by the factor $(-1)^r$, where $r=r(\xi,\s)$ is the number of blocks $\hat B_i$ in $\gD$ such that the corresponding indices satisfy $\de^{=\xi_{2i-1}\xi_{2i}}_{\ \s_{2i-1} \s_{2i}}=1$.
\end{comment}

Next we compute the expectation $\EE\fa_s^{(m)}\bar \fa_s^{(n)}$. 
\begin{comment}
By  definition, the set $\de\cA_s$ consists of vectors $(\xi,\s)$ satisfying \eqref{p1}, \eqref{p4} and $\eqref{de'=}$ such that  there is at least one number $1\leq i\leq 2N$ for which 
$\de^{=\xi_{2i-1}\xi_{2i}}_{\ \s_{2i-1} \s_{2i}}=1$.
\end{comment}
In difference with the computation of $\EE I_s(\gD)$ in Section~\ref{sec:FD},  now we should take into account Wick pairings which couple leaves belonging to the same block of the diagram $\gD$. Indeed, the situation when $\xi_i=\s_j$ for $(\xi,\s)\in\mathcal{B}_s^L(\gD)$ and $i,j$ such that the vertices $c_i$ and $\bar c_j$ belong to the same block $\hat B_k$ (that is $i,j\in\{2k-1,2k\}$) now  is not impossible: it occurs if $\de^{=\xi_{2k-1}\xi_{2k}}_{\ \s_{2k-1} \s_{2k}}=1$. 
Arguing as when proving Lemma~\ref{l:Ea^ma^n}, we get 
\be\non
\EE\fa_s^{(m)}(\tau_1)\bar \fa_s^{(n)}(\tau_2)=\sum_{\gF\in\gF^+_{m,n}}c_\gF\,\mathcal{J}_s(\gF),
\ee 
where the set $\gF^+_{m,n}$ of {\it extended} Feynman diagrams is defined in the same way as the set  $\gF_{m,n}$ of Feynman diagrams from Section~\ref{sec:FD} except that in the diagrams $\gF\in\gF^+_{m,n}$ we allow leaves belonging to the same block to be coupled.
The terms $\cJ_s(\gF)$  are defined similarly to \eqref{Ea^ma^n-J}:
\be\lbl{deJ}
\cJ_s(\tau;\gF)=
\int_{\R^N}
dl\, 
L^{-Nd}\sum_{\xi\in\
	\cB^{\prime L}_s(\gF)}\, \mathcal{F}^\gF\,(\tau, l, \xi)\, 
\Theta(\om^\gF(\xi),\nu^{-1}l).
\ee	
Here the set of  admissible 
multi-indices   $\cB^{\prime L}_s(\gF)\subset(\Z^{d}_L)^{2N+1}$ is given by the relation $\xi_0=\s_0=s$  together
 with \eqref{de'=}, cf. \eqref{p'1}, \eqref{p'2}. 
 From now on  we set $$\s:=\xi_{f_\gF},$$ where the permutation $f_\gF$ is defined by \eqref{pi_F}  in Section~\ref{sec:FD}. 
 The function $\mathcal F^\gF$ is given by 
 \be\label{FF}
 \mathcal F^\gF\,(\tau, l, \xi)=(-1)^{r(\xi,\s)} F^\gF\,(\tau, l, \xi)
 \ee
  (see \eqref{F=}), while the vector $\om^\gF(\xi)= (\om^\gF(\xi)_j)$ is defined as in Section~\ref{sec:FD}. % $\om^\gF(\xi)=\om(\xi,\s)$.

Our next goal is to resolve  relations  \eqref{de'=} with $\s=\xi_{f_\gF}$, entering the definition of the set $\cB^{\prime L}_s(\gF)$. 
In Section~\ref{s:ch} we dealt with an 
analogous problem, but now there is an additional difficulty: the transformation $\xi\mapsto x$ used in Section~\ref{s:ch} is not well defined if in the diagram $\gF$ there are coupled vertices belonging to the same block. That is if $\gF\in\gF^+_{m,n}\sm\gF_{m,n}$.  

Below we call {\it irregular} those blocks of $\gF$ which contain a pair of coupled vertices. 
%while the other blocks are called {\it regular}. 
By definition, an $\gF$ belongs to $\gF^+_{m,n}\sm\gF_{m,n}$ if and only if it has at least one irregular block.   
To overcome the difficulty mentioned above, we will regularize such diagrams by erasing the irregular blocks thus reducing our study to the case $\gF\in\gF_{m,n}$, possibly, with smaller $m,n$.

%Indeed, assume for 
%definiteness  that the irregular  block $\hat B_k$ is such that $c_{2k-1}\sim\bar c_{2k-1}$. Then $\s_{2k-1}=\xi_{2k-1}$, which implies $\de'^{\xi_{2k-1}\xi_{2k}}_{\s_{2k-1} \s_{2k}}=0$. 
%Now  \eqref{del_irr} follows from \eqref{de'=}.

\begin{figure}[t]
	
	a) \parbox{5cm}{
		\begin{tikzpicture}
		\node at (-1,0) (c0) {$c_0^{(3)}$};
		\node at (1,0) (bc0) {$\bar c_{0}^{(0)}$};
		
		\node at (-2.5, -1.2) (c1) {$c_1^{(2)}$};
		\node at (-1.6, -1.2) (c2) {$c_2^{(0)}$};
		\node at (-0.5, -1.2) ( bc1) {$\bar c_1^{(0)}$};
		\node at (0.4, -1.3) (bw2)  {$\bar w_{2}$};  
		
		\node at (-2.5, -2.5) (c3) {$c_3^{(1)}$};
		\node at (-1.6, -2.5) (c4) {$c_4^{(0)}$};
		\node at (-0.5, -2.5) (bc3) {$\bar c_3^{(0)}$};
		\node at (0.4, -2.6) (bw4) {$\bar w_{4}$}; 
		
		\node at (-2.5, -3.8) (c5) {$c_5^{(0)}$};
		\node at (-1.6, -3.8) (c6) {$c_6^{(0)}$};
		\node at (-0.5, -3.8) (bc5) {$\bar c_5^{(0)}$};
		\node at (0.4, -3.9) (bw6) {$\bar w_{6}$}; 
		
		\draw [line width=0.25mm] (-0.8,-0.3)-- (0.35, -1);
		%c1-bw4:
		\draw [line width=0.25mm](c1.south)--(bw4.north);
		%c2-bc0:
		\draw [line width=0.25mm](c2.north)--(-1.6, 0.5)--(1,0.5)--(bc0.north);
		%c3-bw6:
		\draw [line width=0.25mm] (c3.south) --  (bw6.north);
		%c4-bc1:
		\draw [line width=0.25mm](-1.6, -3.5)--(-0.7, -1.5);
		%c4-bc3:
		\draw [line width=0.25mm] (-1.5, -2.6) -- (-0.8, -2.6);
		%c3-bc3:
		\draw [line width=0.25mm] (-2.4, -4.1) .. controls (-1.7,-4.3) .. (-0.7, -4.1);
		
		\end{tikzpicture}
	}
	\qu
	b) \parbox{5cm}{
		\begin{tikzpicture}
		\node at (-1,0) (c0) {$c_0^{(2)}$};
		\node at (1,0) (bc0) {$\bar c_{0}^{(0)}$};
		
		\node at (-2.5, -1.2) (c1) {$c_1^{(1)}$};
		\node at (-1.6, -1.2) (c2) {$c_2^{(0)}$};
		\node at (-0.5, -1.2) ( bc1) {$\bar c_1^{(0)}$};
		\node at (0.4, -1.3) (bw2)  {$\bar w_{2}$};  
		
		\node at (-2.5, -2.5) (c5) {$c_5^{(0)}$};
		\node at (-1.6, -2.5) (c6) {$c_6^{(0)}$};
		\node at (-0.5, -2.5) (bc5) {$\bar c_5^{(0)}$};
		\node at (0.4, -2.6) (bw6) {$\bar w_{6}$}; 
		
		\draw [line width=0.25mm] (-0.8,-0.3)-- (0.35, -1);
		%c1-bw6:
		\draw [line width=0.25mm](c1.south)--(bw6.north);
		%c2-bc0:
		\draw [line width=0.25mm](c2.north)--(-1.6, 0.5)--(1,0.5)--(bc0.north);
		%c2-bc1:
		\draw [line width=0.25mm](c6.north)--(-0.7, -1.5);
		%c6-bc3:
		\draw [line width=0.25mm] (-2.4, -2.8) .. controls (-1.7,-3) .. (-0.7, -2.8);
		
		\end{tikzpicture}
	}
	\hfill
	\\
	
	\bigskip
	
	c)	\parbox{5cm}{
		\begin{tikzpicture}
		\node at (-1,0) (c0) {$c_0^{(1)}$};
		\node at (1,0) (bc0) {$\bar c_{0}^{(0)}$};
		
		\node at (-2.5, -1.2) (c1) {$c_1^{(0)}$};
		\node at (-1.6, -1.2) (c2) {$c_2^{(0)}$};
		\node at (-0.5, -1.2) ( bc1) {$\bar c_1^{(0)}$};
		\node at (0.4, -1.3) (bw2)  {$\bar w_{2}$};  
		
		%c1-bc1:
		\draw [line width=0.25mm] (-2.4, -1.5) .. controls (-1.7,-1.7) .. (-0.7, -1.5); 
		
		\draw [line width=0.25mm] (-0.8,-0.3)-- (0.35, -1);
		%c2-bc0:
		\draw [line width=0.25mm](c2.north)--(-1.6, 0.5)--(1,0.5)--(bc0.north);	
		\end{tikzpicture}
	}
	\qu	
	d)	\parbox{5cm}
	{	
		\begin{tikzpicture}
		\node at (-1,0) (c0) {$c_0^{(0)}$};
		\node at (1,0) (bc0) {$\bar c_{0}^{(0)}$};
		
		\draw [line width=0.25mm](c0.east) -- (bc0.west);
		\end{tikzpicture}
	}
	\hfill
	\caption{a) A diagram $\gF\in\gF_{3,0}^+$. The blocks $\bar B_2$ and $\bar B_3$ are irregular.
		b) The diagram $\gF_1$ is obtained from $\gF$ after the first step of the regularization procedure by erasing the block $\bar B_2$. Its only irregular block is $\bar B_3$.
		c) The diagram $\gF_2$ is obtained by erasing the block $\bar B_3$ in $\gF_1$. The block $\bar B_1$ becomes irregular.
		d) The diagram $\gF_3=\gF_\Rc$ turns out to be trivial: it consists only of the root vertices.}
	\lbl{f:Fext}
\end{figure}

Before explaining the procedure let us note that if $\xi\in\cB'^L_s(\gF)$
then for any $k$ such that the block $\hat B_k$ is irregular, we have 
\be\lbl{del_irr}
\de^{=\xi_{2k-1}\xi_{2k}}_{\ \s_{2k-1} \s_{2k}}=1, \quad \text{i.e.} \quad  \xi_{2k-1} =\xi_{2k} = \s_{2k-1} = \s_{2k}.
\ee
We regularize  a diagram $\gF\in\gF^+_{m,n}\sm\gF_{m,n}$ in several steps. In the first one we take an irregular block $\hat B_k$ 
with the least number $k$.
It contains a pair of coupled leaves, say for definiteness these are $c_{2k-1}\sim\bar c_{2k-1}$.  Also
assume for definiteness that the block $\hat B_k$ is conjugated, i.e. $\hat B_k=\bar B_k$. 
Then, apart from  the leaves $c_{2k-1}$ and $\bar c_{2k-1}$,  it  contains a virtual vertex $\bar w_{2k}$ which is coupled with a parent $c_{k'}$ of the block and a real vertex $c_{2k}$ coupled with a vertex $\bar c_{k''}$  that does not belong to the block $\bar B_k$.
We erase in $\gF$ the block $\bar B_k$, couple directly the parent $c_{k'}$ with the vertex $\bar c_{k''}$ and decrease by $1$ the degree of the parent $c_{k'}$, degree of the parent of the block to which belongs $c_{k'}$, and so on until we arrive at the root vertex $c_0$ or $\bar c_0$, see fig.~\ref{f:Fext}.  Let us denote the resulting diagram by  $\gF_{1}$. 
If we re-numerated the vertices and  blocks of  $\gF_{1}$ to remove the gaps in the
 numerations\,\footnote{
For that we should decrease by one the numbers $j$ of all  blocks $\hat B_{j}$ with $j>k$ and renumber the vertices belonging to these blocks accordingly.}
 (which we do not do!), then we would have that  
$\gF_{1}\in\gF_{m',n'}^+$ with $m'+n'=m+n-1$. So, slightly abusing notation,  we write $\gF_1\in\gF_{m',n'}^+$.
The diagram $\gF_{1}$ has one block less than the initial diagram $\gF$ and in comparison with the latter has a new edge $(c_{k'}, \bar c_{k''})$.
Note that if the vertices $c_{k'}$ and $\bar c_{k''}$ as above  belong to the same block, then the latter 
 cannot be irregular in the diagram $\gF$ but is necessarily 
  irregular in $\gF_1$ (see fig.~\ref{f:Fext}(c)). At the same time, all other blocks of $\gF_1$, irregular or not, are equally irregular or not in $\gF$.
%, so that the number of irregular blocks in $\gF_1$ is less by one or equal to that  of $\gF$. 

Since the vertices $c_{k'}, \bar c_{k''}$ are coupled in $\gF$ to those in  the block  $\bar B_k$, then in view of \eqref{del_irr} we 
get $\xi_{k'}=\s_{k''}$ for $\xi\in\cB'^L_s(\gF)$.
Then the multi-index  $\xi^{1}:=(\xi_i)_{i:\,c_i\in\gF_1}$, associated with the diagram $\gF_1$, satisfies $\xi^1\in\cB'^L_s(\gF_1)$. In particular,  \eqref{del_irr} holds for the indices associated with any irregular block of $\gF_1$.

In the second step we apply the same procedure to the diagram $\gF_1$ and obtain a diagram $\gF_2\in\gF_{m'',n''}^+$ with $m''+n''=m+n-2$ 
(we again slightly abuse the notation). Moreover, we get $\xi^2\in\cB'^L_s(\gF_2)$, where $\xi^2=(\xi_i)_{i:\,c_i\in\gF_2}$. In the third step we apply the procedure to the diagram $\gF_2$, etc. We 
 continue in the same way until we get a final 
diagram $\gF_p$, which we denote $\gF_\Rc$, that does not have irregular blocks. That is $\gF_\Rc\in\gF_{\wt m,\wt n}$, where 
$ \wt m+\wt n= m+n-p$.\footnote{If $\wt m=\wt n=0$ then the diagram $\gF_\Rc$ is trivial, see fig.~\ref{f:Fext}(d).}   Let us  repeat that in $\gF_\Rc$ we keep
the original numeration of vertices and blocks, inherited from $\gF$.

Let us denote by $\Rc\subset\{0,\dots,2N\}$ the set, made by the numbers of vertices in $\gF_\Rc$, 
and by $\Rc_B\subset\{1,\dots,N\}$~-- the set, made by the 
 numbers of blocks in $\gF_\Rc$. So 
$$
\Rc=\{2k-1,2k:\, k\in\Rc_B\}\cup\{0\}.
$$ 
E.g. for the diagram $\gF$ from fig.~\ref{f:Fext}(a), $\Rc_B=\emptyset$ and $\Rc=\{0\}$. Note that the construction of  $\Rc_B$ does not exclude
that for some blocks $\hat B_k$ with $k\in \Rc_B$ holds \eqref{del_irr}.

We will say that vertex $c_i$, $i\in\Rc$, {\it confines} a block $\hat B_k$ of the diagram $\gF$, 
 erased during the regularization procedure (so $k\notin \Rc_B$) if  there exists an edge $(c_i,\bar c_j)$ of $\gF_\Rc$, not contained in $\gF$, which 
 appeared in $\gF_\Rc$ as a result of erasing a set of blocks which includes $\hat B_k$. 
  E.g.,  the vertex $c_0$ confines all blocks of the diagram from fig.~\ref{f:Fext}(a).

Let us denote $\xi^\Rc:=(\xi_i)_{i\in\Rc}$.
From what is told above, by induction we get
\begin{lemma}\lbl{l:set_B}
	(i)\ Every  $\xi\in\cB'^L_s(\gF)$ satisfies \eqref{del_irr} for  $k\notin\Rc_B$.
	
	\smallskip
	
	(ii) \ 
	$
	\cB'^L_s(\gF) = \big\{\xi:\, \xi^\Rc\in\cB'^L_s(\gF_\Rc) \qnd  (\xi)_{j\notin\Rc}= \Xi(\xi^\Rc) \big\},
	$ \\
	where  $\Xi(\xi^\Rc)$ is the multi-index such that for $i \in \Rc$  and $j \notin \Rc$,
	$
	 \Xi(\xi^\Rc)_j = \xi_i
	$
	if the vertex $c_i$ confines the block which contains $c_j$.
\end{lemma}

In view of Lemma~\ref{l:set_B}(ii) in order to understand the set $\cB'^L_s(\gF)$ we should first understand $\cB'^L_s(\gF_\Rc)$.
 To this end we apply 
 to the diagram $\gF_\Rc$ the argument developed in Section~\ref{s:ch}.
Namely, we construct the reduced cycle $\hat\gC_{\gF_\Rc}$, the associated permutation $\pi_{\gF_\Rc}$ and the S-regular 
 incidence matrix $\al^{\gF_\Rc}$ of size $\#\Rc_B$. Then we 
  make the transformation $\xi^\Rc\mapsto x^\Rc:=(x_i)_{i\in\Rc}$,
   defined by formulas \eqref{x,y-xi,x} and \eqref{xi(x)}, and  write $\xi^\Rc=\xi^\Rc(x^\Rc)$.

Similarly to Proposition~\ref{l:change} we get
\begin{proposition}\lbl{al:change}
	(i)\ The set $\cB'^L_{s}(\gF_\Rc)$
	consists of all vectors $\xi^\Rc=\xi^\Rc (x^\Rc)$, where components $x_k\in\Z^d_L$ of $x^\Rc$  satisfy the relations
	\be\lbl{a:ch1} 
	(a) \
	x_0=s, \qquad\qu
	(b) \
	x_{2k}=-x_{2k-1} \quad\forall k\in\Rc_B, 
	\ee
	and $x_{2k-1}= 0$ if and only if 
	$\sum_{i\in\Rc_B}\al_{ki}^{\gF_\Rc} x_{2i-1}= 0$.
	
	(ii) \ If $\xi^\Rc=\xi^\Rc(x^\Rc)$ then $x_{2k-1}=0$ if and only of $\de^{=\xi_{2k-1}\xi_{2k}}_{\ \s_{2k-1}\s_{2k}}=1$, where $\s=\xi_{f_{\gF_\Rc}}$.
\end{proposition}
{\it Proof.} 
Let us decompose 
$$
\cB'^L_s(\gF_\Rc) = \bigcup_{\cK\subset\Rc_B}\, \cB'^L_{s,\cK}(\gF_\Rc), 
$$ 
where the union is taken over all subsets $\cK$  of $\Rc_B$ (including the empty one and $\Rc_B$) and 
the set $\cB'^L_{s,\cK}(\gF_\Rc)$ consists of those vectors $\xi^\Rc$ for which
$$
\de^{=\xi_{2k-1}\xi_{2k}}_{\ \s_{2k-1}\s_{2k}}=1 \qmb{for}\qu k\in\cK \qu\qnd\qu
\de^{\prime\xi_{2k-1}\xi_{2k}}_{\ \s_{2k-1}\s_{2k}}=1 \qmb{for}\qu k\notin\cK
$$
(we recall \eqref{mutual}). 
Fix any $\cK\subset\Rc_B$ and take $k\in\cK$ (if $\cK\ne\emptyset$).
Since we have \eqref{pi=f}, then 
the assumption $\de^{=\xi_{2k-1}\xi_{2k}}_{\ \s_{2k-1}\s_{2k}}=1$ imposed on $\xi^\Rc\in\cB'^L_{s,\cK}(\gF_\Rc)$ is equivalent to 
$\xi_{2k-1}=\xi_{2k}=\xi_{\pi(2k-1)}=\xi_{\pi(2k)}$. Or, to $x_{2k-1}=x_{2k}=0$ together with \be\lbl{ner}
\xi_{2k-1}-\xi_{\pi(2k)}=0.
\ee 
In particular, we get \eqref{a:ch1}(b) for $k\in\cK$. This also proves the  implication $\Leftarrow$ in item {\it (ii)} of the proposition.

Take now $k\notin\cK$ (if $\cK\ne\Rc_B$). Then, as in Proposition~\ref{l:change}, relation $\de^{\prime\xi_{2k-1}\xi_{2k}}_{\ \s_{2k-1}\s_{2k}}=1$ imposed on  $\xi^\Rc\in\cB'^L_{s,\cK}(\gF_\Rc)$ is equivalent to \eqref{de'1} together with \eqref{intersec=empty}, or to \eqref{a:ch1}(b) with $x_{2k-1}\ne 0$ and 
\be\lbl{ravnn}
\xi_{2k-1}-\xi_{\pi(2k)}\ne 0
\ee 
(see \eqref{intersec=empty''}).
In particular, we get \eqref{a:ch1}(b)  for $k\notin\cK$ and an implication $\Rightarrow$ in item {\it (ii)} of the proposition.

 As in the proof of   Proposition~\ref{l:change}, using \eqref{a:ch1}(b), we get that the relation $\xi_0=\s_0=s$, imposed on $\xi^\Rc\in\cB'^L_{s,\cK}(\gF_\Rc)$, is equivalent to \eqref{a:ch1}(a). We also get that
$$
\xi_{2k-1}-\xi_{\pi(2k)}=\sum_{i\in\Rc_B}\al_{ki}^{\gF_\Rc} x_{2i-1},
$$
see \eqref{differ-xi}.
Then, in view of \eqref{ner}, \eqref{ravnn}, we obtain the last assertion in item {\it (i)} of the proposition.
\qed

\smallskip

Similarly to Section~\ref{s:ch}, for $j\in\Rc_B$ we denote $z_j:=x_{2j-1}$  and set $z:=(z_j)_{j\in\Rc_B}$.  
Due to Proposition~\ref{al:change}(i), vector $z$ forms coordinates on the set $\cB'^L_s(\gF_\Rc)$.
Namely, let
\be\lbl{ZZZ-aaa}
\cZ^+(\gF_\Rc)=\big\{z\in\Z_L^{d\#\Rc_B}:\ z_{j}\ne 0 \Leftrightarrow\sum_{i\in\Rc_B}\al_{ji}^{\gF_\Rc} z_i\ne 0 \quad \forall j\in\Rc_B\big\},
\ee
cf. \eqref{ZZZ}. Then 
\be\lbl{xi-z_a}
\cB'^L_{s}(\gF_\Rc)=\{\xi^\Rc(z):\, z\in\cZ^+(\gF_\Rc)\},
\ee
where the function $\xi^\Rc(z)$ is defined via the $s$-dependent transformation $z\mapsto x^\Rc \mapsto \xi^\Rc$, given by a natural 
modification of  formula  \eqref{xi,s(z)}.

According to Lemma~\ref{l:set_B}(ii) the vector $z$ forms coordinates on the set $\cB'^L_s(\gF)$ as well. The functions $\om_j^\gF(\xi)$ from \eqref{deJ}, restricted to the set $\cB_s'^L(\gF)\ni\xi$ and written in the $z$-coordinates, take the form
\be\lbl{om_j^F(z)-aaa}
\om_j^\gF(z)=2z_j\cdot \sum_{i\in\Rc_B} \al_{ji}^{\gF_\Rc} z_i \ \mbox{ for } j\in\Rc_B
\qnd \om_j^\gF(z)\equiv 0 \ \mbox{ for } j\notin\Rc_B.
\ee
Indeed, the first equality follows by literally repeating the proof of Lemma~\ref{l:Om-change}. 
The second identity immediately follows from Lemma~\ref{l:set_B}(i).

Now, using Lemma~\ref{l:set_B}(ii) and \eqref{xi-z_a}, we write $\xi\in\cB'^L_s(\gF)$ as $\xi=\xi(z)$ in the integral \eqref{deJ}, and similarly to Theorem~\ref{p:aman} we get
\begin{proposition}
	\lbl{p:aman-aaa}
	For any integers $m,n\geq 0$ satisfying $N=m+n\geq 1$, $s\in\Z^d_L$ and $\tau_1,\tau_2\geq -T$, we have
	\be\lbl{final_Ea^ma^n-aaa}
	\EE\fa_s^{(m)}(\tau_1)\bar \fa_s^{(n)}(\tau_2)=\sum_{\gF\in\gF^+_{m,n}}c_\gF\,\mathcal{J}_s(\gF),
	\ee
	where 
	\be
	\lbl{J(F)-z-aaa}
	\cJ_s(\tau;\gF)=
	\int_{\R^N}
	dl\, 
	L^{-Nd}\sum_{z\in\cZ^+(\gF_\Rc)}\, \mathcal{F}_s^\gF\,(\tau, l, z)\, 
	\Theta(\om^\gF(z),\nu^{-1}l).
	\ee
	Here $\Theta(\om^\gF,\nu^{-1}l)=\prod_{j=1}^N \theta(\om_j^\gF,\nu^{-1}l_j)$, while components $\om_j^\gF$ of the vector $\om^\gF$ are given by \eqref{om_j^F(z)-aaa}.
	The incidence matrix $\al^{\gF_\Rc}$ from \eqref{ZZZ-aaa}, \eqref{om_j^F(z)-aaa} is S-regular. 
	%skew-symmetric and its components satisfy $\al_{ij}^{\gF_\Rc}\in\{0,\pm 1\}$. 
	The real density function $\mathcal{F}_s^\gF(\tau, l, z):=\mathcal{F}^\gF(\tau, l, \xi(z))$ is given by \eqref{FF} 	 and satisfies  
	\be\lbl{F-prop1-aaa}
	|\p_{s}^\mu\p_{z}^{\ka}\mathcal{F}_s^\gF(\tau,l,z)|\leq C_{\mu,\ka}^{\#}(s)C_{\mu,\ka}^{\#}(z)\,e^{-\de \big(\sum_{i=1}^m|\tau_1-l_i|
		+\sum_{i=m+1}^N|\tau_2-l_i|\big)}
	\ee
	with a suitable $\delta=\de_N>0$, for any vectors $\mu\in\Z^d_+$, $\ka\in\Z_+^{d\#\Rc_B}$, and any $s\in\R^d$, $z\in\R^{d\#\Rc_B}$.
\end{proposition} 
Let us note that the matrix $\al^{\gF_\Rc}$ may have zero lines and rows,
 and in difference with the situation in Section~\ref{s:ch} this does not imply that $\cZ^+(\gF_\Rc)=\emptyset$ (cf. Proposition~\ref{l:Az}). That is why we do not introduce an analogue of the subset of diagrams~$\gF^{\,true}_{m,n}$.

\subsubsection{Proof of Proposition~\ref{l:approx_eq}}\lbl{s:aaa-proof}
To establish the proposition it suffices to estimate  the difference $\EE\fa_s^{(m)}(\tau_1)\bar \fa_s^{(n)}(\tau_2)-\EE a_s^{(m)}(\tau_1)\bar a_s^{(n)}(\tau_2)$
since its modulus equals to the l.h.s. of \eqref{corr_v-y}. 
According to Theorem~\ref{p:aman} and Proposition~\ref{p:aman-aaa},
\be\non
\EE\fa_s^{(m)}\bar \fa_s^{(n)}-\EE a_s^{(m)}\bar a_s^{(n)}
=\sum_{\gF\in\gF^+_{m,n}\setminus \gF_{m,n}}c_\gF\,\mathcal{J}_s(\gF)
+\sum_{\gF\in\gF_{m,n}}c_\gF\,\big(\mathcal{J}_s(\gF)-J_s(\gF)\big).
\ee
Here we have used that, according to Proposition~\ref{l:Az}, $J_s(\gF)=0$ for $\gF\in\gF_{m,n}\sm\gF^{\,true}_{m,n}$, so we can replace $\gF^{\,true}_{m,n}$ by $\gF_{m,n}$ in the formulation of Theorem~\ref{p:aman}.

Let us first simplify the second sum. Since $\gF\in\gF_{m,n}$, we have $\gF_\Rc=\gF$,  $\Rc_B=\{1,\ldots,N\}$,  $\#\Rc_B=N$ and $\xi^\Rc=\xi$. Then, in the integral $\cJ_s(\gF)$ the summation it taken over $z\in\cZ^+(\gF)\subset \Z^{dN}_L $,  while in  $J_s(\gF)$ ~-- over $z\in\cZ(\gF)\cap\Z^{dN}_L$. 
Note that $\cZ(\gF)\cap\Z^{dN}_L\subset\cZ^+(\gF)$; more specifically,
$$
\cZ(\gF)\cap\Z^{dN}_L=\{z\in\cZ^+(\gF):\, z_j\ne 0 \; \forall j\},
$$
see \eqref{ZZZ}. We then write 
\be\non
\cJ_s(\gF)=\cJ_s^1(\gF)+\cJ_s^2(\gF),
\ee
where in the integral $\cJ_s^1(\gF)$ the set $\cZ^+(\gF)$ is replaced by $\cZ(\gF)\cap\Z^{dN}_L$, and in  
$\cJ_s^2(\gF)$~-- by $\cZ^+(\gF)\setminus \cZ(\gF)$.

We claim that $\cJ^1_s(\gF)=J_s(\gF)$. Indeed, 
the function $\xi(z)$, entering the integrand of the integral $\cJ^1_s(\gF)$, coincides with the function $\xi(z)$, entering the integrand of $J_s(\gF)$.
Thus, these integrals differ only in the density functions $\cF_s^\gF$ and $F_s^\gF$, where $\cF_s^\gF=(-1)^{r(z)}F_s^\gF$ and 
$r(z)$ stands for the number of indices $1\leq i\leq N$ for which $\de^{=\xi_{2k-1}\xi_{2k}}_{\ \s_{2k-1}\s_{2k}}=1$, $\xi=\xi(z)$.
But $r(z)=0$ for $z\in\cZ(\gF)$, due to Proposition~\ref{al:change}(ii), since $x_{2k-1}=z_k\ne 0$ for every $k$.

We have seen that the second sum from \eqref{A--a} takes the form $\sum_{\gF\in\gF_{m,n}}c_\gF\cJ_s^2(\gF)$, where
\be\lbl{JJ^2}
\cJ_s^2(\gF)= \int_{\R^N}dl\, L^{-Nd}\sum_{\substack{z\in\cZ^+(\gF):\\z_j=0 \mbox{ {\footnotesize for some} }j}}\, \mathcal{F}_s^\gF\,(\tau, l, z)\, 
\Theta(\om^\gF(z),\nu^{-1}l),
\ee
since the set $\cZ^+(\gF)\setminus \cZ(\gF)$ consists of $z\in\cZ^+(\gF)$ satisfying $z_j=0$ for some~$j$. 
Thus, we get
\begin{proposition}\lbl{l:a----A} For any $m,n\geq 0$ satisfying $m+n\geq 1$, any $s\in\Z^d_L$ and $\tau_1,\tau_2\geq -T$,
	\be\lbl{A--a}
	\EE\big(\fa_s^{(m)}(\tau_1)\bar \fa_s^{(n)}(\tau_2)- a_s^{(m)}(\tau_1)\bar a_s^{(n)}(\tau_2)\big) 
	= \!\!\!  \sum_{\gF\in\gF^+_{m,n}\setminus \gF_{m,n}}c_\gF\,\mathcal{J}_s(\gF)
	+\sum_{\gF\in\gF_{m,n}}c_\gF\,\mathcal{J}^2_s(\gF),
	\ee
	where the integrals $\mathcal{J}_s(\gF)$ and $\mathcal{J}^2_s(\gF)$ are given by \eqref{J(F)-z-aaa} and \eqref{JJ^2} correspondingly.
\end{proposition}
We estimate the two sums from \eqref{A--a} separately.
We first consider the case $\gF\in\gF^+_{m,n}\setminus \gF_{m,n}.$ 
According to \eqref{J(F)-z-aaa} and \eqref{F-prop1-aaa},
\be\non
|\cJ_s(\gF)|\leq C^\#(s)L^{-Nd}\sum_{z\in\Z^{d\#\Rc_B}_L} C^\#(z)\leq C_1^\#(s)L^{-(N-\#\Rc_B)d}\leq C_1^\#(s)L^{-d},
\ee 
since the diagram $\gF$ has at least one irregular block, that is $\#\Rc_B<N$. 
Thus, the first sum from \eqref{A--a} is bounded by $C^\#(s)L^{-d}$. 

Now we consider the case $\gF\in\gF_{m,n}$. According to \eqref{F-prop1-aaa},
\be\non
|\cJ^2_s(\gF)|\leq C^\#(s) L^{-Nd} 
\sum_{\substack{z\in\Z^{dN}_L:\\z_j=0 \mbox{ {\footnotesize for some} }j}} C^\#(z) \leq C_1^\#(s)L^{-d}.
\ee
Thus, the second sum from \eqref{A--a} is bounded by $C^\#(s)L^{-d}$ as well. 
This concludes the proof of 
Proposition~\ref{l:approx_eq}.

\subsubsection{Example} 
In our work  \cite{DKMV} we apply the theory,  developed in this section, to the situation when the function $\theta(x,t)$ is not given by \eqref{theta-def}
(as in Theorem~\ref{th:Ea^ma^n-cont}), but is the indicator function $\mathbb{I}_{\{0\}}(x)$
 of the point $x=0$ (so it is independent from $t$). Below we briefly discuss this case.  Again our goal is to  estimate the difference 
$\EE\fa_s^{(m)}(\tau_1)\bar \fa_s^{(n)}(\tau_2)-\EE a_s^{(m)}(\tau_1)\bar a_s^{(n)}(\tau_2)$
when $\theta(x,t)=\mathbb{I}_{\{0\}}(x)$. 

As before,  we estimate the two sums from \eqref{A--a} separately.
Let first $\gF\in\gF^+_{m,n}\setminus \gF_{m,n}$. According to \eqref{F-prop1-aaa}, 
\be
|\cJ_s(\gF)|\leq C^\#(s)L^{-Nd}
\sum_{\substack{z\in\cZ^+(\gF_\Rc):\\ \om_k(z)=0 \,\forall k\in\Rc_B}}C^\#(z).
\ee
Assume now $\gF\in\gF_{m,n}$. Then, 
\be
|\cJ_s^2(\gF)|\leq C^\#(s)L^{-Nd}
\sum_{\substack{z\in\cZ^+(\gF):\\  z_j=0 \mbox{ {\footnotesize for some} }j, \\
		\om_k(z)=0 \,\forall 1\leq k\leq N}}C^\#(z).
\ee

\subsection{Integrals $\wt J_s(\gF)$ in which the time variable $l$ is integrated out}
\lbl{s:time out}

In this appendix we integrate out the time variable $l$ in the integrals $\wt J_s(\gF)$, which are defined in \eqref{J_c(F)}.
We believe that it is relevant for further study of the energy spectrum \eqref{spec_form} of quasisolutions since for the moment of writing
 we know asymptotical as $\nu\to 0$ behaviour of the integrals $\wt J_s(\gF)$ only in case $N=2$, and to find it we have to write the integrals in such integrated in $l$ form.
See in \cite{DK}, where we found this asymptotic and, using a particular case of Theorem~\ref{th:Ea^ma^n-cont}, deduced from it the asymptotical as $\nu\to 0,\,L\to\infty$ behaviour of the moment $\EE|a_s^{(1)}|^2$.
Note, however, that in case of general $N$, even to get an upper bound for the integrals $\wt J_s(\gF)$ in Theorem~\ref{t:est wt_J} we write them in the form \eqref{J_c(F)} and we do not know how to get the upper bound using their integrated in $l$ form.

For simplicity of computation we assume $\tau_1=\tau_2=:t$ and $T=\infty$. 
We denote by $\cS^N$ the set of all permutations $q$ of the set $\{1,\ldots,N\}$ and let
$$
\Lc_q^N:=\{l\in\R^N:\,-\infty\leq l_{q(N)}\leq \ldots \leq l_{q(1)}\leq t\}, \qmb{so that}\qu \R_l^N=\bigcup_{q\in\cS^N} \Lc_q^N,
$$
and the sets $\Lc_q^N$ with different $q$ can intersect only over sets of zero Lebesgue measure.
Then, in view of \eqref{F=},
\be\label{19.55}
\wt J_s(\gF)=\int_{\R^{dN}}
\prod_{\psi\in E_L(\gF)} \frac{(b^{\psi}(z))^2}{\ga^{\psi}(z)}
\sum_{q\in\cS^N} I_s^q(\gF;z) \, dz\,
,
\ee 
where
\begin{align}\non
I_s^q(\gF,z)=\int_{\Lc_q^N} e^{i\nu^{-1}\Om^\gF(l,z)}  
\prod_{\phi\in E_D(\gF)} e^{-\ga^{\phi}(l^{\phi}_r - l^{\phi}_w)}&\mI_{\{-T\leq l^{\phi}_w \leq l^{\phi}_r\}}(t,l)
\\\lbl{I_g}
&\prod_{\psi\in E_L(\gF)} 
e^{-\ga^{\psi}|l^{c_\psi}-l^{\bar c_\psi}|}\,dl,
\end{align}
where $b^\psi(z)=b^\psi(\xi(z))$, $\ga^\psi(z)=\ga^\psi(\xi(z))$ and $\ga^\phi(z)=\ga^\phi(\xi(z))$.
It remains to compute the integrals $I_s^q$, and for this purpose we introduce the following notation.
For $0\leq i<j\leq N$ we set 
\be\non
\ga_{ij}^\gF:=\sum_{\vartheta:\,\hat B_{i}\volna \hat B_{j}} \ga_\vartheta \geq 0,
\ee
where the sum is taken over all edges $\vartheta\in E(\gF)$ of the diagram $\gF$ which couple a vertex from the block $\hat B_{i}$ with a vertex from the block $\hat B_{j}$.
Here by $\hat B_{0}$ we denote the set of roots $\{c_0,\bar c_0\}$. 

Let $\cS^N_\gF\subset\cS^N$ be the set of permutations $q$ for which 
the relation $l\in\Lc^N_q$ implies 
\be\non
l^{\phi}_w \leq l^{\phi}_r \qmb{for all}\qu \phi\in E_D(\gF), 
\ee
cf. the indicator function in \eqref{I_g}.

Below we denote $q(0):=0$.
\begin{lemma}\lbl{l:int in time}
	Assume $\tau_1=\tau_2=t$ and $T=\infty$. Then for any $m,n\geq 0$ satisfying $N:=m+n\geq 2$   and $\gF\in\gF_{m,n}$ the integral $\wt J_s(\gF)$ is given by \eqref{19.55}, where
	\be\lbl{I_g=}
	I_s^q(\gF;z)=\frac{1}{2\ga_s}\prod_{k=2}^N \left(i\nu^{-1} \sum_{j=k}^N \om^\gF_{q(j)}(z) + \sum_{0\leq r<k\leq j\leq N} \ga^\gF_{q(r)q(j)}(z)  \right)^{-1}
	\ee
	if $q\in\cS^N_\gF$ and $I_s^q(\gF;z)\equiv 0$ otherwise. 
	Moreover, $\sum_{0\leq r<k\leq j\leq N} \ga^\gF_{q(r)q(j)} \geq 1$
	for any $2\leq k \leq N$, so the denominator in \eqref{I_g=} is separated from zero.
\end{lemma}
The lemma can be proven by induction.

Note that outside the intersection of  quadrics
$$
Q:=\{z: \,\om^\gF_{j}(z) =0 \quad\text{for} \;\; 1\le j\le N\}
$$
we have $I_s^q(\gF;z) \sim \nu^{N-1}$, while on $Q$ we have  $I_s^q(\gF;z) \sim 1$. So the integrand 
 in \eqref{19.55} asymptotically degenerates on $Q$ as $\nu\to0$, and the integral  in \eqref{19.55}  is asymptotically singular.

\subsection{Estimation of asymptotically singular 
 integrals of quotients.}
\lbl{s:est int quadr}

In Lemma~\ref{l:int in time} we integrated out the time variable $l$ in 
 integrals $\wt J_s(\gF)$, defined in \eqref{J_c(F)}, and  represented $\wt J_s(\gF)$ as sums of integrals of quotients with large quadratic forms in divisors, see  \eqref{19.55},\eqref{I_g=}. In Theorem~\ref{t:est wt_J} we got an upper bound for a family of integrals of quotients that have more general form than  the integrals $\wt J_s(\gF)$. As a corollary, in this appendix  we obtain an upper bound for a natural family of integrals 
%of quotients with large quadratic forms in divisors 
of more general form than those in  \eqref{19.55}, \eqref{I_g=}.
%This family is natural and we believe that the obtained result is of independent interest.
%For simplicity, we do not follow the dependence on $s$.

As in Section \ref{s_6.1}, we consider the family of quadratic forms
$$
(Q_k z)\cdot z = \sum_{i,j=1}^M q_{ij}^k z_i \cdot z_j, \quad 1\le k \le K, 
$$
where $z$ is the polyvector $z=(z_1, \dots, z_M), z_j \in \R^d$, and $Q_k=(q_{ij}^k)_{1\leq i,j\leq M}$ are real symmetric matrices.
As before, we denote by $R$ the rank of the system of matrices $Q_1, \dots, Q_K$ in the space of $M\times M$-matrices, equal to the rank of the system of
vectors
$$
q_{ij} \in \R^K, \quad 1\le i,j\le M, \;\; q_{ij} = (q_{ij}^1, \dots, q_{ij}^K). 
$$

Let
\be\lbl{J^nu-def}
J^\nu=\int_{\R^{dM}} \frac{G(z)}{\prod_{k=1}^K \big(i\nu^{-1}(Q_k z)\cdot z +\Ga_k(z)\big)}\,dz,
\ee
where $0<\nu\leq 1/2$, $G$ is a Schwartz function, the functions $\Ga_k$ are smooth, have at most polynomial growth at infinity together with their partial derivatives of all orders, and satisfy $\Ga_k(z)\geq C_{\Ga}$ for some constant $C_\Ga>0$ and any $k$ and $z$.

%Denote by $r$ the rank of the linear span  $\lan A_1,\ldots,A_K\ran$ in the space of $M\times M$ real matrices.
\begin{proposition}\lbl{t:est int quadr}
	Assume that $\operatorname{tr} Q_k=0$ for every $1\leq k\leq K$. Then 
	\be\lbl{J^nu}
	|J^\nu|\leq C\nu^{\min(R,d)}\psi^R_d(\nu),
	\ee
	where the function $\psi^R_d$ is defined in \eqref{psi_d^k}.
\end{proposition}
 See \cite{Dym} for a generalization of this result.
In \cite{K} the asymptotic behaviour as $\nu\to 0$ of the integral $J^\nu$ was found in the case when $d\geq2$ is a pair number,
 $K=2$, $\Ga_1=\Ga_2$ and $Q_1=-Q_2$, where the quadratic form $Q_{1}$ was assumed to be non-degenerate and to have the index of the form $(m,m)$, for some $m$.
This result plays a crucial role in \cite{DK} when analysing the asymptotic behaviour of the term $n^{(2)}$ from the decomposition \eqref{spec cut off}.
In particular, it was shown that $J^\nu\sim\nu$ which is in accordance with the estimate \eqref{J^nu}.
The family of integrals \eqref{J^nu-def} is significantly more complicated than that in \cite{K}.
 We get   for them only the upper bound and for the moment do not now how to examine  their asymptotic behaviour.
\smallskip
  
{\it Proof of Proposition \ref{t:est int quadr}.}
Since $\Ga_k(z)>0$ for any $z$ and $k$, we have
\be\lbl{int time}
\frac{1}{i\nu^{-1}Q_k z\cdot z +\Ga_k(z)}=\int_{-\infty}^0 e^{l_k\big(i\nu^{-1}Q_k z\cdot z +\Ga_k(z)\big)}\, dl_k.
\ee
Denote $l=(l_1,\ldots,l_K)$, 
$$Q(l,z)=\sum_{k=1}^K   l_k  Q_k z\cdot z  \,\qnd
F(l,z)=G(z)\, e^{\sum_{k=1}^K\Ga_k(z)l_k}\, \mI_{\{l_k\leq 0\,\forall k\}}(l).$$
Then, using representation \eqref{int time} in \eqref{J^nu-def} for every $k$, we get
$$
J^{\nu}=\int_{\R^K}dl \, \int_{\R^{dM}}dz \; F(l,z)\,e^{i\nu^{-1}Q(l,z)}. 
$$
Since 
%Setting $q_{ij}=(q_{ij}^1,\ldots,q_{ij}^K)$, we find
$$
Q(l,z)=\sum_{k=1}^K l_k\sum_{i,j=1}^M q_{ij}^k \, z_i\cdot z_j=\sum_{i,j=1}^M (q_{ij}\cdot l) \, (z_i\cdot z_j),
$$
then the integral $J^\nu$ has the form \eqref{J===}, where the density $F$ is independent from the parameter $s$ (one can think that in \eqref{J===} we fix some $s=s_0$ and do not follow the dependence of estimates in $s_0$).
It is straightforward to see that assumptions on the functions $G$ and $\Ga_k$ imply that the function $F$ satisfies \eqref{F_in_z} with $l_0=0$ and $h_\ka(x)=C_\ka e^{-\de C_\Ga x}$, where the constant $\de >0$ is sufficiently small and $C_\ka>0$ are sufficiently large.

As $\operatorname{tr} Q_k=0$ for any $k$, then the assumptions of Theorem~\ref{t:est wt_J} are fulfilled and  integral $J^\nu$ satisfies  estimate \eqref{st_ph est} (with $C^\#(s)$ replaced by $C$).
\qed

\end{document}